\documentclass[12pt,a4paper]{article}
\usepackage[latin1]{inputenc}
\usepackage{amsmath}
\usepackage{amsfonts}
\usepackage{amssymb}
\usepackage{mathtools}
\usepackage{mathrsfs}
\usepackage{amsmath}

\usepackage{graphicx,a4wide}
\usepackage{verbatim,cite}

\usepackage{tikz}
\usepackage{color}

\usetikzlibrary{positioning}
\usepackage{float}
\usepackage{braket}

\DeclareMathAlphabet{\mathpzc}{OT1}{pzc}{m}{it}
\usepackage{breqn}

\usepackage{setspace}
\usepackage{simplewick}

\usepackage[mathscr]{euscript}

\usepackage[hidelinks]{hyperref}

\def\cO{{\mathcal O}}
\def\cN{{\mathcal N}}

\DeclareMathOperator{\Tr}{Tr}

\usepackage{booktabs}
\usepackage[vcentermath]{youngtab}
\usepackage{float}

\newcommand{\beq}{\begin{equation}}
\newcommand{\eeq}{\end{equation}}
\newcommand{\bea}{\begin{eqnarray}}
\newcommand{\eea}{\end{eqnarray}}

\def \be  {\begin{equation}}
\def \ee  {\end{equation}}
\def \ba  {\begin{eqnarray}}
\def \ea  {\end{eqnarray}}

\begin{document}

\thispagestyle{empty}

\null\vskip-43pt \hfill
\begin{minipage}[t]{30mm}
\end{minipage}


\vskip.0truecm
\begin{center}
	\vskip 0.2truecm {\Large\bf
		{\Large Single Particle Operators and their Correlators\\[.3cm] 
		in Free 
			 $\mathcal{N}=4$ SYM}
	}\\
	\vskip 1truecm
	{\bf F.~Aprile${}^{1}$, J.~M. Drummond${}^{2}$, P.~Heslop${}^{3}$, \\
	H.~Paul${}^{2}$, F.~Sanfilippo${}^{4}$, M.~Santagata${}^{2}$, A.~Stewart${}^{3}$ \\
	}
	
	\vskip .6truecm
	
	{\it
		${}^{1}$ Dipartimento di Fisica, Universit\`a di Milano-Bicocca \& INFN, 
		Sezione di Milano-Bicocca, I-20126 Milano,\\
		\vskip .2truecm }
	\vskip .15truecm
	{\it
		${}^{2}$ School of Physics and Astronomy and STAG Research Centre, \\
		University of Southampton,
		Highfield,  SO17 1BJ,\\
		\vskip .2truecm                        }
	\vskip .1truecm
	{\it
		${}^{3}$ Mathematics Department, Durham University, \\
		Science Laboratories, South Rd, Durham DH1 3LE, \vskip .2truecm                        }
	\vskip .05truecm
	{\it ${}^{4}$
		INFN Sezione di Roma 3, Via della Vasca Navale 84, I-00146 Roma.
	\vskip .2truecm                       }
\end{center}

\vskip 1truecm 
\centerline{\bf Abstract} 

We consider a set of half-BPS operators in $\mathcal{N}=4$ super Yang-Mills theory which are appropriate 
for describing single-particle states of superstring theory on AdS${}_5 \times S^5$. 
These single-particle operators are defined to have vanishing two-point functions 
with all multi-trace operators and therefore correspond to admixtures of single- and multi-traces.  
We find explicit formulae for all single-particle operators and for their two-point 
function normalisation. We show that single-particle $U(N)$ operators belong to 
the $SU(N)$ subspace, thus for length greater than one they are simply the $SU(N)$ 
single-particle operators. Then, we point out that at large $N$, as the length of the 
operator increases, the single-particle operator naturally interpolates between the 
single-trace and the $S^3$ giant graviton. 
At finite $N$, the multi-particle basis, obtained by taking products of the single-particle operators, 
gives a new basis for all half-BPS states, and this new basis naturally cuts off  when 
the length of any of the single-particle operators exceeds the number of colours. 
From the two-point function orthogonality we prove a multipoint orthogonality 
theorem which implies vanishing of all near-extremal correlators.
We then compute all maximally and next-to-maximally extremal free correlators,
and we discuss features  of the correlators when the extremality is lowered.
Finally, we describe a half-BPS projection  of the operator product expansion on 
the multi-particle basis which provides an alternative construction of four- and 
higher-point functions in the free theory.

\medskip

\noindent

\newpage
\setcounter{page}{1}\setcounter{footnote}{0}
\tableofcontents
\newpage


\section{Introduction}

As the most symmetric QFT in four dimensions, and at the same time the  
paradigmatic example of the AdS/CFT correspondence,  $\cN$=4 super Yang-Mills (SYM) 
has been the subject of a huge amount of interest. The expected 
flow of information uses semiclassical physics in AdS to reconstruct strong 't Hooft coupling 
phenomena in the gauge theory. Instead, recently this flow has been reversed, with  precise 
and concrete investigations of perturbative quantum gravity undertaken
by means of analytic bootstrap techniques at strong coupling.
For example one-loop quantum gravity amplitudes in AdS have been obtained by computing 
$O(1/N^4)$ corrections to strong coupling correlators in SYM~\cite{Aprile:2017bgs,Alday:2017vkk,Aprile:2017qoy,Aprile:2019rep,Alday:2019nin}.
  
One aspect of these investigations was the need to be very careful about the 
precise definition of the gauge theory operators dual to single-particle supergravity states, 
in order to properly account for the OPE at tree level and one loop. 
These single-particle operators are protected half-BPS operators, but the space of half-BPS operators of given 
charge 
is degenerate, and only in the planar limit can the single-particle operators be identified  with the 
single-trace half-BPS operators Tr$(\phi^p)$. This identification indeed was known to receive 
$O(1/N)$ multi-trace corrections (see eg \cite{Arutyunov:1999en,Arutyunov:2000ima,Rastelli:2017udc}),
and the first order double trace corrections have recently been fully worked out directly 
from supergravity in~\cite{Arutyunov:2018neq,Arutyunov:2018tvn}. On the other hand, the 
non-perturbative nature of the AdS/CFT correspondence strongly points towards  
a non-perturbative (i.e.~valid to all orders in $N$) definition of the single-particle states.
In fact, a deceptively simple non-perturbative definition  was formulated in~\cite{Aprile:2018efk}:
\begin{itemize}
	\item Single-particle operators are {half-BPS operators which have vanishing two-point functions with all multi-trace operators}. 
\end{itemize} 
This definition fixes the single-particle operators uniquely, up to normalisation, in terms of a 
certain admixture of single- and multi-trace operators. In particular, the leading operator in the 
planar limit is the single-trace operator, but the admixture is new. 
Our definition has been already used, and indeed is crucial in correctly determining  one-loop  
$O(1/N^4)$ SUGRA correlators of operators with charges higher than four, 
for example,  arbitrary charge correlators in position space  [4], where a number of subtle features have been solved, 
and $22pp$ correlators in Mellin space \cite{Alday:2019nin}.

Our first result in this paper is to obtain explicit formulae for the multi-trace components of the 
single-particle operators and examine some of their nice properties. Then, we will study some 
of their correlators and show that compared to the single-trace cousins, the single-particle 
operators have a number of very surprising and nice properties. This is slightly counter-intuitive at first, 
because now we have to deal with an admixture  of single and multi-trace operators, but nevertheless 
it is true in many ways, as we will demonstrate. 

One of the nice properties is that the operators in the $U(N)$ theory and the $SU(N)$ theory are the same.  
Indeed, in the $U(N)$ theory the single-particle operators of 
charge 
greater equal than two must be orthogonal  to all multi-trace operators involving also Tr$(\phi)$, 
and this automatically makes them the $SU(N)$ operators. To formalise this statement we 
introduce the $SU(N)$ projection on the space of the $U(N)$ operators,  and show that 
the $U(N)$ single-particle operators belong to the $SU(N)$ subspace, which is orthogonal 
to the span of multi-trace operators in which at least one trace is Tr$(\phi)$. It follows that 
correlators of $U(N)$ single-particle operators are equal to correlators of $SU(N)$ single-particle operators.

Another nice property of the single-particle operators is that they automatically vanish as the
charge 
of  the operators exceeds the number of colours $N$. This should be contrasted for example with 
the single trace basis which do not vanish, but rather become complicated linear combinations 
of products of lower trace operators.  This property of the operators dual to single particle states 
has long been expected from AdS/CFT and follows from the string exclusion principle~\cite{Maldacena:1998bw}. 
As the angular momentum of the gravitons increases they become less and less pointlike, 
eventually  growing  into giant gravitons, D3 branes wrapping an $S^3 \subset S^5$~\cite{McGreevy:2000cw} 
which can not grow bigger than the size of the $S^5$ sphere. In~\cite{Balasubramanian:2001nh} 
(sub)-determinant half-BPS operators were defined as duals to these predicted sphere giants and 
shortly later the Schur polynomial basis of half-BPS operators was defined and the sphere giant 
gravitons associated with the completely antisymmetric (single column Young tableau) 
Schur polynomials~\cite{Corley:2001zk}. We find that at large $N$, the single-particle operators 
with charge close to $N$ indeed approach these (sub)-determinant operators.

In fact these operators have appeared before in the literature, and not identified 
as operators dual to single particle states.  A beautiful orthogonal 
basis for all half-BPS operators in the $U(N)$ theory  was given in \cite{Corley:2001zk} in terms of 
Schur polynomials, labelled by Young tableaux. These also automatically truncate with $N$ since a 
$U(N)$ Young tableau with height larger than $N$ vanishes. However this basis does not project  
onto an orthogonal  basis for $SU(N)$ (and indeed the operators are not even linearly independent 
in the $SU(N)$ theory). In~\cite{deMelloKoch:2004crq} a non-orthogonal but linearly independent basis 
of all $SU(N)$ half-BPS operators was defined. This basis was later identified as the dual (via the metric 
defined by the two point function) to the trace basis in~\cite{Brown:2007bb} and a group theoretic 
expression for this dual  basis was given. The single particle operators we discuss here are a subset 
of the dual basis: the operators  dual to single trace operators.

It is the purpose of this paper to unpack the definition of single-particle operators, turn it into an explicit formula, 
and investigate its properties. The outline will be as follows.

In section \ref{section_2} we discuss the details of the multi-trace admixture which defines the single-particle operators. 
We first give explicit examples at low charge, and then we use group theory techniques to obtain a general formula, 
valid for any single-particle operator of any dimension. This leads us to very explicit formulae for two-point functions 
of single-particle operators. 

In  section \ref{section_3} we uplift the defining two-point function orthogonality to a multi-point orthogonality theorem, 
which in turn implies vanishing of a large class of diagrams in correlators. We call these near-extremal 
$n$-point functions, where extremality will be defined as a measure of how much the diagram is connected 
w.r.t.\! the heaviest operator (see \eqref{intro_near_extr}). This is the first instance of hidden simplicity 
of multi-point single-particle correlators versus the single-trace correlators, 
and very interestingly, a similar feature was noticed on the (super-)gravity side in \cite{DHoker:2000xhf}. 

In section \ref{section_4}, 
we consider the first non vanishing correlators, and we study
maximally-extremal (ME) and next-to-maximally extremal (NME) $n$-point functions.
Both are simple. The ME correlators are computed by trees and two point functions. The NME are mostly 
computed by weighted sums of ME correlators, which we know in general. 
When we compute these correlators by using Wick contractions techniques on the trace basis, 
the combinatorics is hard in the intermediate steps. Instead, the final result is way much simpler. 
We provide more evidence about this mechanism  mentioning also the case of NNME three-point functions.

Finally, in section \ref{hbpsope} we discuss the half-BPS OPE on the single-particle basis, 
and we use it as an alternative computational tool to bootstrap, and provide constraints, 
on the correlators of interest. We exemplify the procedure at four-points for 
next-to-next- and next-to-next-to-next extremal correlators.

\section{Single-particle half-BPS operators (SPOs)}\label{section_2}

Half-BPS operators in $\mathcal{N}=4$ super Yang-Mills theory can be described via products 
of single-trace scalar operators in the symmetric traceless representations of $SO(6)$,
\be
T_p(x) = {\text{Tr}}\, \phi(x)^p\ \qquad;\qquad {\phi}(X,Y)=Y^{R}\phi_{R}(X)
\ee
Here $\phi_{R}$ is the elementary field of the $\mathcal{N}=4$ supermultiplet, 
and we have introduced an auxiliary $SO(6)$ vector $Y^R$ which is null, $Y\cdot Y=0$,
to project onto the symmetric traceless representation. An insertion point $x_i$ corresponds 
to both a space time $X_i$ coordinate and an $SO(6)$ vector $Y^R$.

In addition to the single-trace operators $T_p$ we obtain other half-BPS operators from products of the form
\be\label{multi_trace_def}
T_{p_1 \ldots p_m}(x) \equiv T_{p_1}(x) \ldots T_{p_m}(x)\,, \qquad p_1 \geq \ldots \geq p_m\ge 1\,.
\ee
The scaling dimension of $T_{p_1 \ldots p_m}$ is given by $(p_1 + \ldots + p_m)$. 
The case $m=1$ reduces obviously to the single trace operator.

We will refer to the basis of half-BPS operators made of all possible $T_{p_1 \ldots p_m}(x)$ 
as the \emph{trace basis}, and we will denote the basis elements with the symbol $T_{ \underline{p} }$, 
where $\underline{p}$ stands for a partition of $p$, so if the partition has length greater than 
two the operator is multi-trace. 

To compute correlation functions in free field theory we use elementary propagators. 
Here we will take the gauge group to be $U(N)$ or $SU(N)$ so that the propagator takes the form,
\begin{align}
\Big\langle \phi_{r}{}^{{s}}(X_1,Y_1) \phi_{t}{}^{{u}}(X_2,Y_2)\Big\rangle &= \delta_r^{ u} \delta_t^{ s}\,  g_{12}\qquad  && U(N) \,\label{unprop}\\
\Big\langle \phi_{r}{}^{{s}}(X_1,Y_1) \phi_{t}{}^{{u}}(X_2,Y_2)\Big\rangle &= \Bigl(\delta_r^{ u} \delta_t^{ s} - \frac{1}{N} \delta_r^{ s} \delta_t^{ u} \Bigr)\, g_{12} \qquad  && SU(N)\,
\label{sunprop}
\end{align}
where
\be
\quad\quad g_{12} = \frac{ Y_1 \cdot Y_2 }{(X_1-X_2)^2}\, 
\ee
A correlation function of operators in the trace basis will have the schematic form
\begin{align}
\langle T_{ \underline{p_1} }(x_1)\ldots T_{ \underline{p_n} }(x_n) \rangle =  \sum_{  \{ b_{ij} \}  } \prod_{i,j} g_{ij}^{b_{ij} }\ \mathcal{C}_{\{ b_{ij} \},\, \underline{p_1} \ldots \underline{p_n}  }(N)
\end{align}
where $b_{ij}$ counts the number of propagators from insertion point $i$ to $j$, and the collection of these bridges, 
$ \{ b_{ij} \}_{i<j}$, thus labels the propagator structure. With $\mathcal{C}_{\{ b_{ij} \} }(N)$ we
denote the corresponding color factor. 

\subsection{Definition and low charge examples}

The AdS/CFT correspondence maps the spectrum of operators in $\mathcal{N}=4$ super 
Yang-Mills theory to the spectrum of IIB superstring theory on 
the AdS${}_5\times S^5$ background. The superstring can be found in unexcited states 
(giving the IIB supergravity multiplet) or excited states. The half-BPS operators correspond 
to the supergravity states (and their multi-particle products). 

In the natural basis of scattering states, the multi-particle states should be orthogonal to single-particle ones. 
A key point of our discussion is that the trace basis of half-BPS operators 
is not an orthogonal basis with respect to the inner product given by the two-point functions. In general 
\be
\langle T_p(x_1) \,T_{q_1 \ldots q_n}(x_2) \rangle \neq 0 \qquad n\geq 2\ .
\ee

To align with the AdS/CFT intuition, in~\cite{Aprile:2018efk,Aprile:2019rep} we gave 
a prescription for identifying which half-BPS operators correspond to the single-particle states.  
Our definition of the relevant operators $\mathcal{O}_p$ is simply to take those operators 
which are orthogonal to all multi-trace operators,
\be
\label{orthogonality}
\text{Single particle operators } \equiv \left\{ \cO: \ \langle \mathcal{O}_p(x_1) \,T_{q_1\ldots q_n}(x_2) \rangle = 0\,, \quad (n \geq 2)\right\}\,.
\ee
This in turn implies $\langle \mathcal{O}_{p}(x_1) [\cO_{q_1}\ldots \cO_{q_n}](x_2)\rangle=0$,
 i.e.~single-particle operators are orthogonal to multi-particle states.

From the Gram-Schmidt orthogonalisation we immediately obtain 
a formula for $\cO_p$ in terms of the (color factor of) two-point functions.
In fact, consider the space of operators of charge $p$ and label them by corresponding partitions 
$\{ \lambda_{i}=q_1\ldots q_n\}_{i=1,\ldots P}$, with $\lambda_P\equiv\{ p\}$. 
Then, 
\be\label{det_rep}
\cO_p(x)=  \frac{1}{\mathcal{N}_p} \det\left( \begin{array}{llcl}  
\mathcal{C}_{\lambda_1| {\lambda_1} }			& 	\mathcal{C}_{\lambda_2| {\lambda_1} }				& \ldots 		\ \	&\mathcal{C}_{ {p}|\lambda_1 }  \\ 
		\vdots  							& 	\vdots										& \ldots 		\ \	& \vdots \\  
\mathcal{C}_{{\lambda_1}|\lambda_{P-1}  }		&  	\mathcal{C}_{\lambda_{2}| {\lambda_{P-1}} }			& \ldots 		\ \	&   \mathcal{C}_{p|\lambda_{P-1} } \\[.2cm]
\hline
\rule{0pt}{.5cm}
T_{\lambda_1} (x)								& 	T_{\lambda_2} (x)									& \ldots			& T_{p} (x)\end{array}
\right)   
\ee
where 
\be
\mathcal{N}_p=\det \left(  \mathcal{C}_{\lambda_i\lambda_j} \right)_{1\leq i,j\leq P-1} \qquad;\qquad  \langle T_{\lambda_i}(x_1)T_{\lambda_j}(x_2)\rangle= g_{12}^p \, \mathcal{C}_{\lambda_i\lambda_j}
\ee

The determinant \eqref{det_rep} is understood as a Laplace expansion  about the last row, 
which contains the list of all allowed operators of charge $p$. Upon using the determinant
 for computing $\langle \cO_p(x_1) T_{q_1\ldots q_n}(x_2)\rangle$, it becomes obvious that the 
 last row of the matrix would now coincide with another row, leading to a vanishing result.  

In our normalisation $\cO_p$ coincides with the single-trace operators $T_p$ up to 
multi-trace admixtures. Each multi-trace contribution is suppressed at large $N$, 
and the single-particle operator reduces to single-trace operators in the strict large $N$ limit.  
However, the novelty of our SPO is precisely the determination of the multi-trace admixture, 
which we will now exemplify with a number of computations, for low charge operators, 
before discussing our general formulas in the next section.

With  $SU(N)$ gauge group, $T_p$ and $\mathcal{O}_p$ coincide for $p=2,3$ 
since there are no multi-trace operators for charges $p<4$,
\begin{align}\label{sun23}
{\mathcal{O}}_p &= T_p \,\qquad \text{ for } p =2,3 \qquad \qquad &&SU(N)\,.
\end{align}
In terms of supergravity states the $p=2$ case corresponds to the superconformal 
primary for the energy-momentum multiplet which is dual to the graviton multiplet in 
AdS${}_5$. The $p=3$ case is the first Kaluza-Klein mode arising from reduction of 
the IIB graviton supermultiplet on $S^5$.

For 	gauge group $U(N)$ on the other hand there exists a $p=1$ operator, since the
 trace of the fundamental scalar no longer vanishes. At weight 1 there is only one operator 
 and thus $\cO_1=T_1$.  However using Wick contractions with the $U(N)$ propagator~\eqref{unprop} 
 one can easily verify that for $p>1$, as well as the single-trace contributions, we have non-trivial 
 multi-trace admixtures. At weight 2 there are two operators, $T_2$ and $T_{11}$ with  
 the single particle operator defined to be orthogonal to the double-trace $T_{11}$.
Similarly for $p=3$ the single particle operators is defined to be orthogonal to both 
the double-trace $T_{21}$ and the triple trace $T_{111}$. Explicitly we obtain
\begin{align}\label{un_23}
	\begin{array}{ll}\cO_2 &= T_2 -\tfrac 1N T_{11} \\[.2cm]
	\cO_3 &= T_3 -\tfrac{3 }{N}T_{21}+\tfrac{2 }{N^2}T_{111}\end{array} \qquad \qquad U(N)\,.
\end{align}
where for example the coefficient of the double-trace contribution to $\cO_2$ 
is determined from the orthogonality condition $ \langle \mathcal{O}_2(x_1) T_{11}(x_2) \rangle = 0$.
The additional terms compared to the $SU(N)$ operators~\eqref{sun23} in fact simply project out 
the trace part of the fundamental scalar $\phi$ and so the $SU(N)$ and $U(N)$ operators in fact coincide. 

For $p>3$ we have non-trivial multi-trace admixtures even in the $SU(N)$ case as can be 
easily verified using $SU(N)$  Wick contractions (i.e.~with propagator~\eqref{sun23}). 
This was discussed in\cite{Aprile:2018efk} (see also previous discussions in \cite{Arutyunov:1999en,DHoker:1999jke,Rastelli:2017udc}). 
For example, the single-particle operator for $p=4$ is given by
\begin{align}\label{sun_2}
\mathcal{O}_4 &= T_4 - \frac{2N^2-3}{N(N^2+1)} T_{22} &&SU(N)\,,
\end{align}
where the coefficient of the double-trace contribution determined from 
the orthogonality condition $\langle \mathcal{O}_4 T_{22} \rangle = 0$.
It is now interesting to consider the single particle operator in the $U(N)$ theory. 
It is given by 
\begin{align}\label{un_4}
	\cO_4 = T_4 - \frac{\left(2
		N^2-3\right)}{N
		\left(N^2+1\right)} T_{22}+ \frac{10 }{N^2+1}T_{211}
	-\frac{4 }{N}T_{13}-\frac{5
		}{N
		\left(N^2+1\right)}T_{1111} \qquad U(N)\,,
\end{align}
where the coeffcients are determined by demanding orthogonality with 
all higher trace operators $T_{22},T_{211},T_{13},T_{1111}$.

We see that the $U(N)$ operators given above, explicitly reduce to 
the $SU(N)$ operators upon imposing $T_1=0$. This pattern goes on, and 
for illustration we give the next few higher weight examples,  
%
%

\begin{align}
&
\mathcal{O}_5=T_5-\frac{5 \left(N^2-2\right) T_{32}}{N \left(N^2+5\right)}+\mathcal{U}_5 \label{un_5}\\
&
\mathcal{O}_6=
	T_6-\frac{\left(3 N^4-11 N^2+80\right) T_{33}}{N \left(N^4+15
	N^2+8\right)}-\frac{6 (N-2) (N+2) \left(N^2+5\right) T_{42}}{N \left(N^4+15
	N^2+8\right)}+\frac{7 \left(N^2-7\right) T_{222}}{N^4+15 N^2+8}+ \mathcal{U}_6 \notag
\end{align}
where
\begin{align}
&
\mathcal{U}_5=\frac{15
	\left(N^2-2\right) T_{221}}{N^2 \left(N^2+5\right)}+\frac{5 \left(3 N^2+8\right)
	T_{311}}{N^2 \left(N^2+5\right)}-\frac{35 T_{2111}}{N \left(N^2+5\right)}+\frac{14
	T_{11111}}{N^2 \left(N^2+5\right)}-\frac{5 T_{41}}{N}\\
&
 \mathcal{U}_6=\frac{42 (N-1) (N+1)
	T_{321}}{N^4+15 N^2+8}+\frac{21 \left(N^2+11\right) T_{411}}{N^4+15 N^2+8}-\frac{42
	\left(2 N^2-5\right) T_{2211}}{N \left(N^4+15 N^2+8\right)}+ \notag\\
	&
	\qquad
	-\frac{56 \left(N^2+5\right)
	T_{3111}}{N \left(N^4+15 N^2+8\right)}+\frac{126 T_{21111}}{N^4+15 N^2+8}-\frac{42
	T_{111111}}{N \left(N^4+15 N^2+8\right)}-\frac{6 T_{51}}{N}
\end{align}
For $SU(N)$ the contributions denoted by $\mathcal{U}$ vanish in each case.


The statement that $U(N)$ single particle operators reduce to the $SU(N)$ single 
particle operator upon imposing $T_1=0$ is true in general, as we will prove in section 
\ref{UN_SUN}, and it has very nice consequences. The correlators of $U(N)$ operators 
are much simpler to compute, and can be directly related to group theoretic quantities, 
essentially since the propagator~\eqref{unprop} is so simple. 
For this reason, there is a long history of studying half-BPS operators in the $U(N)$ theory, eg~\cite{Corley:2001zk}.  
Here we see that although correlators in the  $U(N)$ theory are simpler to compute, the trade-off 
is that the SPOs themselves are more complicated. In the end the computation of a correlator 
involving SPOs is equivalent whether using the $U(N)$ or $SU(N)$ theory.

\subsection{General Formulae for SPOs}\label{sec_gen_formulas}

So far we have uniquely defined SPOs (up to normalisation) as operators orthogonal 
to all multi-trace operators and illustrated with some examples at low charges. 
We will now give three different formulae for the single particle operators, which 
precisely capture the multi-trace admixture.  The plan will be the following:

In section \ref{trace_sec} we will give a formula in the trace basis, see \eqref{coeff}, 
which is perhaps the most familiar basis. This formula is equivalent to \eqref{det_rep} 
but uses more powerful group theory techniques to resolve  for the expansion of the 
SPO in terms of multi-traces.  We quote it here 
\begin{align}
	\cO_p(x)& = \sum_{\{q_1..q_m\}\vdash p}  C_{q_1,\ldots ,q_m}T_{q_1,\ldots, q_m}(x) \\
	C_{q_1,..q_m} &= 
	\frac{|[\sigma_{q_1..q_m}]|}{(p-1)!} \sum_{s \in \mathcal{P}(\{q_1,..,q_m\})}\frac{{(-1)^{|s|+1}}{(N+1-p)_{p-\Sigma(s)}(N+p- \Sigma(s))_{\Sigma(s)}}}{{(N)_{p}}-{(N+1-p)_{p}}}\notag
	\end{align}
It is very non trivial. The group theory data consists of $\mathcal{P}(\{q_1,\ldots,q_m\})$, 
the powerset of the traces $T_{q_1,\ldots,q_m}$, then $|s|$ is the cardinality of $s$ and $\Sigma(s)= \sum_{s_i \in s} s_i$.
Finally, $|[\sigma_{q_1..q_m}]|$ is the size of the conjugacy classes of $\sigma$ with length cycles $q_1\ldots q_m$.

In section \ref{eigen_sec} we give a much simpler formula, see \eqref{eq:SPOEigenBasisGeneral}, 
directly in terms of the eigenvalues $E_k(z_{i})$ of the elementary fields $\phi$,
\begin{align}
   \cO_p (x)&= \sum_{k=1}^p   d_{k}(p,N)   E_k(z_{i})(x)\\
d_k(p,N)&=\frac{(-1)^{k+1} p\,  (N-p+1)_{p-k} (p-1)_k}{(N)_p-(N-p+1)_p}
\end{align}

In section \ref{schur_sec} we give another simple formula in terms of the Schur 
polynomial basis (for operators), see~\eqref{sposinchi}, where we note that only hook representations appear.

If we think that half-BPS operators are symmetric functions of the eigenvalues of the scalar matrix $\phi_r^s$, 
the three basis for gauge invariant operators, that we used to expand the SPOs above, correspond to 
three well-known bases for symmetric polynomials: 
1) The trace basis corresponds to the power sum symmetric polynomials. 
2) The explicit eigenvalue monomials (after summing over permutations) are called the monomial symmetric polynomials. 
3) The Schur polynomials are named directly in terms of the corresponding symmetric polynomial basis of the same name. 

There are a number of well known formulae relating 1), 2) and 3) to each other, 
known as Newton identities, and it would be interesting to explore these relations further in this context.

\subsubsection{Formula in terms of products of traces}\label{trace_sec}

We first observe that the single-particle operators $\mathcal{O}_p$ 
must be proportional to the dual $\xi_p$ field of the single-trace operators $T_p$
\be\label{casem_1}
\langle \xi_p(x_1)\, T_{q_1\ldots q_n}(x_2) \rangle = 0\,, \qquad n \geq 2
\ee
where the dual fields $\xi_p$ were introduced by Tom Brown in~\cite{Brown:2007bb}.

Indeed, group theoretic formulae for SPOs were given in~\cite{Brown:2007bb} 
(albeit without the explicit physical description as single particle operators).
He defined a more general basis of operators: the dual of the trace basis. 
This is given by operators $\xi_{p_1\ldots p_n}$ (with $p_1 \geq \ldots \geq p_n$) which obey%
\footnote{In this section we will focus only on two-point functions and we 
will always drop the trivial space-time dependent part, $(g_{12})^{p}$, 
referring the discussion to the normalisation/color factor.}
\be
\langle \xi_{p_1\ldots p_m}(x_1) T_{q_1 \ldots q_n}(x_2) \rangle = 
\begin{cases}
	1 \quad \text{ if } (p_1,\ldots,p_m) = (q_1,\ldots,q_n)\,,\\
	0\quad \text{ otherwise.}
\end{cases}
\label{dualbasis}
\ee
In other words each element of the dual to the trace basis is orthogonal to
 (i.e.~it has vanishing two-point function with) all elements of the trace basis but one, 
 and we then normalise it to have unit two-point coefficient with this element. 
 Note that an orthogonal basis is its own dual. We will refer to the basis given 
 by the $\xi_{p_1,\ldots,p_n}$ as the \emph{dual basis}.

For a single index (i.e.\! $m=1$) the definition of the dual basis~~(\ref{dualbasis}) reduces to \eqref{casem_1}, which 
is the defining property of single particle operators~\eqref{orthogonality}.
So the $\xi_p$ are orthogonal to all multi-trace operators and thus equal to $\mathcal{O}_p$ up to normalisation. 
The normalisation can then be determined from \eqref{det_rep}, i.e.~$\cO_p=T_p+$ multi-traces,  which implies
\be
\langle \xi_p \mathcal{O}_p \rangle = 1
\ee
and hence%
\be
\xi_p(x) = \frac{\mathcal{O}_p(x)}{\langle \mathcal{O}_p \mathcal{O}_p \rangle}\, \qquad \text{and} \qquad  
\label{eq:SingleParticleDual}
\mathcal{O}_p(x) = \frac{\xi_p(x)}{\langle \xi_p \xi_p \rangle}\,.
\end{equation}

By definition (as the dual basis with respect to the inner product defined by the two-point function) 
the change of basis matrix from the dual to the trace basis is simply the two point function:  
\begin{equation}
\label{dualbasisgroup}
\xi_{p_1..p_n} (x)= \sum_{\{q_1,..q_m\}\vdash p } \langle \xi_{p_1..p_n} \xi_{q_1..q_m} \rangle T_{q_1..q_m}(x)\ 
\end{equation}
where the sum is over all partitions of $p$, that is all
sets of integers  $q_1\geq\dots \geq q_m$ such that $q_1+..+q_m=p$.
Brown in \cite{Brown:2007bb} gives group theoretic expressions for these two-point functions as
\begin{align}\label{xixi}
\langle \xi_{p_1..p_n} \xi_{q_1..q_m} \rangle &= 
			\frac{|[\sigma_{p_1..p_n}]|}{p!}\frac{|[\sigma_{q_1..q_m}]|}{p!}	
					\sum_{R\vdash p}\frac{1}{f_R}\chi_R(\sigma_{p_1..p_n})\chi_R(\sigma_{q_1..q_m}) \notag \\
\qquad p&=p_1+\ldots +p_n=q_1+\ldots+q_m\ .
\end{align}
where
\begin{itemize}
\item
$\sigma_{p_1..p_n}\in S_p$ is a permutation \footnote{For example the one given by 
$(1,\ldots, p_1)(p_1+1,\ldots p_1+p_2)\ldots (p_1+\ldots + p_{n-1}+1,\ldots ,p)$.}  
made of disjoint cycles of lengths $p_1,p_2,..$ and $|[\sigma_{p_1..p_n}]|$ is the size  
of the corresponding conjugacy class of the permutation (i.e.\! the number of  permutations 
with that disjoint cycle structure). Similarly for $\sigma_{q_1..q_m}$. 
\item the sum is over all partitions $R$ of $p$. One can view $R$ as a Young diagram 
	with $p$ boxes, labelling a representation of the permutation group $S_p$.
\item the expression $\chi_R(\sigma)$ is the character for this  $S_p$ representation. 
\item the sum is weighted by 
\begin{align}
	f_R = \prod_{(r,c) \in R} (N-r+c)
\end{align}
where $(row,cln)$ are the coordinates of the boxes of the Young diagram.\footnote{Group 
theoretically $f_R$ is proportional to the ratio between the dimensions of $R$ as a 
rep of $U(N)$ and the dimension of $R$ as a rep of $S_p$, i.e.\! $f_R= {p! d_R[U(N)]}/{d_R[{S_p}]}$.}
\end{itemize}

Since we are focusing on single particle operators we can assume $n=1$ in~\eqref{dualbasisgroup} 
so $\xi_{p_1..p_n}=\xi_p$ and the permutation $\sigma_{p_1..p_n}$  consists of a single cycle 
of length $p$ of  which there are $(p-1)!$ possibilities, so  $|[\sigma_p]| = (p-1)!$. Furthermore 
we observe that for such a permutation, only Hook reprentations have a non vanishing $S_p$. 
Indeed we note that:
\begin{align}\label{chiR}
	\chi_R(\sigma_p) =\left\{  \begin{array}{ll}
	(-1)^{h_R-1} & R = \text{ hook YT of height }h_R\\
	0 & \text{otherwise}
	\end{array}      \right.
\end{align}
 where ``hook YT'' means that $R$ has  a hook-shaped  diagram: all non-zero rows 
 but the first have length 1 and all non-zero columns but the first have height 1.
  Thus the coefficient of a given multi-trace operator in $\xi_p$ is
 \begin{align}\label{xip}
 	\langle \xi_p \,\xi_{q_1..q_m}\rangle= \frac{1}{p}\frac{|[\sigma_{q_1..q_m}]|}{p!}	
							\sum_{R\in\, \text{hooks}}\frac{1}{f_R}(-1)^{1+h_R} \chi_R(\sigma_{q_1..q_m}) \ .
 \end{align}
We have found that this sum over hook representations is always given by  
the following explicit formula 
\begin{align}\label{xipxiqqq}
\begin{split}
\langle \xi_p \,\xi_{q_1..q_m}\rangle &= \frac{1}{p}\frac{|[\sigma_{q_1..q_m}]|}{p!} \frac{1}{
	p-1}\sum_{s \in \mathcal{P}(\{q_1,..,q_m\})}\frac{(-1)^{|s|+1}}{(N+1- \Sigma(s))_{p-1}}\,.
\end{split}
\end{align}
The sum $s \in \mathcal{P}(\{q_1,..,q_m\})$ is over all subsets $s$ of the set 
$\{q_1,..,q_m\}$ including the empty set and the  full set $\{q_1,..,q_m\}$. 
Note that the set of all subsets of  a set $S$ is known as the powerset of $S$ 
and denoted ${\mathcal P} (S)$.\footnote{For example
$\mathcal{P}(\{3,2,1\}) = \{ \{\}, \{1\}, \{2\}, \{3\}, \{2,1\}, \{3,1\}, \{3,2\}, \{3,2,1\}\}$.}
Then $|s|$ denotes the number of elements of the subset $s$ and $\Sigma(s)$ 
the total of all the elements in subset $s$,  $\Sigma(s)= \sum_{s_i \in s} s_i$.

An important special case of~\eqref{xipxiqqq} is the case $m=1$ giving the two-point function 
of the dual of the single trace operator. In this case the sum is over just two elements $s=\{\}$ 
and $s=\{p\}$ since ${\mathcal P}(\{p\})=\{\{\},\{p\}\}$. The expression~\eqref{xipxiqqq} thus 
simplifies to
 \begin{align}\label{xipxip}
 \langle \xi_p \,\xi_{p}\rangle &= \frac{1}{p^2} \frac{1}{
 	p-1}\left(\frac{1}{(N{+}1{-}p)_{p-1}}-\frac{1}{(N{+}1)_{p-1}}\right)\,.
\end{align}

Finally inserting \eqref{xipxiqqq} into~\eqref{dualbasisgroup} and in turn 
into~\eqref{eq:SingleParticleDual} together with~\eqref{xipxip} gives an explicit 
expression for the single particle operator as a sum of multi-trace operators
\begin{align}\label{gentrb}
	\cO_p(x) = \sum_{\{q_1..q_m\}\vdash p} C_{q_1,..,q_m}T_{q_1,..q_m}(x)
\end{align}
with coefficients
\begin{align}
	C_{q_1,..q_m} &= \frac{\langle \xi_p \, \xi_{q_1..q_m} \rangle}{\langle\xi_p\,\xi_p\rangle} \notag \\
	&= \frac{|[\sigma_{q_1..q_m}]|}{(p-1)!} \sum_{s \in \mathcal{P}(\{q_1,..,q_m\})}\frac{(-1)^{|s|+1}}{(N+1- \Sigma(s))_{p-1}}  \left(\frac{1}{(N+1- p)_{p-1}}-\frac{1}{(N+1)_{p-1}}\right)^{-1}
	\notag \\
	&= \frac{|[\sigma_{q_1..q_m}]|}{(p-1)!} \sum_{s \in \mathcal{P}(\{q_1,..,q_m\})}\frac{{(-1)^{|s|+1}}{(N+1-p)_{p-\Sigma(s)}(N+p- \Sigma(s))_{\Sigma(s)}}}{{(N)_{p}}-{(N+1-p)_{p}}}\ .\label{coeff}
	\end{align}
The second equality is more useful computationally and is obtained by 
multiplying and dividing by $(N+1-p)_{2p-1}$ and using the identity
\begin{align}
	\frac{(N+1-p)_{2p-1}}{(N+1- \Sigma(s))_{p-1}}=(N+1-p)_{p-\Sigma(s)}(N+p- \Sigma(s))_{\Sigma(s)}\ .
\end{align}
The only final ingredient is the size of the conjugacy classes. If all cycle lengths 
are distinct, i.e.\! $p=q_1+\ldots +q_m$ with $q_i\neq q_j$, 
the size of the conjugacy classes is simply $p!/\prod q_i$.
However if there are multiple cycles of the same length, i.e.\! some $q_i=q_j$ in the partition of $p$, 
then these cycles are interchangeable and we have to divide by this symmetry in addition. 
To deal with this case,  let $\lambda_1, \lambda_2, ..., \lambda_r$ be {\em distinct} 
so that $\prod_j \lambda_j^{k_j}= \prod_i q_i$ and $\sum_j k_j \lambda_j = p$. 
Then 
\begin{equation}
\label{eq:PermutationGroupConjugacySize}
|[\sigma_{q_1..q_m}]| = \frac{p!}{\prod_{i=1}^{r} k_i! \lambda_i^{k_i}}.
\end{equation}

We thus have a formula for computing the coefficients $C_{q_1\ldots q_m}$ 
which give the representation of the SPOs in the trace basis. Very nicely, 
this formula is explicit in $p$ and $\{q_1,\ldots q_m\}$, and depends only 
on the group theory data associated to such partition. 

Notice that the value of $m$ counts the splitting of $\cO_p$ into $m$ traces. 
In appendix \ref{trace_sector_app} we provide some extra examples for 
generic $q_1\ldots q_m$ when $m=2$, i.e.\! double traces, and $m=3$, i.e.\! triple traces.

\subsubsection{Formula in terms of eigenvalues}\label{eigen_sec}

In fact the formula for the SPOs is much simpler when expressed directly in 
terms of the eigenvalues of the  adjoint scalar $\phi_r^s$ which we call $z_1, z_2, .., z_N$.
For this purpose consider
\begin{align}
&
m_{[\lambda_1,\ldots, \lambda_N]}(z_1,\ldots z_N)= \sum_{\sigma \in S_N}  z_{\sigma(1)}^{\lambda_1} z_{\sigma(2)}^{\lambda_2}\ldots z_{\sigma(N)}^{\lambda_N} \\[.2cm]
&
E_{p,k}(z_1,\ldots z_N) =\sum_{\substack{ q_1+\ldots + q_k=p \\[.1cm] q_1\ge q_2 \ge \ldots q_k >0} } m_{[q_1,\ldots,q_k,0^{N-k} ]}(z_1,\ldots z_N)
\end{align}
which respectively are: $m_{\underline{\lambda}}$ the monomial symmetric 
polynomial indexed by $\underline{\lambda}$, and $E_{p,k}$ the sum over all 
monomials indexed by a partition of $p$ in $k$ parts.

So $E_{p,1} = T_p(x)$ is an obvious. Other examples are
\begin{align}
	 E_{p,p}(z_1\ldots z_N) &= z_1\ldots z_p +\ldots \notag\\[.2cm]
	 E_{4,2}(z_1\ldots z_N) &= (z_1^3z_2 + \ldots )+(z_1^2z_2^2+\ldots )\
\end{align}

We find that the single particle operators can be written as
\begin{align}
\label{eq:SPOEigenBasisGeneral}
   \cO_p &= \sum_{k=1}^p    
 d_{k}(p,N)   E_{p,k}(z_{i})\,,
\end{align}
where the coefficient $d_k(p,N)$ is
\begin{align}
\label{eq:dpN}
d_k(p,N)
&=\frac{(-1)^{k+1} p\,  (N-p+1)_{p-k} (p-1)_k}{(N)_p-(N-p+1)_p}
\end{align}

Note that interestingly the coefficient of a monomial in this formula only depends 
on the number of different eigenvalues appearing in the monomial and not on 
any other details of the monomial.

\subsubsection{Formula in terms of Schur polynomials}\label{schur_sec}

Finally we consider the SPOs written in terms of the Schur polynomial formula. 
In~\cite{Brown:2007bb} a group theoretic formula for the dual of the trace basis 
was also given in terms of the Schur polynomial basis, $\chi_R(\phi)$. The Schur 
polynomial basis is  an orthogonal (in $U(N)$ ) basis for all half-BPS operators 
introduced in~\cite{Corley:2001zk}. The formula given in~\cite{Brown:2007bb} is:
\begin{align}
\xi_{p_1..p_n} = \frac{|[\sigma_{p_1..p_n}]|}{p!}\sum_{R\vdash p} \frac 1{f_R}\chi_R(\sigma_{p_1..p_n}) \chi_R[\phi]\ . 
\end{align}

The relation between the dual basis operators and single particle operators~\eqref{eq:SingleParticleDual} 
together with the observation about $S_p$ characters of cycle permutations~\eqref{chiR} 
therefore gives the following explicit formula for SPOs directly in terms of the Schur polynomials 
of Hook Young tableaux of height $k$, with $p$ boxes in total,  $R_{k}^p$: 
\begin{align}\label{sposinchi}
	\cO_p 
&=   \sum_{k=1}^p \tilde d_k(p,N)   \chi_{R_k^p}[\phi]\notag \\
\tilde d_k(p,N)&= {p(p{-}1)}  {(-1)^{k-1}}\frac{(N{-}p{+}1)_{p-k}(N{+}p{-}k{+}1)_{k-1}} {{(N)_{p}}-{(N{+}1{-}p)_{p}} } 
\end{align}

\begin{center}
	\begin{tikzpicture}[scale=.4]
	\node at (-7,4) { $R_k^p  =  [p-k+1,1^{k-1}]= $};
	\draw (0,8) -- (20,8) -- (20,7) -- (0,7) -- (0,8);
	\node at (10,7.5) {\scriptsize $\leftarrow p{-}k\rightarrow$};
	\draw (-1,8) -- (0,8) -- (0,0) -- (-1,0) -- (-1,8);
	\node at (-.45,4.5) {\scriptsize $\begin{array}{c}\uparrow\\ k\\ \downarrow \end{array}$};
	\end{tikzpicture}
\end{center}

\subsubsection{Examples in the three basis}

We find useful at this point to consider some low charges SPOs, and show 
their representation in the three basis we just constructed.

For $\cO_2$,  
\begin{align}
\begin{array}{l}
	\mathcal{O}_2=T_2-\frac{1}{N}T_{11} \\[.2cm]
	\mathcal{O}_2=\frac{ (N-1)}{N}E_{2,1}-\frac{2 }{N}E_{2,2} \\[.2cm]
	\mathcal{O}_2=\frac{(N-1)  }{N}\chi_{R_1^2}-\frac{(N+1) }{N} \chi_{R_2^2} 
\end{array}
\end{align}

For $\cO_3$,  
\begin{align}
\begin{array}{l}
	\mathcal{O}_3=\frac{2 }{N^2}T_{111}-\frac{3 }{N}T_{21}+T_3 \\[.2cm]
	\mathcal{O}_3=\frac{ (N-2) (N-1)}{N^2}E_{3,1}-\frac{3  (N-2)}{N^2}E_{3,2}+\frac{12 }{N^2}E_{3,3} \\[.2cm]
	\mathcal{O}_3=\frac{(N-2) (N-1) }{N^2}\chi_{R_1^3} -\frac{(N-2) (N+2)  }{N^2}\chi_{R_2^3}+\frac{(N+1) (N+2)  }{N^2}\chi_{R_3^3} 
\end{array}
\end{align}

For $\cO_4$
\begin{align}
\begin{array}{l}
	\mathcal{O}_4=-\frac{\left(2 N^2-3\right) }{N \left(N^2+1\right)}T_{22}+\frac{10 }{N^2+1}T_{211}-\frac{5 }{N \left(N^2+1\right)}T_{1111}-\frac{4 }{N}T_{31}+T_4 \\[.3cm]
	\mathcal{O}_4=\frac{ (N-3) (N-2) (N-1)}{N \left(N^2+1\right)}E_{4,1}-\frac{4  (N-3) (N-2)}{N \left(N^2+1\right)}E_{4,2}+\frac{20  (N-3)}{N \left(N^2+1\right)}E_{4,3}-\frac{120 }{N
		\left(N^2+1\right)}E_{4,4} \\[.3cm]
	\mathcal{O}_4=\frac{(N-3) (N-2) (N-1) }{N \left(N^2+1\right)} \chi_{R_1^4}-\frac{(N-3) (N-2) (N+3)  }{N \left(N^2+1\right)}\chi_{R_2^4}+\frac{(N-3) (N+2) (N+3) }{N
		\left(N^2+1\right)}\chi_{R_3^4} -\frac{(N+1) (N+2) (N+3)  }{N \left(N^2+1\right)}\chi_{R_4^4} \\[.3cm]\\
\end{array}
\end{align}

A feature of the expansion of the single-particle operator $\cO_p$ in the Schur basis is the 
homogeneous degree in $N$ of its coefficients w.r.t~the partitions of $p$, i.e.\! the different basis elements. 
%
%
%
%

\subsection{SPOs interpolate between single-trace and giant gravitons } \label{largeN}

In the large $N$ limit holding the dimension of the operator $p$ fixed,  
the single particle operator becomes equivalent to the single trace operator 
\begin{align}
	\cO_p\rightarrow T_p+O(1/N)\ .
\end{align}
This is less obvious from the general  formula in terms of traces~\eqref{gentrb}, 
but upon inspection, the coefficients of multi-trace corrections with $m$ traces, $C_{q_1..q_m}$, are indeed $O(1/N^{m-1})$. 
What happens is that each term of the sum in $C_{q_1..q_m}$ is actually $O(N)$,  
but the alternating sum provides $m$ cancellations,  one at each order in $N$, this 
brings it down to $O(1/N^{m-1})$.  It is much more direct to see this  from the formula 
in terms of eigenvalues~\eqref{eq:SPOEigenBasisGeneral}.  The coefficients of terms 
with $k$ different eigenvalues $d_k(p,N)$ is $O(1/N^{k-1})$. More precisely 
\begin{align}
	d_k(p,N) \rightarrow \frac{(-1)^{k-1}p_{k-1}}{N^{k-1}} \qquad \text{as} \qquad N\rightarrow \infty\ 
\end{align}
so the $k=1$ term ($z_1^p+...z_N^p$ i.e.\! the single trace term) dominates.

As the charge of the single-particle operator increases, approaching the number 
of colours $N$, it is clear that the single-particle operator is very different to the single trace operator. 
For example for $p>N$ the operator vanishes
\begin{align}\label{spo_no_exist}
	\cO_p=0 \qquad p>N\ ,
\end{align}
 as can be seen directly from the explicit formulae in the previous subsection.
 Note that this behaviour is very different from the single-trace operators which do not vanish for $p>N$, 
 but rather become linear combinations of multi-trace operators.  This explains why the single-particle 
 operators vanish: for $p>N$ {\em all} operators are multi-trace, therefore by definition the 
 single-particle operators are orthogonal to all operators and hence must vanish.

In the large $N$ limit, with charge $p$ also growing large but with $N-p=p'$ fixed, we find 
that the single-particle operators become proportional to (sub)-determinant operators. These 
operators were proposed in~\cite{Balasubramanian:2001nh} as duals to the sphere giants 
predicted in~\cite{McGreevy:2000cw} as  Kaluza Klein states with masses approaching the sphere radius. 
They are the Schur polynomials associated with totally  antisymmetric Young tableau 
($R_p^p$ in the notation of~\eqref{sposinchi}) and in terms of eigenvalues take the form 
$z_1..z_p + ..$ where the dots denote similar terms obtained by permutations.

We find that at large $N$, 
  the coefficient appearing in the eigenvalue formula~\eqref{eq:dpN} has the large $N$ limit
\begin{align}
d_{N-p'-k'}(N{-}p',N) \rightarrow  (-1)^{N}(-2)^{-1-k'-p'}(1{+}p)_k /N^{k'-1}\qquad \text{as} \qquad N\rightarrow \infty\ .
\end{align}

Thus at large $N$
 we obtain:
 \begin{align}
 \frac1N \cO_{N-p'} \rightarrow   2^{-1-p'}  \times (z_1z_2..z_{N-p'} + ..) +O(1/N) \ .
 \end{align}
 where the dots denote all possible terms with $N-p'$ different $z$s  all with coefficient one. This is precisely the subdeterminant operator identified as the sphere giant gravitons in~\cite{McGreevy:2000cw}.

\subsection{SPOs in $U(N)$ are SPOs in $SU(N)$}\label{UN_SUN}

Single-particle operators can be defined for both $SU(N)$ and $U(N)$ gauge group.   
As we observed in the low charge examples  \eqref{un_23}, \eqref{un_4} and \eqref{un_5}, 
the two are closely related. 
The $SU(N)$ elementary fields are of course different from the $U(N)$ elementary field, 
and the two theories in general are different. 
However,  the fact that $U(N)$ SPOs represented in the trace
basis were shown to give the $SU(N)$ SPOs  upon setting $\Tr(\phi)$ to zero,
suggests that SPOs in $U(N)$ are SPOs in $SU(N)$.  
In this section we make this statement rigorous. 
 
The elementary field $\hat\phi_r{}^{ s}$ in $SU(N)$ is the traceless part of the elementary field $\phi_r{}^{ s}$ in $U(N)$, namely
\begin{align}\label{phinonhat}
\hat \phi_r{}^{ s} =\phi_r{}^{ s}- \tfrac 1N \delta_r^{ s} \phi_t{}^{ t}\ ,
\end{align}
For any operator $O[\phi]$ in $U(N)$ we define the $SU(N)$ projection as the map, 
\begin{align}
\Pi:\ O[\phi]\ \rightarrow\ \hat O[\phi]\equiv O[\hat\phi(\phi)]
\end{align}
The operator $ \hat O[\phi]$ is then a new operator in $U(N)$.
For example, the $SU(N)$ projection of $O[\phi]= \Tr( \phi^2)$ is 
\begin{align}
	\hat O[ \phi] = \Tr(\, \hat \phi(\phi).\hat \phi(\phi)\,) =\underbracket{ 
	\Tr(\phi^2)
	}_{\rule{0pt}{.3cm} O[\phi]}
	-\tfrac1N \Tr(\phi)\Tr(\phi)\ .
\end{align}
More generally, we find that an operator $\hat O[ \phi]$ has the form
\begin{align}
	\hat O[\phi] = O[\phi] - [ T_1 \tilde O[\phi]]\ \qquad ;\qquad T_1=\Tr[\phi]
\end{align}
for some operator $\tilde O[\phi]$. Notice that $ O[\phi]$ is also the leading 
term in the $1/N$ expansion of $\hat O[\phi]$, with the fields $\phi$ and 
$\hat \phi$ treated as formal variables. 

The  $\hat O[\phi]$ operators in $U(N)$ span a subspace, since they 
are independent of the trace.  The map $\Pi: O \rightarrow \hat O$ decomposes
the space of $U(N)$ operators as $Im(\Pi)\oplus Ker(\Pi)$, with the 
kernel everything of the form $[ T_1 \tilde O[\phi]]$. 
To show that the $SU(N)$ projection is orthogonal we have to show that 
\begin{align}
\langle \hat O (x_1) \, [T_1 \tilde O] (x_2) \rangle _{U(N)}= 0\ .
\end{align}
The claim follows from the fact that the operator $\hat O[\phi]$ is constructed 
only from the traceless part $\hat \phi^r_s$, then by applying Wick's theorem to 
compute this two-point function we always find a contraction between 
$\hat \phi^r_s(x_1)$ within $\hat O(x_1)$ and  $T_1(x_2)$. In the $U(N)$ 
theory with propagator \eqref{unprop} we have  
\begin{align}
\langle \hat \phi_r^s(x_1) T_1(x_2) \rangle  = \langle \phi_r^s(x_1) \phi^t_t(x_2) \rangle - \tfrac{1}{N} \delta_{r}^s \langle   \phi_u^u (x_1) \phi^t_t(x_2) \rangle =0
\end{align}

Thus {\em any} operator $\hat O[\phi]$ constructed only from the traceless part 
$\hat \phi^r_s$ is orthogonal to {\em any} operator involving the trace as a factor $ [T_1 \tilde O]$. 

Single-particle operators $\cO$ in $U(N)$ are precisely defined to be orthogonal 
to all multi-traces, in particular to the operators of the form  $[T_1\tilde O]$ that we 
discussed above.\footnote{Notice that in the examples \eqref{un_23}, \eqref{un_4} and \eqref{un_5}, 
we did not use the basis $Im(\Pi)\oplus Ker(\Pi)$ that instead we are constructing here.}
Thus single particle operators in $U(N)$ automatically live in 
the $SU(N)$ subspace. The $SU(N)$ single-particle operator is now candidate 
to give the $U(N)$ single-particle operator. 

Since the two-point orthogonality in $U(N)$ and $SU(N)$ is obtained via Wick's 
theorem with two different propagators, the only thing remaining to check is that 
the $U(N)$ inner product restricted on the $SU(N)$ subspace is the same as the 
inner product of the $SU(N)$ theory. A short computation shows that,
\begin{align}
	\langle \hat \phi_r^s (\phi)\hat \phi_t^u(\phi) \rangle_{U(N)} = 
	\Big \langle \Big(\phi_r^s-\tfrac 1N \delta_r^s \phi_v^v \Big) 
	\Big( \phi_t^u-\tfrac1N\delta_t^u\phi_v^v \Big)\Big \rangle_{U(N)} = \delta_r^u \delta_t^s - \tfrac1N \delta_r^s \delta_t^u 
\end{align}
which is precisely the $SU(N)$ propagator~\eqref{sunprop}.

We conclude that $U(N)$ single-particle operators are SPOs in $SU(N)$ ,
\begin{align}
	  \cO^{U(N)}_p[\phi]=\hat \cO_p^{SU(N)}[\hat \phi] \qquad p\geq 2 
\end{align}
and any correlators of SPOs of charge 2 or higher computed in either $U(N)$ 
or $SU(N)$ will be identical.

Our result here manifests, purely in the free theory,  the well known feature of $\cal N$=4 SYM 
in the context of AdS/CFT that the $U(1)$ part of the gauge group $U(N)$ decouples and 
only $SU(N)$ remains  in the interacting theory. Here we find  that the sector of single-particle 
operators with length greater than one in the $U(N)$ theory is that of the $SU(N)$ theory, 
and furthermore it is orthogonal to anything we could construct with a $T_1$.

\subsection{Two-point function normalisation }

From our study of the SPOs in the trace basis, we can obtain readily 
their two-point function normalisation. Indeed, it follows from the relation (\ref{eq:SingleParticleDual})
 that such a normalisation is just the inverse
 of the $\xi_p$ normalisation from the dual basis, namely
\be
\langle \mathcal{O}_p(x_1) \mathcal{O}_p(x_2) \rangle =  
					R_p\, g_{12}^p \qquad;\qquad {\langle \xi_p(x_1) \xi_p(x_2) \rangle}= \frac{1}{R_p} g_{12}^p \,
\ee
%
where we use the notation $R_p$ to denote the $N$-dependent 
color factor. From~\eqref{xipxip} it takes the form
\be
\label{OpOpexp}
R_p = p^2 (p-1)\biggl[ \frac{1}{(N-p+1)_{p-1}} - \frac{1}{(N+1)_{p-1}} \biggr]^{-1}\,.
\ee

Note that $R_p$ has zeros at $N=1,\ldots p-1$, reflecting the fact that the operator 
vanishes when $N<p$, and since it is also symmetric in $N\rightarrow -N$ it must 
contain explicit factors of the form $(N^2-r^2)$, for $r=1,\ldots p-1$.

It is useful to make these facts manifest by rewriting
\be
\label{twoptfns}
R_p = \frac{p}{N^{p-2}} \times  {\prod_{r=1}^{p-1}(N^2-r^2)}\times \frac{1}{ Q_p(1/N^2)}  \,,
\ee
where $Q_p(1/N^2)$  is a polynomial of degree $\lfloor\tfrac{p-2}{2}\rfloor$ in $1/N^2$. 
The first few cases are given by
\be
Q_2(1/N^2) = 1\,, \qquad Q_3(1/N^2) = 1\,, \qquad Q_4(1/N^2) = 1 + \frac{1}{N^2}\,.
\ee
whereas the general form of  $Q_p(1/N^2)$ is given by
 \begin{align}
 	Q_{p-2}(N)=\frac{(N+1)_{p-1}-(N{-}p{+}1)_{p-1}}{p(p-1)}\ .
 \end{align}
 
One can manifest the $N \rightarrow -N$ symmetry by using  
rising and lowering factorials instead of always the rising factorial 
(Pochhammer). With the notation $x^{\overline{p}}$ for the Pochhammer 
(rising factorial) and $x^{\underline{p}}$ for the falling factorial, we find 
\begin{align}\label{2pnt}
R_p &={p^2} (p-1) \frac{{(N{-}1)^{\underline{p{-}1}}} {(N+1)^{\overline{p{-}1}}}  }{   \rule{0pt}{.5cm} {(N+1)^{\overline{p{-}1}}}-{(N{-}1)^{\underline{p{-}1}}}}
\end{align}

In the form \eqref{2pnt} it is clear that $R_p$ is the simplest possible  
rational function  of $N^2$ with the above zeros and of order $O(N^p)$ at large $N$.

\subsection{On multi-particle operators }

A complete basis of half-BPS operators in the theory is obtained by taking arbitrary 
products of the single-particle operators. We call this basis the multi-particle basis. 

The statement above can be true also for single-trace operators, modulo the fact that 
single-traces overcount. Any $T_{p}$ with $p>N$ is not an independent operator both 
in $SU(N)$ and $U(N)$, and yet it has non trivial two point functions with other operators. 
On the other hand, the multi-particle basis does not have this issue. Indeed, if $p>N$ there 
are only multi-trace operators, and the single-trace operator is not independent. Then, 
by definition, the single-particle operators are orthogonal to all operators and hence must vanish. 
Very remarkably, this feature is automatically implemented in the two-point function normalisation \eqref{2pnt}.

The multi-particle basis will not be orthogonal, although of course the single-particle sector 
will be orthogonal to the multi-particle sector. This is to be contrasted with the Schur polynomial 
basis of~\cite{Corley:2001zk} which is both complete and orthogonal, for all half-BPS operators. 
In particular, operators are in 1-1 correspondence with Young tableau of height no more than $N$.

Unlike the Schur polynomial basis, the multi-particle basis has the advantage of being 
a basis both for $SU(N)$ and $U(N)$, depending simply on whether $\cO_1$ is included or not.

Requiring orthogonality of multiparticle states is however subtle.
To obtain an orthogonal basis one can of course simply implement the Gram Schmidt procedure:
Start with an ordered list of basis elements, then leave the first element unchanged, 
then run over $n>1$ by considering the $n$-th element and add linear combinations 
of previous elements such that the new element is orthogonal to all previous ones.  
So for example at weight 6 we could choose the ordering
\begin{align}
(T_{111111},T_{21111},T_{2211},T_{3111},T_{222},T_{321},T_{411},T_{33},T_{42},T_{51}
,T_6) \qquad &\ U(N)\notag\\
(T_{222},T_{33},T_{42}
,T_6) \qquad & SU(N)
\ 
\end{align}
Then performing Gram-Schmidt orthogonalisation, would provide an orthogonal 
set of operators. In sum this can be done by cutting appropriately the matrix in \eqref{det_rep}. 
However at this point, the single-particle operator is singled out as the last one. In the example above $\cO_6$, 
is the one with greater admixture,  in contrast to $T_{111111}$ or $T_{222}$, but then it is not clear if there is a canonical choice for the ordering 
of the other operators. 


As long as the single-particles are fixed, a completely equivalent way of obtaining the same 
orthogonal basis is to start with the dual basis, list the operators in reverse order, and perform 
Gram-Schmidt orthogonalisation. This process yields exactly the same orthogonal basis 
(up to normalisation). Referring to the same example as above,
\begin{align}
	(\xi _6,\xi _{51},\xi _{42},\xi _{33},\xi _{411},\xi _{321},\xi
	_{222},\xi _{3111},\xi _{2211},\xi _{21111},\xi _{111111}) \qquad &\ U(N)\notag\\
		(\xi _6,\xi _{42},\xi _{33},\xi
	_{222}) \qquad & SU(N)\ 
\end{align}

In the first approach  the operator with the most traces $T_{111111}$ (or $T_{222}$) remains unchanged 
whereas the single-particle operator is the most complicated from the point of view of the trace-basis. 
In the second approach, this is turned on its head. 
In terms of the dual basis, the single particle operator is -- just $\xi_p$ -- whereas an 
operator labelled by a partition with increasing length becomes more intricate from the 
point of view of the $\xi$ basis. In the end, the most intricate  of such operators - 
a linear combination of all dual operators - must equal $T_{11111}$ (or $T_{222}$).

We leave the task of extending the single-particle operators to a full orthogonal basis 
to a future work. Perhaps an AdS/CFT understanding of multi-particle KK modes will help 
us figuring out a canonical way to fix the multi-particle states in $\mathcal{N}=4$ SYM.



\section{Multipoint orthogonality} \label{section_3}

In the previous section we obtained explicit expressions for SPOs.
From the two-point orthogonality it followed that SPOs are given by a specific admixture 
of single- and multi-trace operators. In a sense, the SPOs are composites of traces and
the richness of this structure would suggest that multipoint correlation functions 
are rather more complicated than multipoint single-trace correlation functions. 
However this expectation is too naive and as a first counter-example  
we show in this section that the defining two-point function orthogonality 
uplifts to a multipoint orthogonality theorem, which in turn implies vanishing 
of a large class of diagrams. 
Let us state our main result. 

{\bf Multipoint orthogonality Theorem.} 
Any propagator structure, which can contribute to any half-BPS correlator, with a single-particle operator $\cO_p$  
connected to two sub-diagrams, themselves disconnected from each other, has a vanishing color factor. 
The statement holds for both $SU(N)$ and $U(N)$ free theories.

We are considering any propagator structure $\mathcal{F}_{{p}\mid\underline{q_1}\ldots \underline{q_{n-1} } }$ 
which becomes disconnected upon removing $\cO_p$, and thus has the shape of a dumbbell:\\  

\be\label{fig1}
             \begin{tikzpicture}
		\def\lato 	{3.5}
		
		\def\xuno	{-.5}
		\def\yuno	{1.5}
		
		\def\xdue {\xuno+\lato}
		\def\ydue {1.5}
		
		\def\xtre {\xuno-\lato}
		\def\ytre {1.5}

		\def\xquat{\xuno-\lato}
		\def\yquat {0.5}


		\draw[thick] (\xuno,\yuno+.1)  --  (\xdue,\ydue+.1);
		\draw[thick] (\xuno,\yuno)  --  (\xdue,\ydue);
		\draw[thick] (\xuno,\yuno-.1)  --  (\xdue,\ydue-.1);

		\draw[thick] (\xuno,\yuno+.1)  --  (\xtre,\ydue+.1);
		\draw[thick] (\xuno,\yuno)  --  (\xtre,\ydue);
		\draw[thick] (\xuno,\yuno-.1)  --  (\xtre,\ydue-.1);


		\draw[fill=red!25,draw=black] (\xuno,\yuno) circle (.3cm);
		\draw[ fill=green!25,draw=black] (\xdue,\ydue) circle (1.2cm);
		\draw[ fill=green!25,draw=black] (\xtre,\ytre) circle (1.2cm);
		
				
		\draw (\xuno,\yuno+.3) 				 node[above] 							{$\mathcal{O}_{p}(x)$};
		
		\draw (\xdue+.4,\ydue-.5) 			 	node[left,font=\footnotesize] 				{$T_{ \underline{ q_{n-1} } }$};
		\draw[fill=black] (\xdue+.4,\ydue-.5)			circle (.05cm);
		
		\draw (\xdue+.4,\ydue) 			 	node[left,font=\footnotesize] 				{$\ldots\ \ $};
		
		\draw (\xdue+.4,\ydue+.5) 			 	node[left,font=\footnotesize] 				{$T_{\underline{ q_{r+1} } }$};
		\draw[fill=black] (\xdue+.4,\ydue+.5)			circle (.05cm);
		
		
		\draw (\xtre-.4,\ydue-.5) 			 	node[right,font=\footnotesize] 				{$\ \ T_{ \underline{ q_r  }  }$};
		\draw[fill=black] (\xtre-.4,\ydue-.5)			circle (.05cm);
		
		\draw (\xtre-.4,\ydue) 			 	node[right,font=\footnotesize] 				{$\ \ldots$};
		
		\draw (\xtre-.4,\ydue+.5) 			 	node[right,font=\footnotesize] 				{$\ \ T_{\underline{q_1 }  }$};
		\draw[fill=black] (\xtre-.4,\ydue+.5)			circle (.05cm);
		
		\draw (\xtre-1.5,\ydue)						node[left] 								{$\mathcal{F}_{{p}\mid\underline{q_1}\ldots \underline{q_{n-1} } }=$};

		\draw (\xdue+1.6,\ydue)  node {$\qquad$};
\end{tikzpicture}
\ee

A propagator structure of this sort will have a number of bridges going between 
points in the two (green) blobs, left and right, and a number of bridges connecting $\cO_p(x)$ 
with points belonging to the left and right blobs. No bridges between left and right blobs. 

Our theorem states that
\begin{align}
\mathcal{F}_{{p}\mid\underline{q_1}\ldots \underline{q_{n-1} } }=0.
\end{align}

We can gain some intuition about the multipoint orthogonality by
considering the fact that $\mathcal{F}_{p|\underline{q_1}\ldots \underline{q_{n-1}} }$
can be rearranged as a determinant. This follows directly from the representation 
of $\cO_{p}(x)$ as a determinant, that we mentioned in \eqref{det_rep}. 
All rows of this determinant, but the last one, are given by the (color factor of the) 
two points functions $\langle T_{\lambda_i}T_{\lambda_j}\rangle$, where $\lambda_i$ and $\lambda_j$ 
denote partitions of $p$. The last row is given by (the color factors of) 
$\langle T_{\lambda_i}(x) T_{\underline{q_1}}(x_1) \dots T_{\underline{q_{n-1}}}(x_{n-1})  \rangle$ in the 
given propagator structure. 
The determinant will vanish as long as this last row reduces to a linear 
combination of the others. But this linear combination might be $N$-dependent and 
thus hard to chase. We give some examples in Appendix \ref{app_determinant_multipoint}. 
However, a crucial point 
is that for \emph{any} two point functions 
$\langle T_{\lambda_i}T_{\lambda_j}\rangle$ at least one  partition has 
length greater than two. Therefore we might displace a part of it on a fictitious 
point, and since there cannot be propagators inside a $T_{\lambda}$, this configuration 
is actually a dumbbell. We learn in this way that, for the determinant to have 
a chance to vanish, the topology of the $\langle T_{\lambda_i}(x) T_{\underline{q_1}}(x_1) 
\dots T_{\underline{q_{n-1}}}(x_{n-1})  \rangle$  has to have the same feature of the two point functions $\langle T_{\lambda_i}T_{\lambda_j}\rangle$. 
We can then argue that $\mathcal{F}_{p|\underline{q_1}\ldots \underline{q_{n-1}} }$ 
is a dumbbell, as we draw in \eqref{fig1}.

The reasoning above leaves a free parameter. In fact, the assignment 
$\underline{q_1},\ldots ,\underline{q_{n-1}}$ can be such that 
\begin{align}\label{intro_deg_extr}
	  \tfrac12 \left(-p + \sum_{i=1}^{n-1} q_i\right)=k\ge 0
\end{align}
and yet the diagram disconnects as soon as we remove $\cO_p$. The value 
of $k$ measures the excess of $\sum q_i$ to be a partition of $p$. This is possible 
precisely because  differently from a two-point function, in a multipoint function 
there can be $k\ge 0$ Wick contractions  distributed among the 
$T_{\underline{q_1}}(x_1) \dots T_{\underline{q_{n-1}}}(x_{n-1})$,
such that when $\cO_p$ is removed, the diagram still disconnects.

In general, the combinatorics which leads to $\mathcal{F}_{{p}\mid\underline{q_1}\ldots \underline{q_{n-1} } }=0$
is hard. It would be very interesting to have a combinatorial proof of our theorem, by using
Wick contractions and double line notation, but our proof in the next section will go through a nice alternative route.

Some readers might wish to jump directly to section \ref{nEvan}  at this point. There we 
use multipoint orthogonality to show that near-extremal $n$-point functions, defined 
by the constraint $k\leq n-3$, vanish.

\subsection{Proof of the theorem}\label{proof}

As long as the topology of $\mathcal{F}_{{p}\mid\underline{q_1}\ldots \underline{q_{n-1} } }$ is fixed as in \eqref{fig1},
we can work with a fixed propagator structure.

To show that $\mathcal{F}_{{p}\mid\underline{q_1}\ldots \underline{q_{n-1} } }=0$, 
we will first make an argument for the part of  the diagram, consisting of $\cO_p$ and everything to the right. 
Take $\cO_p(x) T_{\underline{q_{r+1}}}(x_{r+1}), \dots , T_{\underline{q_{n-1}} }(x_{n-1})$ 
and consider Wick contracting the fundamental fields in all possible ways consistent with 
the propagator structure $\mathcal{F}_{{p}\mid\underline{q_1}\ldots \underline{q_{n-1} } }$. 
Since the number of bridges going from $\cO_p(x)$ to the right is fixed and less than $p$,
some of the fundamental fields inside  $\cO_p(x)$ will remain unlinked,  
thus resulting in a new half-BPS operator of lower charge, say $R$, inserted at $x$.
For each arrangement of Wick contractions, we can write this new half-BPS operator of 
charge $R$ in the trace basis, as illustrated below, 
    \be\label{hepfig1}
             \begin{tikzpicture}
		\def\lato 	{3.5}
		
		\def\xuno	{-.5}
		\def\yuno	{1.5}
		
		\def\xdue {\xuno+\lato}
		\def\ydue {1.5}
		
		\def\xtre {\xuno-\lato*0.5}
		\def\ytre {1.5}

		\def\xquat{\xuno-\lato}
		\def\yquat {0.5}


		\draw[thick] (\xuno,\yuno+.1)  --  (\xdue,\ydue+.1);
		\draw[thick] (\xuno,\yuno)  --  (\xdue,\ydue);
		\draw[thick] (\xuno,\yuno-.1)  --  (\xdue,\ydue-.1);

		\draw[thick] (\xuno,\yuno+.1)  --  (\xtre,\ydue+.1);
		\draw[thick] (\xuno,\yuno)  --  (\xtre,\ydue);
		\draw[thick] (\xuno,\yuno-.1)  --  (\xtre,\ydue-.1);


		\draw[fill=red!25,draw=black] (\xuno,\yuno) circle (.3cm);
		\draw[ fill=green!25,draw=black] (\xdue,\ydue) circle (1.2cm);

		\draw  (\xtre-.55,\ytre-.2)  				node[left]			
													 {$\displaystyle \prod g_{ij}^{d_{ij}}
													\sum_{\underline R \, \vdash\, R}\ C_{\underline R}\ T_{\underline R}(x)\ =$};
		
				
		\draw (\xuno,\yuno+.3) 				 node[above] 							{$\mathcal{O}_{p}(x)$};
		
		\draw (\xdue+.4,\ydue-.5) 			 	node[left,font=\footnotesize] 				{$T_{ \underline{ q_{n-1} } }$};
		\draw[fill=black] (\xdue+.4,\ydue-.5)			circle (.05cm);
		
		\draw (\xdue+.4,\ydue) 			 	node[left,font=\footnotesize] 				{$\ldots\ \ $};
		
		\draw (\xdue+.4,\ydue+.5) 			 	node[left,font=\footnotesize] 				{$T_{\underline{ q_{r+1} } }$};
		\draw[fill=black] (\xdue+.4,\ydue+.5)			circle (.05cm);
		
		\draw (\xdue+1.6,\ydue)  node {$\qquad$};

\end{tikzpicture}
\ee 
The sum is over operators $T_{\underline{R}}(x)$, indexed by partitions of $R$, 
and multiplied by a coefficient $C_{\underline{R}}$. 
The (partial) propagator structure $ \prod g_{ij}^{d_{ij}}$ is the one we assigned and it is factored out. 

The initial propagator structure is now computed as

\be\label{figgg1}
             \begin{tikzpicture}
		\def\lato 	{1.5}
		
		\def\xuno	{-.5}
		\def\yuno	{1.5}
		
		\def\xdue {\xuno+\lato}
		\def\ydue {1.5}
		
		\def\xtre {\xuno-\lato}
		\def\ytre {1.5}

		\def\xquat{\xuno-\lato}
		\def\yquat {0.5}


		\draw[thick] (\xuno,\yuno+.1)  --  (\xdue,\ydue+.1);
		\draw[thick] (\xuno,\yuno)  --  (\xdue,\ydue);
		\draw[thick] (\xuno,\yuno-.1)  --  (\xdue,\ydue-.1);

		\draw[thick] (\xuno,\yuno+.1)  --  (\xtre,\ydue+.1);
		\draw[thick] (\xuno,\yuno)  --  (\xtre,\ydue);
		\draw[thick] (\xuno,\yuno-.1)  --  (\xtre,\ydue-.1);


		\draw[ fill=green!25,draw=black] (\xtre,\ytre) circle (1.2cm);
		\draw[ fill=black,draw=black] (\xdue,\ydue) circle (.2cm);	
		

		\draw (\xdue,\ydue+.25)				 node[above] 							{$T_R(x)$};

		\draw (\xtre-.4,\ydue-.5) 			 	node[right,font=\footnotesize] 				{$\ \ T_{ \underline{ q_r  }  }$};
		\draw[fill=black] (\xtre-.4,\ydue-.5)			circle (.05cm);
		
		\draw (\xtre-.4,\ydue) 			 	node[right,font=\footnotesize] 				{$\ \ldots$};
		
		\draw (\xtre-.4,\ydue+.5) 			 	node[right,font=\footnotesize] 				{$\ \ T_{\underline{q_1 }  }$};
		\draw[fill=black] (\xtre-.4,\ydue+.5)			circle (.05cm);
		
		\draw (\xtre-1.5,\ydue)						node[left] 			{$\displaystyle \mathcal{F}_{{p}\mid\underline{q_1}\ldots \underline{q_{n-1} } }
															\ = \prod g_{ij}^{d_{ij}} \sum_{\underline R \, \vdash\, R} C_{\underline R}$};
		\draw (\xdue+.5,\ydue)  node {$\qquad$};

\end{tikzpicture}
\ee

The next step is to find the coefficients $C_R$, and we do so by sandwiching the equation \eqref{hepfig1} 
with operators $T_{R'}(x')$ at some auxiliary location $x'$. 
Graphically we obtain
    \be\label{hepfig2}
             \begin{tikzpicture}
		\def\lato 	{3.5}
		
		\def\xuno	{-.5}
		\def\yuno	{1.5}
		
		\def\xdue {\xuno+\lato}
		\def\ydue {1.5}
		
		\def\xtre {\xuno-\lato*0.5}
		\def\ytre {1.5}

		\def\xquat{\xuno-\lato}
		\def\yquat {0.5}


		\draw[thick] (\xuno,\yuno+.1)  --  (\xdue,\ydue+.1);
		\draw[thick] (\xuno,\yuno)  --  (\xdue,\ydue);
		\draw[thick] (\xuno,\yuno-.1)  --  (\xdue,\ydue-.1);

		\draw[thick] (\xuno,\yuno+.1)  --  (\xtre,\ydue+.1);
		\draw[thick] (\xuno,\yuno)  --  (\xtre,\ydue);
		\draw[thick] (\xuno,\yuno-.1)  --  (\xtre,\ydue-.1);


		\draw[fill=red!25,draw=black] (\xuno,\yuno) circle (.3cm);
		\draw[ fill=green!25,draw=black] (\xdue,\ydue) circle (1.2cm);
		\draw[ fill=black,draw=black] (\xtre,\ytre) circle (.2cm);

		\draw  (\xtre-.5,\ytre-.2)  				node[left]			
													 {$\displaystyle   \prod g_{ij}^{d_{ij}}
													\sum_{ \underline{ R'}, \underline{R}  \, \vdash\, R}\ C_{\underline R}\ 
																\langle T_{\underline R'}(x') T_{\underline R}(x)\rangle =\ $};
		
				
		\draw (\xuno,\yuno+.3) 				 node[above] 							{$\mathcal{O}_{p}(x)$};
		
		\draw (\xdue+.4,\ydue-.5) 			 	node[left,font=\footnotesize] 				{$T_{ \underline{ q_{n-1} } }$};
		\draw[fill=black] (\xdue+.4,\ydue-.5)			circle (.05cm);
		
		\draw (\xdue+.4,\ydue) 			 	node[left,font=\footnotesize] 				{$\ldots\ \ $};
		
		\draw (\xdue+.4,\ydue+.5) 			 	node[left,font=\footnotesize] 				{$T_{\underline{ q_{r+1} } }$};
		\draw[fill=black] (\xdue+.4,\ydue+.5)			circle (.05cm);
		

		\draw (\xtre,\ytre+.25) 						 node[above] 							{$T_{\underline{R'}}(x')$};

%
%
%
%

\end{tikzpicture}
\ee 
This is a vector equation for $C_{\underline{R}}$, which we can solve by inverting 
the matrix of two point functions. It follows that the coefficients $C_{\underline R}$ 
are given by computing another dumbbell diagram. The two-point functions $\langle T_{\underline R'}(x') T_{\underline R}(x)\rangle$
are given by $g_{x' x}^{R}$, which we can factor out,  times the color factor, 
$\mathcal{C}_{\underline{R},\underline{R'}}$. For clarity we will keep using the notation 
$\langle T_{\underline R'}(x') T_{\underline R}(x)\rangle$.

We can now replace the $C_{\underline R}$ and rewrite the original 
$\mathcal{F}_{p\mid\underline{q_1}\ldots \underline{q_{n-1} } }$ as follows, 

\be\label{factor}
             \begin{tikzpicture}

		\def\lato 	{1}
		
		\def\xuno	{-.5}
		\def\yuno	{-1.5}

		\def\xduesecond {\xuno+3*\lato}

		\def\ydue {-1.5}
		
		\def\xtre {\xuno-\lato}
		\def\xtreprime {\xuno-3.5*\lato}		
		\def\xtresecond {\xuno-5.5*\lato}
		
		\def\ytre {-1.5}


		\draw[thick] (\xuno,\yuno+.1)  --  (\xduesecond,\ydue+.1);
		\draw[thick] (\xuno,\yuno)  --  (\xduesecond,\ydue);
		\draw[thick] (\xuno,\yuno-.1)  --  (\xduesecond,\ydue-.1);

		\draw[thick] (\xuno,\yuno+.1)  --  (\xtre,\ydue+.1);
		\draw[thick] (\xuno,\yuno)  --  (\xtre,\ydue);
		\draw[thick] (\xuno,\yuno-.1)  --  (\xtre,\ydue-.1);

		\draw[thick] (\xtreprime,\yuno+.1)  --  	(\xtresecond,\ydue+.1);
		\draw[thick] (\xtreprime,\yuno)  --  		(\xtresecond,\ydue);
		\draw[thick] (\xtreprime,\yuno-.1)  --  		(\xtresecond,\ydue-.1);


		\draw[fill=red!25,draw=black] (\xuno,\yuno) circle (.3cm);

		\draw[ fill=green!25,draw=black] (\xduesecond,\ydue) circle (1.2cm);

		\draw[ fill=black,draw=black] (\xtre,\ytre) circle (.2cm);	
		\draw (\xtre-.2,\ydue+.25) 				 node[above] 							{$T_{\underline{R'}}$};
		\draw (\xtre-1.2,\ydue) 					 node[below, font=\footnotesize]		{$\langle T_{\underline{R}} T_{\underline{ R'} } \rangle^{-1}$};

		\draw[ fill=black,draw=black] (\xtreprime,\ydue) circle (.2cm);	
		\draw[ fill=green!25,draw=black] (\xtresecond,\ydue) circle (1.2cm);
		\draw (\xtreprime,\ydue+.25) 				 node[above] 							{$T_{\underline{R}}$};

				
		\draw (\xuno,\yuno+.3) 				 node[above] 								{$ \mathcal{O}_{p}$};
		
		\draw (\xduesecond+.4,\ydue-.5) 			 	node[left,font=\footnotesize] 			{$T_{ \underline{ q_{n-1} } }$};
		\draw[fill=black] (\xduesecond+.4,\ydue-.5)			circle (.05cm);
		
		\draw (\xduesecond+.4,\ydue) 			 	node[left,font=\footnotesize] 				{$\ldots\ \ $};
		
		\draw (\xduesecond+.4,\ydue+.5) 			 	node[left,font=\footnotesize] 			{$T_{\underline{ q_{r+1} } }$};
		\draw[fill=black] (\xduesecond+.4,\ydue+.5)			circle (.05cm);
		
		
		\draw (\xtresecond-.4,\ydue-.5) 			 	node[right,font=\footnotesize] 				{$\ \ T_{ \underline{ q_r  }  }$};
		\draw[fill=black] (\xtresecond-.4,\ydue-.5)			circle (.05cm);
		
		\draw (\xtresecond-.4,\ydue) 			 	node[right,font=\footnotesize] 				{$\ \ldots$};
		
		\draw (\xtresecond-.4,\ydue+.5) 			 	node[right,font=\footnotesize] 				{$\ \ T_{\underline{q_1 }  }$};
		\draw[fill=black] (\xtresecond-.4,\ydue+.5)			circle (.05cm);

		\draw (\xtresecond-1.25,\ydue-.1)						node[left] 			{$\displaystyle
																					\sum_{\underline{R'}, \underline{R}\,\vdash\, {R}  }$};
		\draw (\xtresecond-1.25,\ydue+1.5)						node[left] 			{$\mathcal{F}_{p\mid\underline{q_1}\ldots \underline{q_{n-1} } }\simeq$};

\end{tikzpicture}
\ee

Repeating a similar discussion for the left hand side of $\mathcal{F}_{p\mid\underline{q_1}\ldots \underline{q_{n-1} } }$ 
we conclude then that the original propagator structure can be rewritten as   

\be\label{factor}
             \begin{tikzpicture}

		\def\lato 	{1}
		
		\def\xuno	{-.5}
		\def\yuno	{-1.5}
		
		\def\xdue {\xuno+\lato}
		\def\xdueprime {\xuno+3.5*\lato}		
		\def\xduesecond {\xuno+5.5*\lato}

		\def\ydue {-1.5}
		
		\def\xtre {\xuno-\lato}
		\def\xtreprime {\xuno-3.5*\lato}		
		\def\xtresecond {\xuno-5.5*\lato}
		
		\def\ytre {-1.5}


		\draw[thick] (\xuno,\yuno+.1)  --  (\xdue,\ydue+.1);
		\draw[thick] (\xuno,\yuno)  --  (\xdue,\ydue);
		\draw[thick] (\xuno,\yuno-.1)  --  (\xdue,\ydue-.1);
		
		\draw[thick] (\xdueprime,\yuno+.1)  --  	(\xduesecond,\ydue+.1);
		\draw[thick] (\xdueprime,\yuno)  --  		(\xduesecond,\ydue);
		\draw[thick] (\xdueprime,\yuno-.1)  --  	(\xduesecond,\ydue-.1);

		\draw[thick] (\xuno,\yuno+.1)  --  (\xtre,\ydue+.1);
		\draw[thick] (\xuno,\yuno)  --  (\xtre,\ydue);
		\draw[thick] (\xuno,\yuno-.1)  --  (\xtre,\ydue-.1);

		\draw[thick] (\xtreprime,\yuno+.1)  --  	(\xtresecond,\ydue+.1);
		\draw[thick] (\xtreprime,\yuno)  --  		(\xtresecond,\ydue);
		\draw[thick] (\xtreprime,\yuno-.1)  --  		(\xtresecond,\ydue-.1);


		\draw[fill=red!25,draw=black] (\xuno,\yuno) circle (.3cm);
		
		\draw[ fill=black,draw=black] (\xdue,\ydue) circle (.2cm);	
		\draw (\xdue+.2,\ydue+.25) 				 node[above] 							{$T_{\underline{L'}}$};
		\draw (\xdue+1.3,\ydue) 					 node[below, font=\footnotesize]							{$\langle T_{\underline{L'}} T_{\underline{ L} } \rangle^{-1}$};

		\draw[ fill=black,draw=black] (\xdueprime,\ydue) circle (.2cm);	
		\draw[ fill=green!25,draw=black] (\xduesecond,\ydue) circle (1.2cm);
		\draw (\xdueprime,\ydue+.25) 				 node[above] 							{$T_{\underline{L}}$};

		\draw[ fill=black,draw=black] (\xtre,\ytre) circle (.2cm);	
		\draw (\xtre-.2,\ydue+.25) 				 node[above] 							{$T_{\underline{R'}}$};
		\draw (\xtre-1.2,\ydue) 					 node[below, font=\footnotesize]		{$\langle T_{\underline{R}} T_{\underline{ R'} } \rangle^{-1}$};

		\draw[ fill=black,draw=black] (\xtreprime,\ydue) circle (.2cm);	
		\draw[ fill=green!25,draw=black] (\xtresecond,\ydue) circle (1.2cm);
		\draw (\xtreprime,\ydue+.25) 				 node[above] 						{$T_{\underline{R}}$};

				
		\draw (\xuno,\yuno+.3) 				 node[above] 							{$ \mathcal{O}_{p}$};
		
		\draw (\xduesecond+.4,\ydue-.5) 			 	node[left,font=\footnotesize] 		{$T_{ \underline{ q_{n-1} } }$};
		\draw[fill=black] (\xduesecond+.4,\ydue-.5)			circle (.05cm);
		
		\draw (\xduesecond+.4,\ydue) 			 	node[left,font=\footnotesize] 				{$\ldots\ \ $};
		
		\draw (\xduesecond+.4,\ydue+.5) 			 	node[left,font=\footnotesize] 			{$T_{\underline{ q_{r+1} } }$};
		\draw[fill=black] (\xduesecond+.4,\ydue+.5)			circle (.05cm);
		
		
		\draw (\xtresecond-.4,\ydue-.5) 			 	node[right,font=\footnotesize] 				{$\ \ T_{ \underline{ q_r  }  }$};
		\draw[fill=black] (\xtresecond-.4,\ydue-.5)			circle (.05cm);
		
		\draw (\xtresecond-.4,\ydue) 			 	node[right,font=\footnotesize] 				{$\ \ldots$};
		
		\draw (\xtresecond-.4,\ydue+.5) 			 	node[right,font=\footnotesize] 			{$\ \ T_{\underline{q_1 }  }$};
		\draw[fill=black] (\xtresecond-.4,\ydue+.5)			circle (.05cm);

		\draw (\xtresecond-1.25,\ydue-.1)						node[left] 			{$\displaystyle \sum_{\underline{L }, \underline{L'}\,\vdash\, \underline{L}  }
																					\sum_{\underline{R}, \underline{R'}\,\vdash\, \underline{R}  }$};
		\draw (\xtresecond-1.25,\ydue+1.5)						node[left] 			{$\mathcal{F}_{p\mid\underline{q_1}\ldots \underline{q_{n-1} } }\simeq$};

\end{tikzpicture}
\ee

Crucially, the central connected diagram in~\eqref{factor} 
is a three-point function $\langle \cO_{p} T_{\underline{Q}}  T_{\underline{R}} \rangle $ 
with $p=Q+R$, thus extremal. To conclude we then have to show that any  
extremal three-point function vanishes, i.e.\!
 \begin{align}
 	\langle \cO_{p} T_{\underline{Q}}  T_{\underline{R}} \rangle = 0 \qquad ;\qquad p=Q+R\ .
 \end{align}
This is a direct consequence of the two-point function orthogonality, 
which we used in  (\ref{orthogonality}) to define the SPOs, together with 
the fact that extremal 3-point functions are directly related to the corresponding 
two-point function obtained by bringing  points 2 and 3 together (since  there are 
no propagators between $T_{\underline{Q}}$ and $T_{\underline{R}}$)
\begin{align}\label{pQR}
\langle \mathcal{O}_{p} T_{\underline{Q}}T_{\underline{R}} \rangle  
					= \left(\frac{g_{13}}{g_{12}}\right)^{\!\!R} \langle \mathcal{O}_{p}[T_{\underline{Q}}T_{\underline{R}}] \rangle=0\,,
\end{align}

This concludes our proof. We thus have shown that any propagator structure involving 
a single $\cO_p$, which becomes disconnected on removing $\cO_p$ vanishes.

\subsection{Near-Extremal correlators vanish}\label{nEvan}

When discussing $n$-point functions of half-BPS operators it is
useful to introduce the degree of extremality $k$ defined as follows. 
Let $p$ be the largest charge, and $q_{i=1,\ldots n-1}$ the others, then 
\begin{align}\label{intro_near_extr}
 k = \tfrac12 \left(-p + \sum_{i=1}^{n-1} q_i\right)\ .
\end{align}

This definition of $k$ should be now familiar from \eqref{intro_deg_extr}. 

For $k<0$  correlators vanish purely by $SU(4)$ symmetry. Extremal $(k=0)$ 
and next-to-extremal ($k=1)$ correlators, they all were shown to be non-renormalised 
in~\cite{DHoker:1999jke,Bianchi:1999ie,Eden:1999kw,Erdmenger:1999pz,Eden:2000gg}.

In~\cite{DHoker:2000xhf}  the concept of ``near-extremal'' correlators was introduced. 
These are correlators in which the degree of extremality $k$  is not too close to the 
number of points.  Specifically
\begin{align}\label{near_extr_sp}
\text{near extremal correlator: } \qquad \qquad   k \leq n-3\ .
\end{align}
 Notice that for $n\ge 5$, the near-extremal correlators go beyond the 
 extremal- and next-to-extremal cases mentioned above. 


We claim that  any near-extremal $SU(N)$ correlator where the largest charge 
operator is a single-particle operator vanishes:
\begin{align}\label{NE0}
	\langle \cO_{p}(x) T_{\underline{q_1}}(x_1) \dots T_{\underline{q_{n-1} }}(x_{n-1})  \rangle = 0 \qquad k \leq n-3\ .
\end{align}
As usual, $T_{\underline{q_i}}$ stands for any half-BPS operator 
with total charge $q_i$. We remark that {\em any} refers to any  single- or 
multi-trace operator or any combination of these. A corollary of this is that any 
near-extremal correlator involving only SPOs vanishes.
\begin{align}
\langle \cO_{p}(x) \cO_{q_1}(x_1) \dots \cO_{q_{n-1} }(x_{n-1})  \rangle = 0 \qquad k \leq n-3\ .
\end{align}

A similar statement can also be made in the  $U(N)$ theory, with the caveat that then 
it is only true for connected correlators: any {\em connected} near-extremal $U(N)$ 
correlator where the largest charge operator is a single-particle operator vanishes.

In order to prove \eqref{NE0} we will now show that every propagator structure contributing 
to this correlator has the dumbbell shape, as in~\eqref{fig1}, namely it becomes 
disconnected on removing $\cO_p$.

Let us first show that there are two few propagators between the operators 
$T_{\underline{q_i}}$ to connect them all together, and therefore there will be 
two green blobs as in~\eqref{fig1}.
Indeed, since there are $p$ legs coming out of  $\cO_{p}$, and $k$ is the 
excess of $\sum q_i$ to be a partition of $p$, there are  $k$ propagators among 
the $T_{\underline{q_i}}$s. But $k\leq n-3$ and there are $n-1$ operators $T_{\underline{q_i}}$. 
 The minimal scenario to link all the $T_{\underline{q_i}}$ would be to have a 
 necklace with $k=n-1$ bridges, which is not possible, both for $SU(N)$ and $U(N)$.  
 We would need at least two more points.

We understood that the topology of $\cO_{p}(x) T_{\underline{q_1}}(x_1) \dots T_{\underline{q_{n-1} }}(x_{n-1})$ 
is such that it is not possible to connect all $T_{\underline{q_i}}$. This would imply the diagram 
is dumbbell, but for the case in which the diagram is actually made of two disconnected 
parts, i.e.\! one green blob is actually on its own. For concreteness let's say 
$T_{\underline{q}_1}(x_1)\ldots T_{\underline{q}_r}(x_r)$ is disconnected from 
$\cO_p(x)T_{\underline{q}_{r+1}}(x_1)\ldots T_{\underline{q}_n-1}(x_{n-1})$, 
for some value of $r$. 
Let us see when this can happen: The number of bridges among the 
$T_{\underline{q_{r+1}}},\dots,T_{\underline{q_{n-1}} }$  which are not connecting with $\cO_p$ is 
\begin{align}
k_R= \tfrac12\left(-p+\sum_{i=r+1}^{n-1} q_i \right)
\end{align}
We can assume $\cO_p(x) T_{\underline{q_{r+1}}}(x_{r+1}),\dots,T_{\underline{q_{n-1}} }(x_{n-1})$ 
to remain connected on removing $\cO_p$, and since there are $n-r-2$ operators 
$T_{\underline{q_{r+1}}}(x_{r+1}),\dots,T_{\underline{q_{n-1}} }(x_{n-1})$, 
the minimum scenario would be to have them forming a tree, thus 
$k_R\ge n-r-2$.  If we now recall the inequality on $k$ we find
\begin{align}\label{around_95}
k=\tfrac12\left(-p+\sum_{i=r+1}^{n-1} q_i + \sum_{i=1}^{r} q_i\right)\leq n-3 \quad \Rightarrow \quad  \tfrac12 \sum_{i=1}^{r} q_i\leq n-3 -k_R\leq r-1
\end{align}
But $\tfrac12 \sum_{i=1}^{r} q_i$ is the total number of bridges among 
the fields $T_{\underline{q_1}}(x_1)  \ldots T_{\underline{q_r}}(x_r)$, which recall 
are disconnected from the rest of the diagram. Therefore, we learn that at most 
they form a tree. In a tree, some operators will necessarily be connected to others 
just with a single bridge. For the $U(N)$ theory this is possible if a number 
of $\cO_1$ fields are inserted in the original correlator. 
For $SU(N)$ theory there is not such a possibility.

We conclude that all $SU(N)$ diagrams contributing to near extremal correlators 
have a dumbbell shape ~\eqref{fig1} and hence vanish. Similarly for  all connected $U(N)$ diagrams.

\subsubsection{More cases: coincident points and multi-trace splitting}\label{more_cases}

A useful corollary of the vanishing of near-extremal correlators occurs for lower 
point correlators, say $m$-points, which are not near-extremal but have a number of multi-trace operators,
\begin{align}\label{more_point_corollary}
	\langle \cO_{p}(x) T_{\underline{q_1}}(x_1) \dots T_{\underline{q_{m-1} }}(x_{m-1})  \rangle \qquad ;\qquad k=\tfrac{1}{2}(-p+\sum q_i) \geq  m-3\ .
\end{align}
The idea is that when a number $r$ of operators are genuinely multi-trace,  
we can think of \eqref{more_point_corollary} as a correlator $\mathcal{F}$ with $n\ge m$ 
points (with specific propagator structure since there cannot be propagators 
inside a multi-trace). Namely
\begin{align}
\langle \cO_{p}(x) T_{\underline{q_1}}(x_1) \dots T_{\underline{q_{m-1} }}(x_{m-1})\rangle\quad  \rightarrow \quad \mathcal{F}(p|\underline{q'_1}\ldots \underline{q'_n})
\end{align}
The number of new points $n$ depends on the way we want to think of the 
multi-trace operators. The max we can do is to put all its parts on fictitious points. 
So if $l(\underline{q_i})$ measures the number of parts of $\underline{q_i}$, then
\begin{align}
m \leq n\leq (m-r)+\sum_{i=1}^r l(\underline{q_i})
\end{align}
If there is a value of $n$ such that $k\leq n-3$, the original correlator vanishes, 
since it can be though of as an $n$-point near-extremal correlator.

A simple example. For three-point functions we find
\begin{align}
{\rm if}\ \exists\ i\ {\rm such\ that}\  l(\underline{q_i})\ge 2\ {\rm and}\ k\leq 1\quad 
				\Rightarrow \quad \langle \mathcal{O}_{p}(x) T_{\underline{q_1} }(x_1)T_{\underline{q_2}} (x_2)\rangle =0
\end{align}
The claim follows because in this case we can always think of the 
three-point function as a near-extremal four-point function, which vanishes. 

Another example is
\begin{align}
{\rm if}\ \exists\ i\ {\rm such\ that}\  l(\underline{q_i})\ge 2\ {\rm and}\ k\leq m-2\quad
					\Rightarrow\quad \langle \cO_{p}(x) T_{\underline{q_1}}(x_1) \dots T_{\underline{q_{m-1} }}(x_{m-1})  \rangle=0
\end{align}
There are many possibilities involving splitting of more than one 
operator, but we do not list them all, since it should be clear how to 
generalise the examples above.

\section{Exact results for correlators of SPOs} \label{section_4}

\subsection{Maximally-Extremal correlators} \label{ME_section_main}

Given the vanishing of all near-extrema  correlators of SPOs in \eqref{near_extr_sp},  
the next cases to consider are  
\begin{align}
\langle \cO_{p}(x) \cO_{{q_1}} (x_1)\dots \cO_{{q_{n-1}}}(x_{n-1}) \rangle  \quad;\quad k = n-2\quad;\quad
k=\tfrac12 (-p+\sum q_i) 
\end{align}
which we call Maximally-Extremal (ME).

We will compute all ME correlators, beginning with three-points and generalising 
the argument to $n$-points. From now on we focus on the $SU(N)$ theory. 
The final result will have the following flavor, 
\begin{equation}
\langle \cO_{p}(x)\cO_{q_1}(x_1)\cdots \cO_{q_{n-1}}(x_{n-1}) \rangle_{\text{connected}}
= 
\langle \cO_p \cO_p \rangle\left( \sum _{trees\, \mathcal{T}}  |\mathcal{W}[\mathcal{T}] | \  \mathcal{T}
{\scriptsize \begingroup 
\setlength\arraycolsep{1pt} \left[ \begin{array}{ll} d_1 \ \  &b_{ij}  \\[-.2cm]  \ \ \ddots  \\[-.2cm]  &d_{n-1}\end{array}\right] \endgroup}\right)
\end{equation}
where the sum is over all trees $\mathcal{T}$ on $n-1$ points, each point with 
associated degree $d_i\ge 1$ (number of legs per point) such that $\sum_{i=1}^{n-1} d_i= 2(n-2)$.
The tree is specified by a sub-propagator structure $b_{ij}$ which we arrange 
into a matrix, and finally, 
\be
|\mathcal{W}[\mathcal{T}]|=\prod_{i=1}^{n-1}  q_i (q_i-1) \dots (q_i-d_i+1)
\ee
We will progressively get to this result. 


\subsubsection{3-point functions}\label{NE3}

ME three-points functions are also next-to-extremal three-point functions since $k=1$. 
In order to compute them, notice first that
\be
\label{n}
\langle \mathcal{O}_p(x) \mathcal{O}_{q_1}(x_1) \mathcal{O}_{q_2}(x_2) \rangle = \langle  \mathcal{O}_{p}(x) T_{q_1}(x_1) T_{q_2}(x_2)  \rangle\,, \qquad p=q_1+q_2-2\,,
\ee
In fact, the difference between the two is a sum of three-point functions with at least a multi-trace.
By the results in section \ref{more_cases}, they all vanish.

In the three-point function $\langle   \mathcal{O}_{p}  T_{q_1} T_{q_2} \rangle$ there 
is a single propagator between $T_{q_1}$ and $T_{q_2}$,  and therefore a single Wick 
contraction to do in between $x_1$ and $x_2$. If at the same time we bring together 
these two insertion points, we obtain the following result,

\be\label{s1}
\begin{tikzpicture}
	\draw [thick]       (-2.8,.8) -- (-1,.8);
	\draw [thick]       (-2.8,.8) -- (-2.8,.55);
	\draw[thick]    	  (-1,.8) --(-1,.5);
	\draw[]	(0,0)  node[] { $\lim_{x_1\rightarrow x_2} \Tr(\underbrace{\phi(x_1). \ldots .\phi(x_1) }_{q_1} ) \Tr( \underbrace{ \phi(x_2). \ldots .\phi(x_2)}_{q_2} )  
	\simeq  \Tr(\underbrace{\phi(x_2). \dots .\phi(x_2)}_p)$ + \ldots   };

\end{tikzpicture}
\ee
Intuitively, since each trace is cyclic invariant we are always considering a configuration 
which in double line notation looks like the following drawing (with many more free legs), 
\be
\begin{tikzpicture}

		\def\xuno	{-.5}
		\def\yuno	{2}
		
		\def\step 	{0}
		
		\def\xunoA	{5}
		\def\yunoA	{2}
	
		\def\xunoB	{5}
		\def\yunoB	{.5}

		\def\color{gray!60}
		\def\rad{.05cm}
		\def\spesso{thin}


		\draw (\xunoA-1,\yunoA) 				 node[above,font=\footnotesize] 							{$$};

	
\shadedraw [shading=axis,top color=red!25, draw=white] (\xunoA+2,\yunoA+.25) ellipse (2.5cm and .8cm)	;
\shadedraw [shading=axis, top color=green!25,draw=white ] (\xunoB+1.5,\yunoB) ellipse (2cm and .8cm)	;

%
		\draw[\spesso] (\xunoA,\yunoA) -- (\xunoB,\yunoB);
		\draw[\spesso] (\xunoA+1.1-.2*2,\yunoA) -- (\xunoB+1.1-.2*2,\yunoB) ;
%

		\draw[\spesso] (\xunoA,\yunoA)  .. controls (\xunoA+.2,\yunoA+1) and (\xunoA+4-.2,\yunoA+1) ..  (\xunoA+4,\yunoA);	
		
		\draw[fill=\color] (\xunoA,\yunoA)  circle (\rad);
		\draw[fill=\color] (\xunoA+4,\yunoA)  circle (\rad);
		
		\draw[\spesso] 	(\xunoA+1.1,\yunoA) arc (0:180: .2);
		\draw[fill=\color] (\xunoA+1.1,\yunoA)  circle (\rad);
		\draw[fill=\color] (\xunoA+1.1-.2*2,\yunoA)  circle (\rad);

		\draw[\spesso]  	 (\xunoA+2.1,\yunoA) arc (0:180: .2);				
		\draw[fill=\color] (\xunoA+2.1,\yunoA)  circle (\rad);
		\draw[fill=\color] (\xunoA+2.1-.2*2,\yunoA)  circle (\rad);
		
		\draw[\spesso]  	(\xunoA+3.1,\yunoA) arc (0:180: .2);
		\draw[fill=\color] (\xunoA+3.1,\yunoA)  circle (\rad);
		\draw[fill=\color] (\xunoA+3.1-.2*2,\yunoA)  circle (\rad);

		\draw[\spesso] (\xunoB,\yunoB)  .. controls (\xunoB+.2,\yunoB-1) and (\xunoB+3-.2,\yunoB-1) ..  (\xunoB+3,\yunoB);	
		
		\draw[fill=\color] (\xunoB,\yunoB)  circle (\rad);
		\draw[fill=\color] (\xunoB+3,\yunoB)  circle (\rad);
		
		\draw[\spesso] 	(\xunoB+1.1,\yunoB) arc (0:-180: .2);
		\draw[fill=\color] (\xunoB+1.1,\yunoB)  circle (\rad);
		\draw[fill=\color] (\xunoB+1.1-.2*2,\yunoB)  circle (\rad);

		\draw[\spesso]  	 (\xunoB+2.1,\yunoB) arc (0:-180: .2);				
		\draw[fill=\color] (\xunoB+2.1,\yunoB)  circle (\rad);
		\draw[fill=\color] (\xunoB+2.1-.2*2,\yunoB)  circle (\rad);

		\draw (\xunoA+5.5,\yunoA) 				 node[above,font=\footnotesize] 							{$= {T}_{q_1=4}(x_1)$};
		\draw (\xunoB+4.5,\yunoB-.3) 				 node[above,font=\footnotesize] 							{$= {T}_{q_2=3}(x_2)$};

	   \end{tikzpicture}
\ee 
But in general we find all the contributions of an OPE of scalars on the 
r.h.s of \eqref{s1}, however since we will contract that with the half-BPS 
operator $\cO_p$, any other term does not survive.  Clearly there are 
$q_1 q_2$ ways to perform this Wick contraction and we arrive at 
 \begin{align}
 	\langle  \mathcal{O}_{p}  T_{q_1} T_{q_2}  \rangle = q_1q_2\langle  \mathcal{O}_{p} T_p  \rangle\, 
 \end{align}
 Finally we have that $\langle  \mathcal{O}_{p}  T_p \rangle=\langle\cO_p \mathcal{O}_{p}  \rangle$ 
 from the defining orthogonality of $\cO_p$.

Putting all this together we arrive at the result that next-to-extremal 
three-point functions are simply expressed in terms of two-point functions,
\be
\label{n}
\langle \mathcal{O}_p \mathcal{O}_{q_1} \mathcal{O}_{q_2} \rangle = 
					q_1 q_2 \langle \mathcal{O}_{p} \mathcal{O}_{p} \rangle\qquad; \qquad p=q_1+q_2-2\,.
\ee

\subsubsection{$n$-point functions} \label{NEnpoint}

Now consider maximally extremal $n$-point functions.

As in the case of three-points in \eqref{n}, note that we can replace all operators,
{apart from the one with the largest charge}, by single trace operators,  
\begin{align}\label{ttto}
	\braket{\cO_{p}(x)\cO_{q_1}(x_1)\cdots \cO_{q_n}(x_{n-1}) }= 
						\braket{\cO_{p}(x) T_{q_1}(x_1)\cdots T_{q_{n-1}}(x_{n-1}) } \quad;\quad 
	k=n-2
\end{align}
The difference between the two is a sum of $n$-point functions involving 
at least one multi-trace operator. By the results in section \ref{more_cases}, they all vanish.

A non-vanishing connected diagram contributing to the correlator~\eqref{ttto} is such that, 
upon deleting  $\cO_{p}$ and all the bridges attached to it, there will be $n-2$ propagators 
amongst the $n-1$ operators $T_{q_i}$, just enough to connect the $T_{q_i}$s all 
together in a tree.  We draw a five-point example here below for clarity,

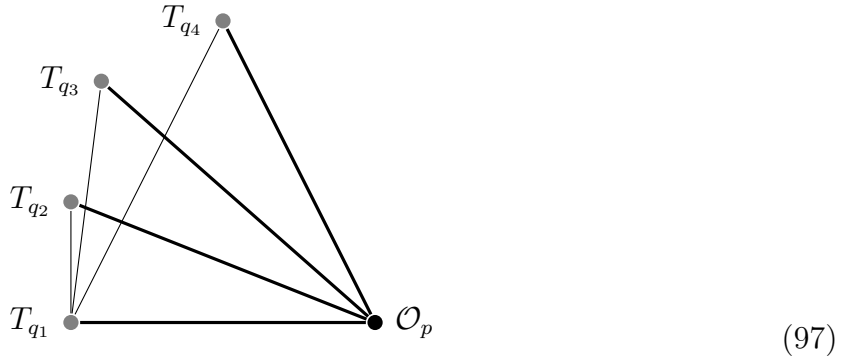
\be\label{example_tree1}
	\begin{tikzpicture}[scale=4,vertex/.style={circle, inner sep=0pt,fill=gray, very thick, minimum size=2mm}	]
	\node[circle, inner sep=0pt,fill=black, very thick, minimum size=2mm,label={right:$\cO_{p}$}]      (Op5) at (1,0)  {};
	\node[vertex,label={left:$T_{q_1}$}]      (Tp1) at (0,0)  {};
	\node[vertex,label={left:$T_{q_2}$}]      (Tp2) at (0,0.4)  {};
	\node[vertex,label={left:$T_{q_3}$}]      (Tp3) at (0.1,0.8)  {};
	\node[vertex,label={left:$T_{q_4}$}]      (Tp4) at (0.5,1)  {};
	\draw[very thick] (Op5)--(Tp1);
	\draw[very thick]  (Op5)--(Tp2);
	\draw[very thick] (Op5)--(Tp3);
	\draw[very thick] (Op5)--(Tp4);
	\draw[black](Tp1)--(Tp2);
	\draw[black](Tp1)--(Tp3);
	\draw[black](Tp1)--(Tp4);
	\end{tikzpicture}
\ee	
In this case the tree consists of $T_{q_1}$ connected to $T_{q_2}$, $T_{q_3}$ and $T_{q_4}$. 
Notice that in a tree there is one and only one bridge which connects pairs of operators. 
On each point we can have at most an $n$-pointed star with some $n\ge 1$. 

We have understood that a propagator structure contributing to a ME $n$-point 
correlator \eqref{ttto} contains a sub-propagator structure which is a tree $\mathcal{T}$. 
We can represent it by using our same notation for propagator structures 
in appendix \ref{Wick_sec}, namely
\begin{align}\label{Ttree_prop}
\mathcal{T}\simeq \left[\begin{array}{ccccc} 
d_1 & b_{12} & b_{13} & \ldots & b_{1n-1} \\  
& d_2 & b_{23} &  \ldots  & \vdots \\ 
& & \ddots  \\  \\ 
& & & & d_{n-1}   \end{array}\right]\qquad;\qquad 
\begin{array}{c}  d_i\ge 1 \\[.4cm]   b_{(ij)}\in\{ 0,1\}  \end{array}
\end{align}
where now $d_i\leq p_i$ is the number of legs of $T_{\underline{q_i}}$ entering the tree, 
and the number of bridges $b_{ij}$ from point $i$ to point $j$ can only take two possible 
values, $b_{(ij)}\in\{ 0,1\}$. The latter does not yet guarantees that \eqref{Ttree_prop} 
is a tree (for example necklace configurations can be arranged with $b_{ij}=1$). 
A unique way to label a tree is to use a Pr\"{u}fer sequence $s=(s_1\ldots )$. 
We explain how the Pr\"{u}fer algorithm works in appendix \ref{app_Prufer}. 
This point of view allows to reduce the pairwise computation 
of Wick contractions to a sequence, and in particular, gives us the result
\begin{align}\label{tree_wick_counting}
|\mathcal{W}[\mathcal{T}]|=\prod_{i=1}^{n-1}  q_i (q_i-1) \dots (q_i-d_i+1)
\end{align}
More details can be found in appendix \ref{app_Prufer}.

At this point, recall that the legs of the $T_{q_i}$s which do not enter the tree, 
form a bridge with the single-particle operator $\cO_p$. By performing the Wick 
contraction on the tree, leaf by leaf as in the Pr\"{u}fer algorithm, each time bringing 
together the points, we can use the same argument of section \ref{NE3} to infer that 
the net result of the tree is to glue together the $T_{q_i}$s to form a new single trace operator 
of charge $T_p$, together with higher trace terms.  These higher-trace terms will 
not contribute in the full ME correlator according to the discussion in section \ref{more_cases}.

The entire connected part of the ME correlator can thus be written as a sum over trees, 
multiplying the two point function $\langle \cO_p\cO_p\rangle$ overall:
\begin{equation}
\label{n-point}
\langle \cO_{p}(x)\cO_{q_1}(x_1)\cdots \cO_{q_{n-1}}(x_{n-1}) \rangle_{\text{connected}}
= \langle \cO_p \cO_p \rangle \left( \sum _{trees\, \mathcal{T}} |\mathcal{W}[\mathcal{T}] | \  \mathcal{T}
{\scriptsize \begingroup 
\setlength\arraycolsep{1pt} \left[ \begin{array}{ll} d_1 \ \  &b_{ij}  \\[-.2cm]  \ \ \ddots  \\[-.2cm]  &d_{n-1}\end{array}\right] \endgroup}\right)
\end{equation}
where the sum is restricted to $d_i\ge 1$ such that $\sum_{i=1}^{n-1} d_i= 2(n-2)$.

Note that the total number of trees contributing to this correlator 
is well known and given by Cayley's formula:    
\begin{equation}
\label{nofvectors}
(n{-}1)^{n-3}\ .
\end{equation}
Note also that many trees correspond to the same arrangement of 
degrees, $d_{i=1,\ldots n-1}$. Therefore many trees have the same value 
of $|\mathcal{W}[\mathcal{T}]|$. (In fact the tree is uniquely specified only 
once the configuration of bridges is also specified.) This degeneracy is counted 
by the multinomial coefficient 
\begin{equation}
\label{degeneracy}
\frac{(n-3)!}{\prod_{i=1}^{n-1} (d_i-1)!}\ .
\end{equation}


\subsubsection*{Disconnected contributions}

The ME correlator can have disconnected contributions even in the $SU(N)$ theory. 
Indeed there can be a disconnected component if and only if at least two of the $T_{q_i}$ 
operators are $\cO_2$. These are precisely necklace configurations of $SU(N)$, and products thereof. 
\footnote{Following a discussion similar to the one around \eqref{around_95},
we conclude that at least one of the connected components has to be made up entirely of $\cO_2$s. 
 }
For example, we might have
\begin{align}
	\langle  \prod_{i=1}^{K} \cO_2(x_i) \cO_{q_{k+1}} \dots \cO_{q_{n-1}} \cO_{p} \rangle_{{\rm discon.}} =  
	\langle \prod_{i=1}^{K} \cO_2(x_i) \rangle_{\rm necklace} \langle \cO_{q_{k+1}} \dots \cO_{q_{n-1}} \cO_{p} \rangle_{\rm conn.}  \qquad 	 \end{align}
Note that the factored expressions on the r.h.s~ are themselves  maximally extremal. 
In fact if there are $K$ $\cO_2$, then $n\rightarrow n'\equiv n-K$, and we find
\begin{align}
n-2=\tfrac{1}{2}( -p+ 2K+\sum_{i=\ell+1}^{n-1} q_i) \qquad \rightarrow \qquad (n'-2)= \tfrac{1}{2}( -p+\sum_{i=\ell+1}^{n-1} q_i)
\end{align}
Any other type of disconnected component will vanish as at least one of the 
connected pieces will be near extremal.

Let us conclude this section with  some  examples:

\subsubsection*{4-pt function}

Here there are three independent arrangements for the possible values of $d_i$, the number of 
legs of  position $i$ in the tree, we write them as $|d_1 d_2 d_3\rangle=\ket{211}$,$\ket{121}$, and $\ket{112}$. 
There is no degeneracy \eqref{degeneracy}, and the correlator is:
\begin{equation}\label{4-point}
\langle \cO_{p}\cO_{q_1}\cO_{q_2}\cO_{q_3}\rangle_{\text{connected}}= q_1 q_2 q_3 \braket{\cO_{p}\cO_{p}} \left[(q_1{-}1) \ket{211}+ (q_2{-}1)\ket{121}+ (q_3{-}1)\ket{112}\right]\ .
\end{equation}

\subsubsection*{5-pt function}

The number of trees is $4^2=16$. This is a case with degeneracy, since the arrangements 
of $|d_1 d_2 d_3 d_4\rangle$ are the following. Four non-degenerate configurations 
$\ket{3111}$,$\ket{1311}$,$\ket{1131}$,$\ket{1113}$. Then twice the configurations 
$\ket{2211}$,$\ket{2121}$,$\ket{2112}$,$\ket{1212}$,$\ket{1221}$,$\ket{1122}$. 
In fact, the degeneracy is two, as counted by \eqref{degeneracy}, and there are 
twelve trees plus the four non-degenerate, in total sixteen. 

Let us also notice that the contribution of some tree might vanish for low charges. For example, 
associated to $\ket{3111}$ is $q_1 q_2 q_3 q_4 (q_1-1)(q_1-2)$ that vanishes in the case $q_1=2$. 
Then, in the case of $\braket{\cO_2(x)\cO_2\cO_2\cO_2\cO_2}$ 
we are left with the twelve degenerate trees, all contributing with coefficient $q_1q_2q_3 q_4 =16$.

\subsection{Next-to-Maximally-Extremal correlators}

Going down in extremality, we consider Next-to-Maximally-Extremal (NME). 
These are $n$-point correlators with $k=n-1$, i.e.\! 
\begin{align}
\langle \cO_{p}(x) \cO_{{q_1}} (x_1)\dots \cO_{{q_{n-1}}}(x_{n-1}) \rangle \quad;\quad k = n-1\quad;\quad
k=\tfrac12 (-p+\sum q_i) 
\end{align}

We will first study three-point functions and them move to $n$-point functions.

\subsubsection{3-point functions}\label{NME_trepoint}

Starting with $n=3$ we are studying next-next-to-extremal three-point functions 
\begin{align}
\langle\cO_p(x) \cO_{q_1}(x_1) \cO_{q_2}(x_2)\rangle\qquad ;\qquad p=q_1+q_2-4.
\end{align} 
As for the NE case, these can be related to the corresponding two point function  
$\langle \mathcal{O}_{p} \mathcal{O}_{p} \rangle$,  but with a more  complicated 
coefficient containing non-factorisable polynomials. 

To evaluate the three-point function, we replace $\cO_{q_1}$ and $\cO_{q_2}$ 
by their respective expansions in the trace basis~\eqref{gentrb}. Differently from what 
happens in the ME case, see section \ref{NE3}, this time the expansion truncates for any term 
involving higher than double-trace operators  at a single point. These higher-traces will result 
in vanishing diagrams with $n\ge 5$ points, according to our general results in section \ref{more_cases}. 
For the same reason we can replace the double trace terms with products of single particle operators. 
Therefore we conclude that:
\begin{align}\label{opoqorexp}
\braket{\cO_p(x) \cO_{q_1} (x_1)\cO_{q_2}(x_2)}= & \ 
\braket{\cO_p(x)T_{q_1}(x_1) T_{q_2}(x_2)}  \\[.2cm]
&
\quad\ +\sum_{p_1=2}^{\frac{q_1}{2}}C_{p_1 (q_1-p_1)}\braket{\cO_p(x) \left[\cO_{p_1}\cO_{ q_1-p_1}\right] (x_1)\cO_{q_2}(x_2)} \\
&
\quad\ +\sum_{p_2=2}^{\frac{q_2}{2}}C_{p_2 (q_2-p_2)}\braket{ \cO_p(x) \cO_{q_1}(x_1) \left[\cO_{p_2} \cO_{q_2-p_2}\right](x_2)}
\end{align}
where $C_{(p_1 p_2)}$ is the mixing coefficient between single-particle states 
and double-trace operators $T_{p_1}T_{p_2}$. (We have written its explicit form 
in appendix, see~\eqref{eq:DoubleTraceCoeffs}).

The terms involving double-traces are equivalent to the four-point ME diagrams 
given in the previous section. Precisely in \eqref{4-point}. 
We find, 
\begin{align}
\braket{\cO_p(x) \left[\cO_{p_1}\cO_{ q_1-p_1}\right] (x_1)\cO_{q_2}(x_2)}= q_2 (q_2-1) p_1(q_1-p_1) |112\rangle \langle\cO_p\cO_p\rangle\\[.2cm]
\braket{ \cO_p(x) \cO_{q_1}(x_1) \left[\cO_{p_2} \cO_{q_2-p_2}\right](x_2)}=q_1(q_1-1) p_2(q_2-p_2)|112\rangle \langle\cO_p\cO_p\rangle
\end{align}
There is one term since $|211\rangle$ and $|121\rangle$ vanish because 
there cannot be a bridge within $\left[\cO_{p_1}\cO_{ q_1-p_1}\right](x_1)$.

The only unknown is therefore $\braket{\cO_p(x)T_{q_1}(x_1) T_{q_2}(x_2)}$. 
This consists of diagrams with two propagators between $T_{q_1}$ and $T_{q_2}$ 
which always count $q_1(q_1-1)q_2(q_2-1)/2$ Wick contractions, according to the 
general results in appendix \ref{Wick_sec} (see \eqref{Wick_contate}). However the 
net color factor has $2^2=1+1+2$ contributions for each arrangement of Wick contraction, 
depending on which part of the $SU(N)$ propagator in \eqref{sunprop} we consider, 
i.e.~whether we use the $U(N)$ part or the other one, $- \frac{1}{N} \delta_r^{ s} \delta_t^{ u}$, which we call the capping part. 

If we consider the two Wick contraction between $T_{q_1}(x_1)T_{q_2}(x_2)$ 
but we use just the capping part since that caps on the matrix indexes on each end of the propagator, 
we find effectively a dumbbell diagram, which vanishes. 

If we consider the two Wick contraction between $T_{q_1}(x_1)T_{q_2}(x_2)$ 
but we use just the $U(N)$ propagator,  we expect to produce effective operators 
of charge $q_1+q_2-4$ labelled by partitions of $4$. In $SU(N)$, we find two, 
the single trace and a double trace. Consider now that \emph{generically} we 
produce an effective double trace operator $[T_{q_1-2}T_{q_2-2}]$ and thus 
another dumbell diagram, which will vanish. 
We illustrate this mechanism with an example in double line notation,
\be\label{ex_doubleline130}
\begin{tikzpicture}

		\def\step 	{0}
		
		\def\xunoA	{5}
		\def\yunoA	{2}
	
		\def\xunoB	{5}
		\def\yunoB	{.5}

		\def\color{gray!60}
		\def\rad{.05cm}
		\def\spesso{thin}

	
\shadedraw [shading=axis,top color=red!25, draw=white] (\xunoA+2,\yunoA+.25) ellipse (2.5cm and .8cm)	;
\shadedraw [shading=axis, top color=green!25,draw=white ] (\xunoB+2,\yunoB-.25) ellipse (2.5cm and .8cm)	;

%
		\draw[\spesso] (\xunoA,\yunoA) -- (\xunoB,\yunoB);
		\draw[\spesso] (\xunoA+1.1-.2*2,\yunoA) -- (\xunoB+1.1-.2*2,\yunoB) ;
		
		\draw[\spesso] (\xunoA+2.1,\yunoA) -- (\xunoB+2.1,\yunoB);
		\draw[\spesso](\xunoA+3.1+.25-2*.2,\yunoA) -- (\xunoB+3.1+.25-2*.2,\yunoB) ;
%

		\draw[\spesso] (\xunoA,\yunoA)  .. controls (\xunoA+.2,\yunoA+1) and (\xunoA+4-.2,\yunoA+1) ..  (\xunoA+4,\yunoA);	
		
		\draw[fill=\color] (\xunoA,\yunoA)  		circle (\rad); 
		\draw[fill=\color] (\xunoA+4,\yunoA)  		circle (\rad);		
		\draw[fill=\color] (\xunoA,\yunoA)  		node[left,font=\scriptsize] {\it 1};
		\draw[fill=\color] (\xunoA+4,\yunoA)  		node[right,font=\scriptsize] {\it 8};
		
		\draw[\spesso] 	(\xunoA+1.1,\yunoA) arc (0:180: .2);
		\draw[fill=\color] (\xunoA+1.1,\yunoA)  circle (\rad);
		\draw[fill=\color] (\xunoA+1.1-.2*2,\yunoA)  circle (\rad);
		\draw[fill=\color] (\xunoA+1.1,\yunoA)  			node[below,font=\scriptsize] {\it 3};
		\draw[fill=\color] (\xunoA+1.1-.2*2,\yunoA)  		node[left,font=\scriptsize] {\it 2};

		\draw[\spesso]  	 (\xunoA+2.1,\yunoA) arc (0:180: .2);				
		\draw[fill=\color] (\xunoA+2.1,\yunoA)  circle (\rad);
		\draw[fill=\color] (\xunoA+2.1-.2*2,\yunoA)  circle (\rad);
		\draw[fill=\color] (\xunoA+2.1,\yunoA)  			node[right,font=\scriptsize] {\it 5};
		\draw[fill=\color] (\xunoA+2.1-.2*2,\yunoA)  		node[below,font=\scriptsize] {\it 4};
		
		\draw[\spesso]  	(\xunoA+3.1+.25,\yunoA) arc (0:180: .2);
		\draw[fill=\color] (\xunoA+3.1+.25,\yunoA)  circle (\rad);
		\draw[fill=\color] (\xunoA+3.1+.25-.2*2,\yunoA)  circle (\rad);
		\draw[fill=\color] (\xunoA+3.1+.25,\yunoA) 			node[below,font=\scriptsize] {\it 7};
		\draw[fill=\color] (\xunoA+3.1+.25-.2*2,\yunoA)  		node[left,font=\scriptsize] {\it 6};

		\draw[\spesso] (\xunoB,\yunoB)  .. controls (\xunoB+.2,\yunoB-1) and (\xunoB+4-.2,\yunoB-1) ..  (\xunoB+4,\yunoB);	
		
		\draw[fill=\color] (\xunoB,\yunoB)  circle (\rad);
		\draw[fill=\color] (\xunoB+4,\yunoB)  circle (\rad);
		\draw[fill=\color] (\xunoB,\yunoB)  node[left,font=\scriptsize] {\it 16};
		\draw[fill=\color] (\xunoB+4,\yunoB)  node[right,font=\scriptsize] {\it 9};
		
		\draw[\spesso] 	(\xunoB+1.1,\yunoB) arc (0:-180: .2);
		\draw[fill=\color] (\xunoB+1.1,\yunoB)  circle (\rad);
		\draw[fill=\color] (\xunoB+1.1-.2*2,\yunoB)  circle (\rad);
		\draw[fill=\color] (\xunoB+1.1,\yunoB)  			node[above,font=\scriptsize] {\it 14};
		\draw[fill=\color] (\xunoB+1.1-.2*2,\yunoB)  		node[left,font=\scriptsize] {\it 15};

		\draw[\spesso]  	 (\xunoB+2.1,\yunoB) arc (0:-180: .2);				
		\draw[fill=\color] (\xunoB+2.1,\yunoB)  circle (\rad);
		\draw[fill=\color] (\xunoB+2.1-.2*2,\yunoB)  circle (\rad);
		\draw[fill=\color] (\xunoB+2.1,\yunoB)  			node[right,font=\scriptsize] {\it 12};
		\draw[fill=\color] (\xunoB+2.1-.2*2,\yunoB)  		node[above,font=\scriptsize] {\it 13};		
		
		\draw[\spesso]  	(\xunoB+.25+3.1,\yunoB) arc (0:-180: .2);
		\draw[fill=\color] (\xunoB+.25+3.1,\yunoB)  circle (\rad);
		\draw[fill=\color] (\xunoB+.25+3.1-.2*2,\yunoB)  circle (\rad);
		\draw[fill=\color] (\xunoB+3.1+.25,\yunoB) 				node[above,font=\scriptsize] {\it 10};
		\draw[fill=\color] (\xunoB+3.1-.2*2,\yunoB)  		node[below,font=\scriptsize] {\it 11};		


			\draw[fill=\color] (\xunoB+5.6,\yunoB+1)		node[below] {$\simeq$};

		\def\xunoC	{\xunoA+4+3}
		\def\yunoC	{2}
	
		\def\xunoD	{\xunoB+4+3}
		\def\yunoD	{.5}

	
\shadedraw [shading=axis,top color=red!25, draw=white] (\xunoC+.6,\yunoC+.3) circle (.8cm)	;
\shadedraw [shading=axis, top color=green!25,draw=white ] (\xunoD+.6,\yunoD-.3) circle (.7cm)	;	

\shadedraw [shading=axis,top color=red!25, draw=white] (\xunoC+3.2,\yunoC+.3) circle (.8cm)	;
\shadedraw [shading=axis, top color=green!25,draw=white ] (\xunoD+3.2,\yunoD-.3) circle (.7cm)	;

%
		\draw[\spesso] (\xunoC,\yunoC) -- (\xunoD,\yunoD);
		\draw[\spesso] (\xunoC+1.1-.2*2,\yunoC) -- (\xunoD+1.1-.2*2,\yunoD) ;
		
		\draw[\spesso] (\xunoC+3.1,\yunoC) -- (\xunoD+3.1,\yunoD);
		\draw[\spesso](\xunoC+4.1+.25-2*.2,\yunoC) -- (\xunoD+4.1+.25-2*.2,\yunoD) ;

		\draw[\spesso] (\xunoC,\yunoC)  arc (180: 80: .8);
		\draw[fill=\color] (\xunoC,\yunoC)  		circle (\rad); 
		\draw[fill=\color] (\xunoC+1,\yunoC+.75)  circle (\rad);
				\draw[fill=\color] (\xunoC+1,\yunoC+.75) 			node[right,font=\scriptsize] {\it 8};

		\draw[\spesso] 	(\xunoC+1.1,\yunoC) arc (0:180: .2);
		\draw[fill=\color] (\xunoC+1.1,\yunoC)  circle (\rad);
		\draw[fill=\color] (\xunoC+1.1-.2*2,\yunoC)  circle (\rad);
		
				\draw[fill=\color] (\xunoC+1.1,\yunoC)			node[below,font=\scriptsize] {\it 3};

		\draw[\spesso]  	 (\xunoC+3.1,\yunoC) arc (0:180: .2);				
		\draw[fill=\color] (\xunoC+3.1,\yunoC)  circle (\rad);
		\draw[fill=\color] (\xunoC+3.1-.2*2,\yunoC)  circle (\rad);
		
						\draw[fill=\color] (\xunoC+3.1-.2*2,\yunoC)			node[below,font=\scriptsize] {\it 4};

		\draw[\spesso] (\xunoC+4.1+.25-.2*2,\yunoC) arc (0: 110: .8);
		\draw[fill=\color] (\xunoC+3.3-.2*2,\yunoC+.75)  circle (\rad);
		\draw[fill=\color] (\xunoC+4.1+.25-.2*2,\yunoC)  circle (\rad);
		
				\draw[fill=\color] (\xunoC+3.3-.2*2,\yunoC+.75)			node[left,font=\scriptsize] {\it 7};

		\draw[fill=\color] (\xunoD,\yunoD)  		circle (\rad); 
		\draw[fill=\color] (\xunoD+1,\yunoD-.75)  circle (\rad);
		\draw[\spesso] (\xunoD,\yunoD)  arc (180:280: .8);
		
						\draw[fill=\color] (\xunoD+1,\yunoD-.75)  			node[right,font=\scriptsize] {\it 9};

		\draw[\spesso] 	(\xunoD+1.1,\yunoD) arc (0:-180: .2);
		\draw[fill=\color] (\xunoD+1.1,\yunoD)  circle (\rad);
		\draw[fill=\color] (\xunoD+1.1-.2*2,\yunoD)  circle (\rad);
		
				\draw[fill=\color] (\xunoD+1.1,\yunoD)			node[above,font=\scriptsize] {\it 14};

		\draw[\spesso]  	 (\xunoD+3.1,\yunoD) arc (0:-180: .2);				
		\draw[fill=\color] (\xunoD+3.1,\yunoD)  circle (\rad);
		\draw[fill=\color] (\xunoD+3.1-.2*2,\yunoD)  circle (\rad);

							\draw[fill=\color]  (\xunoD+3.1-.2*2,\yunoD)			node[above,font=\scriptsize] {\it 13};

		\draw[\spesso] (\xunoD+4.1+.25-.2*2,\yunoD) arc (0: -110: .8);
		\draw[fill=\color] (\xunoD+3.3-.2*2,\yunoD-.75)  circle (\rad);
		\draw[fill=\color] (\xunoD+4.1+.25-.2*2,\yunoD)  circle (\rad);
		
					\draw[fill=\color]  (\xunoD+3.4-.2*2,\yunoD-.55)			node[left,font=\scriptsize] {\it 10};

					\draw[fill=\color] (\xunoD+4.6,\yunoB+1)		node[below] {$$};

	   \end{tikzpicture}
\ee

Instead, there are two ways of achieving a non vanishing result, in which the glued operator looks like the single-trace operator $T_p$. 
We can have $U(N)$ propagators one adjacent to the other, then 
\begin{align}\label{wick}
	\contraction{\Tr( \dots \phi_{ab} }{\phi_{bc} }{\dots ) \Tr( \dots }{\phi_{de} }
	\contraction[2ex]{\Tr( \dots }{\phi_{ab} }{\phi_{bc}\dots )\Tr(\dots \phi_{de}}{\phi_{ef}}
	\Tr(\underbrace{ \dots  \phi_{ab} \phi_{bc}\dots }_{q_1})\Tr(\underbrace{\dots\phi_{de} \phi_{ef} \dots }_{q_2})\delta^a_f\delta^e_b\delta^b_e\delta^d_c&=N\Tr(\underbrace{\phi \dots \phi}_{p})
\end{align}
In this case only $pq$ Wick contractions are non zero.
Finally, we can take a connected and a disconnected part, resulting in
\begin{align}	\label{wick2}
	\contraction{\Tr( \dots \phi^a_b \dots }{\phi^c_d  }{\dots) \Tr(\dots }{\phi^e_f}
	\contraction[2ex]{\Tr(\dots }{\phi^a_b }{\dots\phi^c_d)\Tr(\dots \phi^e_f\dots\ \  \ }{\phi^g_h}
	\Tr(\underbrace{ \dots  \phi^a_b \dots \phi^c_d\dots}_{q_1})\Tr(\underbrace{\dots \phi^e_f \dots \phi^g_h \dots}_{q_2})\Big( -\delta^a_h\delta^g_b\  \frac 1N \delta^c_d\delta^e_f-  \frac 1N\delta^a_b\delta^g_h\  \delta^c_f\delta^e_d\Big)&=-\frac 2N\Tr(\underbrace{\phi \dots  \phi  }_{p})
\end{align}

Inputting \eqref{wick} and \eqref{wick2} into a vev with $\cO_p$ and using that 
$\langle T_p \cO_p \rangle =\langle \cO_p \cO_p \rangle$ yields the final result, 
\begin{equation}
\label{nn}
\braket{T_{q_1} T_{q_2} \mathcal{O}_p}=q_1 q_2 \left[N-\frac{(q_1-1)(q_2-1)}{N} \right] \braket{\mathcal{O}_p \mathcal{O}_p}\ .
\end{equation}

We then obtain an explicit formula for the next-next-to-extremal three-
point functions of single-particle operators 
\begin{align}
&
\braket{\cO_p(x) \cO_{q_1} (x_1)\cO_{q_2}(x_2)}= \braket{\mathcal{O}_p \mathcal{O}_p} \Bigg[ 
q_1 q_2 \Big[N-\frac{(q_1-1)(q_2-1)}{N} \Big] + \notag\\
&
\qquad\sum_{p_1=2}^{\lfloor\frac{q_1}{2}\rfloor}C_{p_1 (q_1-p_1)}q_2 (q_2-1) p_1(q_1-p_1) 
+\sum_{p_2=2}^{\lfloor\frac{q_2}{2}\rfloor}C_{p_2 (q_2-p_2)}q_1(q_1-1) p_2(q_2-p_2) \Bigg]
\label{conclusion1}
\end{align}
The two sums are symmetric and can be performed with the explicit 
knowledge of $C_{p_1(q_1-p_1)}$ given in appendix (see \eqref{eq:DoubleTraceCoeffs}). 
We find 
\begin{align}
&
\sum_{p_1=2}^{\lfloor\frac{q_1}{2}\rfloor}C_{p_1 (q_1-p_1)}p_1(q_1-p_1) =\notag\\
&
 \frac{q_1}{2N(q_1-2)} \biggl[2N^2 + (q_1-1)_2 N + 2(q_1-2)_2 +  \frac{2(q_1-1)_2N (N)_{q_1}}{(N-q_1+1)_{q_1} - (N)_{q_1}} \biggr]
 \label{conclusion2}
\end{align}
%

\subsubsection{$n$-point functions}

NME $n$-point functions can be obtained in a way similar to the three-point 
functions and in this section we sketch how the computation goes. 

The definition of $k=\tfrac12 (-p+\sum q_i)=2$ selects an operator $\cO_p$, and
we call the others ``light".  Start by expanding all these ``light" operators 
in terms of the trace basis,  and note that almost all terms in the expansion vanish since 
they produce correlators equivalent to near-extremal higher point diagrams. 
The result is the following generalisation of the three-point expansion in ~\eqref{opoqorexp}, namely
\begin{equation}
\begin{split}
\braket{\mathcal{O}_{q_1} \dots \mathcal{O}_{q_{n-1}} \mathcal{O}_{p} } = & \braket{T_{q_1}\dots T_{q_{n-1}} \mathcal{O}_p(x_n)}+ \\ 
& \qquad \sum_{i=1}^{n-1} \sum_{p_i=2}^{\lfloor{\frac{q_i}{2}}\rfloor}C_{p_i (q_i-p_i)}\braket{\cO_{q_1} \dots \cO_{q_{i-1}}\left[O_{p_i}\mathcal{O}_{q_i -p_i}\right] \cO_{q_{i+1}} \dots  \mathcal{O}_p(x_n)}\ .
\end{split}
\end{equation}
This equation can be thought of as a separate equation for each contributing Feynman
 diagram independently.  Note that the correlators in the sum are all contributions to 
 $n+1$-point $N^{n-1}$-extremal correlators which are maximally extremal and given 
 in the previous section. The first term can then be  computed by doing the partial Wick 
 contractions $n-1$ in total on the singe trace operators $T_{p_1} \dots T_{p_{n-1}}$,  
 only keeping the relevant terms just as in~\eqref{wick}.

\subsection{3-point functions as multi-particle 2-point functions}

The work done in the previous sections has been to start from ME and NME three-point functions, compute them, 
and understand how to generalise the technique to $n$-points. In order to deal with the three-point functions
we substituted two out of three SPOs with their corresponding expansion in the trace basis,
and we reduced part of the job to compute a three point function with one single-particle 
and two traces. In this section instead we note an interesting feature of three-point function 
of SPO which does not require passing to the trace basis, and it is valid for any extremality. 
The relation works as follows, 
\begin{equation}
\label{eq:threepointfunction}
\left< \mathcal{O}_p \mathcal{O}_q \mathcal{O}_r\right> = \frac{1}{2^k k!} \langle\left[\mathcal{O}_p \mathcal{O}_q \right] [ \mathcal{O}_r \overbrace{\mathcal{O}_2 \cdots \mathcal{O}_2}^{k} \,]\rangle_{\text{connected}}, \qquad p + q - r = 2k\ .
\end{equation}
where the equality is for the color factor of the l.h.s.\! being the same as that of a 
two-point functions of multi-particle operators of SPOs on the r.h.s.

Let us first illustrate the case $k=3$ with the following picture, 

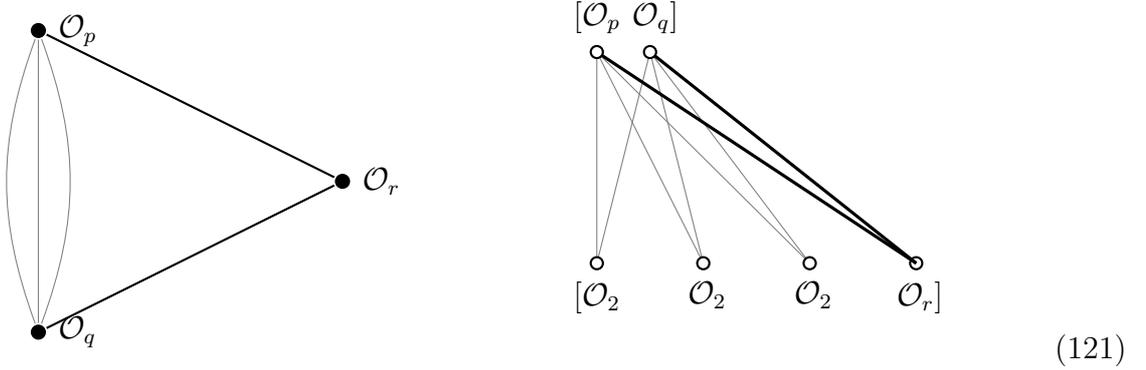
\be
	\begin{tikzpicture}[scale=4,vertex/.style={circle, inner sep=0pt,fill=black, very thick, minimum size=2mm}	]
	\node[vertex,label={right:$\cO_{p}$}]      (Op) at (0,1)  {};
	\node[vertex,label={right:$\cO_{q}$}]      (Oq) at (0,0)  {};
	\node[vertex,label={right:$\cO_{r}$}]      (Or) at (1,1/2)  {};
	\draw [bend left=20][gray] (Op) to (Oq);
	\draw [bend right=20][gray] (Op) to (Oq);
	\draw[gray] (Op) -- (Oq);
	\draw [thick] (Op) -- (Or);
	\draw [thick] (Or) -- (Oq);
	\end{tikzpicture}
\rule{2cm}{0pt}
\begin{tikzpicture}

\def\xuno {0}
\def\yuno {0}
\def\latouno{1.4}
\def\rad{.08cm}


		\draw[thin,gray] (\xuno,\yuno)--(\xuno,\yuno+2*\latouno)	;
		\draw[thin,gray] (\xuno,\yuno)--(\xuno+0.5*\latouno,\yuno+2*\latouno);	
		\draw[thin,gray] (\xuno+\latouno,\yuno)--(\xuno,\yuno+2*\latouno)	;
		\draw[thin,gray] (\xuno+\latouno,\yuno)--(\xuno+0.5*\latouno,\yuno+2*\latouno);
		\draw[thin,gray] (\xuno+2*\latouno,\yuno)--(\xuno,\yuno+2*\latouno)	;
		\draw[thin,gray] (\xuno+2*\latouno,\yuno)--(\xuno+0.5*\latouno,\yuno+2*\latouno);


		\draw[fill=white,thick] (\xuno,\yuno)					circle (\rad);
		\draw[fill=white,thick] (\xuno+\latouno,\yuno)			circle (\rad);
		\draw[fill=white,thick] (\xuno+2*\latouno,\yuno)			circle (\rad);
				\draw  (\xuno,\yuno-.1)								node[below] 				{$[\mathcal{O}_2$};
				\draw  (\xuno+.05+\latouno,\yuno-.1)						node[below] 				{$\mathcal{O}_2$};
				\draw  (\xuno+.05+2*\latouno,\yuno-.1)					node[below] 				{$\mathcal{O}_2$};
		
		\draw[fill=white,thick] (\xuno+3*\latouno,\yuno)			circle (\rad);
		\draw  (\xuno+.05+3*\latouno,\yuno-.1)					node[below] 				{$\mathcal{O}_r]$};

		\draw[very thick] (\xuno+3*\latouno,\yuno)--(\xuno,\yuno+2*\latouno)	;
		\draw[very thick] (\xuno+3*\latouno,\yuno)--(\xuno+0.5*\latouno,\yuno+2*\latouno);

		\draw[fill=white,thick] (\xuno,\yuno+2*\latouno)					circle (\rad);
		\draw[fill=white,thick] (\xuno+0.5*\latouno,\yuno+2*\latouno)		circle (\rad);	
		
			\draw  (\xuno,\yuno+.1+2*\latouno)							node[above] 					{$[\mathcal{O}_p$};
			\draw  ((\xuno+0.5*\latouno,\yuno+.1+2*\latouno)					node[above] 				{$\,\,\mathcal{O}_q]$};

		\draw (\xuno,\yuno-.8*\latouno)			node[right] 				{$$};
		\draw (\xuno+2*\latouno,\yuno)			node[right] 				{$$};

\end{tikzpicture}
\ee

The diagram on the left is the single diagram contributing to $\left< \mathcal{O}_p \mathcal{O}_q \mathcal{O}_r \right>$, 
whereas the one on the right is the only type of diagram contributing to 
$\langle\left[\mathcal{O}_p \mathcal{O}_q \right] [ \mathcal{O}_r {\mathcal{O}_2 \mathcal{O}_2 \mathcal{O}_2} \,]\rangle$.

To show \eqref{eq:threepointfunction}, consider where the two legs out of an $\cO_2$ can end. 
They can not go to $\cO_r$ as they are at the same point. If they both went to $\cO_p$ , 
this would result in a diagram of the form of a dumbbell in~\eqref{fig1}  (centered around $\cO_p$) 
which thus vanishes. Similarly if both legs go to $\cO_q$. The only exception to this is if $p$ or $q$ 
equals two in which case you can have a completely disconnected contribution - which also is 
absent since we specify the connected component. The only remaining possibility is then that 
one leg goes to $\cO_p$ and one to $\cO_q$ resulting in the diagram shown on the right. 
%
There are clearly $2^k k!$ different but equivalent diagrams of this sort,  arising from 
the $k!$ possibilities of swapping the propagators from $\cO_p$ to the $\cO_2$s 
and from the cyclic symmetry around each $\cO_2$.

The color factor of $\left< \mathcal{O}_p \mathcal{O}_q \mathcal{O}_r \right>$ 
is the same of one of the equivalent configuration of $\langle\left[\mathcal{O}_p \mathcal{O}_q \right] 
[ \mathcal{O}_r {\mathcal{O}_2 \ldots \mathcal{O}_2} \,]\rangle$ described above. This follows from the fact that 
\begin{align}
\begin{tikzpicture}[baseline={([yshift=-3pt]phiab.base)},scale=1,vertex/.style={circle, inner sep=0pt,fill=black, very thick, minimum size=2mm}	]
\node[vertex,label={left:$\phi^a_b$}]      (phiab) at (0,0)  {};
\node[vertex,label={right:$\phi^c_d$}]      (phicd) at (1,0)  {};
\draw [very thin,gray] (phiab) to (phicd);
\end{tikzpicture}
&=	
\contraction{}{\phi^a_b}{}{\phi^c_d}
	\phi^a_b\phi^c_d  = \delta^a_d\delta^c_b-\tfrac1N \delta^a_b\delta^c_d,\notag\\[5pt]
\begin{tikzpicture}[baseline={([yshift=-3pt]phiab.base)},scale=1,vertex/.style={circle, inner sep=0pt,fill=black, very thick, minimum size=2mm}	]
		\node[vertex,label={left:$\phi^a_b$}]      (phiab) at (0,0)  {};
		\node[vertex,label={right:$\phi^c_d$}]      (phicd) at (2,0)  {};
		\node[vertex,label={above:$\cO_2$}]      (O2) at (1,0)  {};
		\draw [very thin,gray] (phiab) to (O2);
\draw [very thin,gray] (O2) to (phicd);
		\end{tikzpicture}
&=	
\contraction{}{\phi^a_b}{}{\phi^e_f}
\contraction{\phi^a_b   \phi^e_f}{\phi^f_e}{}{\phi^c_d }
\phi^a_b   \phi^e_f \phi^f_e      \phi^c_d 
=\delta^a_d\delta^c_b-\tfrac1N \delta^a_b\delta^c_d
\end{align}
which conclude our proof of \eqref{eq:threepointfunction}.\footnote{The ideas of our proof here can be generalised 
to multipoint correlators as well. For example those which are equivalent to the l.h.s.\! of \eqref{eq:threepointfunctionmulti}.}

While the right hand side of~\eqref{eq:threepointfunction} is restricted to the connected part of 
the two point function (thinking of it as a limit of a higher point function), rather than the full two-point function,
we need this distinction 
only 
when  $p$ or $q$ equals $2$ 
and $k=1$, where instead we have
\begin{equation}
\label{eq:connectedmulti2point}
\left< \mathcal{O}_q \mathcal{O}_2 \mathcal{O}_q \right> = \frac{1}{2} \left( \left< \left[\mathcal{O}_q \mathcal{O}_2 \right] \left[ \mathcal{O}_q \mathcal{O}_2 \right] \right> - \left< \mathcal{O}_q \mathcal{O}_q \right> \left< \mathcal{O}_2 \mathcal{O}_2 \right> \right).
\end{equation}

Note that the condition giving the value of $k$ in \eqref{eq:threepointfunction} is dependent 
on the order of $\mathcal{O}_p$, $\mathcal{O}_q$ and $\mathcal{O}_r$; in particular it distinguishes 
$\mathcal{O}_r$ from the other two operators.  However the color factor of the 
three-point function \emph{does not depend on this ordering}.
%
For example, consider the three point function of $\mathcal{O}_3$, $\mathcal{O}_4$ 
and $\mathcal{O}_5$, we can have three multi-particle two-point functions with 
$k=1,2,3$ respectively,  
\newcommand{\veq}{\mathrel{\rotatebox{90}{$=$}}}
\be
\begingroup 
\setlength\arraycolsep{1pt}
\begin{array}{ccccl}
\left<\mathcal{O}_3 \mathcal{O}_4 \mathcal{O}_5 \right>&=&\left<\mathcal{O}_3 \mathcal{O}_5 \mathcal{O}_4 \right>&=&\ \ \ \, \left<\mathcal{O}_4 \mathcal{O}_5 \mathcal{O}_3 \right>\ \ =\ \ 
\displaystyle \frac{60\prod_{i=1}^4 (N^2-i^2)}{N(5+N^2)}\\
\veq & & \veq& &\rule{1.2cm}{0pt}\veq\\
 \frac{1}{2} \left< \left[ \mathcal{O}_3 \mathcal{O}_4 \right] \left[ \mathcal{O}_5 \mathcal{O}_2 \right] \right>& &
 \frac{1}{8}\left< \left[ \mathcal{O}_3 \mathcal{O}_5 \right] \left[ \mathcal{O}_4 \mathcal{O}_2 \mathcal{O}_2 \right] \right>& &
\frac{1}{48} \left< \left[ \mathcal{O}_4 \mathcal{O}_5 \right] \left[ \mathcal{O}_3 \mathcal{O}_2 \mathcal{O}_2 \mathcal{O}_2\right] \right>
\end{array}
\endgroup
\ee
All three multi-particle two-point functions are then equal!.

We conclude that for a triplet of single-particle operators all multi-particle two-point 
functions  which correspond to different dispositions of the three SPOs give the same 
color factor up to a multiplicity counted by $2^k k!$. 

Also note that while the discussion above required $\cO_p$ and $\cO_q$ to be SPOs, 
it nowhere relied on $\cO_r$ to be an SPO. Thus the following more general relation 
holds for any half-BPS operator $T_{r_1,\dots, r_l}$:
\begin{equation}
\label{eq:threepointfunctionmulti}
\left< \mathcal{O}_p \mathcal{O}_q T_{r_1,\dots, r_l} \right> = 
				\frac{1}{2^k k!} \langle\left[\mathcal{O}_p \mathcal{O}_q \right] [T_{r_1,\dots, r_l} \overbrace{\mathcal{O}_2 \cdots \mathcal{O}_2}^{k} \,]\rangle|_{\text{connected}},
\end{equation}
with $r_1 + ... + r_l = r$ and $p + q - r = 2k$.


\subsection{On correlators with lower extremality}\label{sec_on_lower_extremality}

We understood how to classify free theory correlators according to their degree 
of extremality w.r.t.~ME correlator.  Lowering this degree increases the complexity 
of the computation. ME are the simplest non vanishing correlators and are computed 
in terms of tree graph. As function of the charges $p,q_{i=1,\ldots ,n-1}$,  the charge dependence 
is fully factorised for each tree graph, with the factor having a clean interpretation.  
NME showed more structure. We expect that NNME will have more, and so on so forth. 

The complexity of NNME is already evident in the three-point functions. For example
$\braket{\cO_p \cO_{q_1} \cO_{q_2}}$ with $r=p+q-6$ will have a contribution of the form 
$\braket{\mathcal{O}_p T_{q_1}T_{q_2}}$, with three bridges between $T_{q_1}$ and $T_{q_2}$.  
We show few cases are
\begin{align}
\label{3ptex_1}
\langle \cO_6 T_6 T_6 \rangle &= \left( 300 + \frac{7200}{N^2} + 36 N^2 \right) \langle \cO_6 \cO_6 \rangle \\
\label{3ptex_2}
\langle \cO_7 T_7 T_6 \rangle&=\left( 840 + \frac{12600}{N^2}+ 42 N^2\right)\langle \cO_7 \cO_7 \rangle \\
\label{3ptex_3}
\langle \cO_8 T_8 T_6 \rangle&=\left( 1680+\frac{20160}{N^2}  + 48 N^2 \right) \langle \cO_8 \cO_8\rangle\\
\label{3ptex_4}
\langle \cO_8 T_7 T_7 \rangle&=\left( 1911+ \frac{22050}{N^2}  + 49 N^2 \right) \langle \cO_8 \cO_8\rangle\\
\label{3ptex_5}
\langle \cO_9 T_8 T_7 \rangle&=\left( 3528+ \frac{35280}{N^2}  + 56 N^2 \right) \langle \cO_9 \cO_9\rangle
\end{align}
and we attach an ancillary file to show the reader how complicated are the correlators 
$\langle T_{\underline{p}}T_{q_1}T_{q_2}\rangle$ before we convert those to the 
single-particle operator $\cO_p$.  About \eqref{3ptex_1}-\eqref{3ptex_5}, 
we can say a number of things based on the possible scenarios in double-line notation:
\begin{itemize}
\item[(1)] $O(1/N^2)$ contributions can only arise from picking the three $SU(N)$ 
propagators in the configuration: two capping parts ($- \frac{1}{N} \delta_r^{ s} \delta_t^{ u}$) and a $U(N)$ part in \eqref{sunprop}. 
When the two capping parts are used we are left with $T_{q_1-2}(x_1)T_{q_2-2}(x_2)$ 
with no link in between. The $U(N)$ propagator has the effect of producing an 
effective $T_p$ operator. 
\item[(2)] $O(1/N)$ contributions can only arise from picking the propagators 
in the configuration:  one capping part and two $U(N)$ parts, and these two $U(N)$ 
propagators have to be generic, i.e.\! not consecutive. When the capping part is used, 
we reduce the number of legs to $T_{q_1-1}(x_1)T_{q_2-1}(x_2)$ with no link in between. 
When we attach the two $U(N)$ propagators we are in a situation like NME 
(see \eqref{ex_doubleline130}) and the result will vanish. 
\item[(3)] $O(N)$ contributions can only arise from picking the propagators in the 
configuration:  three $U(N)$ parts,  but two $U(N)$ propagators have to be consecutive 
and one generic. The consecutive propagators produce a factor of $N$, reduce the number 
of legs to $T_{q_1-2}(x_1)T_{q_2-2}(x_2)$ and link the operators. Considering the other 
generic $U(N)$ propagators we are again in a situation like NME (see \eqref{ex_doubleline130}) 
and the result will vanish. 
\item[(4)]  $O(N^2)$ contributions can only arise from picking the propagators in 
the configuration:  three $U(N)$ parts, but all consecutive. 
\item[(5)] $O(1)$ contributions can arise from two configurations: 
a) one capping part and two $U(N)$ parts, but these two have to be consecutive. 
b) three $U(N)$ parts, generic.  Notice that a) is necessarily negative, whereas b) is positive.  
\end{itemize}
The number of Wick contractions is $(q_1-2)_3(q_1-2)_3/3!$, and for each arrangement 
we have a multiplicity $2^3=1+1+3+3$, depending on the way we pick the propagators, 
i.e.~whether we consider the $U(N)$ or the capping part.  It is clear that item $(1)$ contributes fully, 
and item $(4)$ contributes only with $q_1q_2$. 
Thus we have found
\begin{align}
\braket{\mathcal{O}_p T_{q_1}T_{q_2}} = 
 \biggl[ q_1 q_2 {N^2}+ 3\times\frac{(q_1 - 2)_3 (q_2 - 2)_3}{ 3!N^2} + \left( -\frac{1}{N} (Nc_a) + c_b \right) \biggr]\braket{\cO_p \cO_p} \notag
\end{align}
where $c_a,c_b>0$ and depend on the charges $q_1$ and $q_2$.\footnote{A simple guess is $c_b-c_a=(q_1-1)_2 (q_2-1)_2 ( \frac{1}{3}(q_1 - 5)(q_2 - 5) - \frac{1}{4}(q_1-6) (q_2-6))$.}   

It is not straightforward at this point to extract further information from the combinatorics, 
and in practise the dependence on the charges becomes hidden in the combinatorics. 
Nevertheless, the lesson we learn from our considerations about NNME three point function above
is the surprising simplicity of the final result, that once more points to the possibility of 
understanding single-particle multipoint correlators in a way that eschews from a brute force 
computation. We hope to make this observation more concrete in the future.

\section{The half-BPS OPE}\label{hbpsope}

In this section we illustrate an alternative approach to obtaining multi-point 
correlation functions, the half-BPS OPE, which among many interesting features
offers a different understanding of the colour dependence of the correlators. 
The idea of the half-BPS OPE is  simply to bootstrap the free theory correlators 
by projecting it onto the half-BPS states.\footnote{This approach has been pursued 
also in \cite{Caron-Huot:2018kta} at order $1/N^2$. }

The half-BPS OPE follows directly from the full super OPE in $\cal N$=4 analytic 
superspace~\cite{Heslop:2001gp}, and can be defined as the limit,  
\begin{align}
\lim_{x_1 \rightarrow x_2} g_{12}^{P}\Big[  \cO_{p_1}(x_1) \cO_{p_2}(x_2)\Big] =   
			\sum_{\underline t\, \vdash t} C_{p_1 p_2}^{\underline{t}}\  \mathcal{O}_{\underline{t}}(x_2), \qquad t=p_1+p_2-2P  .
\label{halfBPSOPE}
\end{align}
The symbol $g_{12}^{P}[\cO_{p_1}(x_1) \cO_{p_2}(x_2)]$  specifies that out of 
the full super OPE on the l.h.s.~we are picking the term with $Y$ dependence of the 
form $Y_{12}^{2P}$.\footnote{Multiplying  by $X_{12}^{2P}$ we can take  
$X_1 \rightarrow X_2$, and this limit will now project on  
$\mathcal{O}_{\underline{t}}$ }  
The result on the r.h.s.~is intuitive, what happens is that by fusing the $P$ 
bridges between $\cO_{p_1}$ and $\cO_{p_2}$, we find an expansion 
over the half-BPS operators with $t=p_1+p_2-2P$ legs, i.e.~$t$ is the twist of the 
operator. We refer to this projection as the `half-BPS OPE at twist $t$'.

One can pick any basis of half-BPS operators to express the r.h.s.~of (\ref{halfBPSOPE}). 
For definiteness we have written it in terms of the basis generated by products of SPOs, 
$\mathcal{O}_{\underline{t}} = [\mathcal{O}_{t_1} \ldots \mathcal{O}_{t_m}]$ where 
$\underline{t}=(t_1,\ldots,t_m)$ is a partition of $t$.  

The coefficients $C_{p_1 p_2}^{\underline{t}}$ can be related more precisely to 
three-point functions, by taking the vev with other half BPS operators. We first find
\be\label{threepoint_halfbps}
\langle \mathcal{O}_{p_1} \mathcal{O}_{p_2} \mathcal{O}_{\underline{t}'} \rangle = \sum_{\underline{t}\, \vdash t} C_{p_1 p_2}^{\underline{t}} g_{\underline{t} \underline{t}'}\,,
\ee
where $g$ is the matrix of two-point functions of half-BPS operators of twist $t$,
\be
g_{\underline{t} \underline{t}'} = \langle \mathcal{O}_{\underline{t}} \mathcal{O}_{\underline{t}'} \rangle\,.
\ee
Inverting \eqref{threepoint_halfbps} as a vector equation we obtain 
\be 
C_{p_1 p_2}^{\underline{t}} = \sum_{\underline{t}' \, \vdash t} \langle \mathcal{O}_{p_1} \mathcal{O}_{p_2} \mathcal{O}_{\underline{t}'} \rangle (g^{-1})_{\underline{t}' \underline{t}}\,.
\ee

Free theory correlators decompose into propagator structures: If we arrange the 
operators at the corners of a square and as usual we draw a line between point $i$ 
and point $j$ to represent a propagator $g_{ij}$, we can in fact represent the
the correlator as a sum over all possible propagator structures, 
\begin{align}
\langle \mathcal{O}_{p_1} \ldots \mathcal{O}_{p_n} \rangle = 
				\sum_{\{b_{ij}\}} \alpha_{\{b_{ij}\}} \prod_{i<j} g_{ij}^{b_{ij}}\,, \qquad \sum_{i\neq j} b_{ij} = p_i\,, \quad b_{ji}=b_{ij}\,, \quad b_{ii}=0\,.
\end{align}
The color factors $\alpha_{\{b_{ij}\}}$ can in principle be computed by Wick contractions. 
However, a brute force computation is quite expensive, since the single particle operators 
are admixture of single and multi-trace operators. Therefore intermediate steps are cumbersome. 
Nevertheless, we already saw for the NNME three-point functions in section \ref{sec_on_lower_extremality} that
the final result is much simpler than the intermediate steps. Thus, the idea of the half-BPS OPE 
is to constrain the $\alpha_{\{b_{ij}\}}$ and compute them by using the consistency of the general 
OPE, projected onto the simpler sector of half-BPS operators. To illustrate the power and simplicity 
of this procedure we will now consider four-point functions 
$\langle \mathcal{O}_{p_1} \mathcal{O}_{p_2} \mathcal{O}_{p_3} \mathcal{O}_{p_4} \rangle$.

In the following analysis we will always take $p_4$ to be the largest charge. 
We have three `channels' to perform the OPE, depending on which operator 
$\mathcal{O}_{p_{i=1,2,3}}$ we bring close to $\mathcal{O}_{p_4}$. Schematically, 
$(12)\leftrightarrow (34)$, $(23)\leftrightarrow (14)$ and $(13)\leftrightarrow (24)$.
Let us consider the first channel, for illustration. The reasoning will be similar for the others. 

If we perform the half-BPS OPE $(\mathcal{O}_{p_1} \times \mathcal{O}_{p_2})$ 
we find the twist $t$ to lie in the range ${ max}(|p_{12}|,|p_{43}|) \leq t \leq { min}(p_1 + p_2,p_3 + p_4)$. 
The extrema are special, and will be discussed separately. For any other value of 
$t$ in ${ max}(|p_{12}|,|p_{43}|) < t < { min}(p_1 + p_2,p_3 + p_4)$, we obtain a relation 
for a linear combination of the coefficients $\alpha$, of the form
\be
\label{coeffcomb}
\sum_{b_{12}+b_{13}+b_{23}+b_{24}=t} \alpha_{\{b_{ij}\}} = 
\sum_{\underline{t},\underline{t}' \, \vdash t} \langle \mathcal{O}_{p_1} \mathcal{O}_{p_2} \mathcal{O}_{\underline{t}}\rangle (g^{-1})_{\underline{t} \underline{t}'} \langle \mathcal{O}_{\underline{t}'} \mathcal{O}_{p_3} \mathcal{O}_{p_4} \rangle\,.
\ee
The sum on the l.h.s.~above is specified by the condition $b_{12}+b_{13}+b_{23}+b_{24}=t$ so 
that it involves only diagrams with $t$ propagators  between the pair 
$(\mathcal{O}_{p_1},\mathcal{O}_{p_2})$ and the pair $(\mathcal{O}_{p_3},\mathcal{O}_{p_4})$.

A feature of using the basis generated by products of SPOs is that the single-particle 
space is orthogonal to the multiparticle sector (by the definition of SPOs). Therefore 
the r.h.s~ of (\ref{coeffcomb}) can be split into a single particle contribution and a multiparticle contribution,
 \begin{align}
\label{coeffcomb1}
\sum_{b_{12}+b_{13}+b_{23}+b_{24}=t} \alpha_{\{b_{ij}\}}  =&\ \langle \mathcal{O}_{p_1} \mathcal{O}_{p_2} \mathcal{O}_{t} \rangle \langle \mathcal{O}_t \mathcal{O}_t \rangle^{-1} \langle \mathcal{O}_{t} \mathcal{O}_{p_3} \mathcal{O}_{p_4} \rangle  \\
&\label{coeffcomb2}
+\sum_{\underline{t},\underline{t}' \, \vdash t} \langle \mathcal{O}_{p_1} \mathcal{O}_{p_2} \mathcal{K}_{\underline{t}}\rangle (\tilde{g}^{-1})_{\underline{t} \underline{t}'} \langle \mathcal{K}_{\underline{t}'} \mathcal{O}_{p_3} \mathcal{O}_{p_4} \rangle\,.
\end{align}
Here the sum in the second term is over only partitions $\underline{t},\underline{t}'$ of 
length at least two. We use the notation $\mathcal{K}$ to emphasise that there is always 
more than one factor in the operator $\mathcal{K}_{\underline{t}} = [\mathcal{O}_{t_1} \ldots \mathcal{O}_{t_m}]$, 
i.e. $(m\geq 2)$ and we write $\tilde{g}_{\underline{t} \underline{t}'} = \langle \mathcal{K}_{\underline{t}} \mathcal{K}_{\underline{t}'} \rangle$ 
for the metric projected on the multi-particle sector. 

Referring to \eqref{coeffcomb1} and \eqref{coeffcomb2}, we can appreciate that 
considering the lowest possible twist, i.e.~$t=max(|p_{12}|,|p_{43}|)>0$, we find in any case extremal three point functions, 
which vanish according to our general discussion in section \ref{nEvan}. For the highest value instead,   
$t=min(p_{1}+p_{2},p_{3}+p_4)$, we find that \eqref{coeffcomb1} is again extremal, thus 
vanishing, but \eqref{coeffcomb2} instead gives a sum over three point functions with 
multi-particle states. However, these three-point functions are of the same kind of the 
original four-point function we want to bootstrap, therefore do not lead to useful constraints. 
Rather, we will show that can be used to obtain multi-particle two-point functions.

\subsection{N${}^2$E correlators}

Let us fix ideas by re-considering the simplest ME four-point functions first addressed 
in section \ref{ME_section_main}.  These are the next-to-next-to-extremal (N${}^2$E)
 correlators, and obey,
\be
s =p+q+r - 4\,,
\label{4ptME}
\ee
where we always take $s$ to be the largest charge.
The condition (\ref{4ptME}) implies that all but two propagators are connected
 to $\mathcal{O}_s$, leaving six possible topologies which can contribute to 
 the correlator, depicted below 
\begin{align}
\braket{\mathcal{O}_p \mathcal{O}_q \mathcal{O}_r \mathcal{O}_s} = 
\quad& \alpha_1\,\,
\vcenter{\hbox{\begin{tikzpicture}[scale=1.4]
		\draw[very thick] (1,0) -- (0,0);
		\draw[very thick] (1,0) -- (0,1);
		\draw[very thick] (1,0) -- (1,1);
		\draw [very thin,gray] (0,0) -- (0,1) ;
		\draw [very thin,gray] (0,0) -- (1,1) ;
		\end{tikzpicture}
}}+
\alpha_2\,\,
\vcenter{\hbox{\begin{tikzpicture}[scale=1.4]
		\draw[very thick] (1,0) -- (0,0);
		\draw[very thick] (1,0) -- (0,1);
		\draw[very thick] (1,0) -- (1,1);
		\draw [very thin,gray] (0,0) -- (0,1) ;
		\draw [very thin,gray] (1,1) -- (0,1) ;
		\end{tikzpicture}
}}+
\alpha_3\,\,
\vcenter{\hbox{\begin{tikzpicture}[scale=1.4]
		\draw[very thick] (1,0) -- (0,0);
		\draw[very thick] (1,0) -- (0,1);
		\draw[very thick] (1,0) -- (1,1);
		\draw [very thin,gray] (0,0) -- (1,1) ;
		\draw [very thin,gray] (1,1) -- (0,1) ;
		\end{tikzpicture}
}} \notag \\[.2cm]
+\,&  \tilde{\alpha}_1\,\,
\vcenter{\hbox{\begin{tikzpicture}[scale=1.4]
		\draw[very thick] (1,0) -- (0,0);
		\draw[very thick] (1,0) -- (0,1);
		\draw[very thick] (1,0) -- (1,1);
		\draw [very thin,gray,bend right=20] (0,0) to (0,1);
		\draw [very thin,gray] (0,0) to (0,1);
		\end{tikzpicture}
}}+
\tilde{\alpha}_2\,\,
\vcenter{\hbox{\begin{tikzpicture}[scale=1.4]
		\draw[very thick] (1,0) -- (0,0);
		\draw[very thick] (1,0) -- (0,1);
		\draw[very thick] (1,0) -- (1,1);
		\draw [very thin,gray,bend right=10] (0,0) to (1,1);
		\draw [very thin,gray,bend left=10] (0,0) to (1,1);
		\end{tikzpicture}
}}+
\tilde{\alpha}_3\,\,
\vcenter{\hbox{\begin{tikzpicture}[scale=1.4]
		\draw[very thick] (1,0) -- (0,0);
		\draw[very thick] (1,0) -- (0,1);
		\draw[very thick] (1,0) -- (1,1);
		\draw [very thin,gray,bend right=20] (0,1) to (1,1);
		\draw [very thin,gray] (0,1) to (1,1);
		\end{tikzpicture}
}}
\label{ME4ptdiags}
\end{align}
The operator $\mathcal{O}_s$ sits at right corner on the bottom of each square. 
Here the thin lines correspond to single propagators not connected to $\mathcal{O}_s$ 
whereas  the thick lines represent multiple propagators (as many as needed to match 
the charges $p$, $q$, $r$, $s$). 

Let us consider the half-BPS OPE $(\mathcal{O}_p \times \mathcal{O}_q)$ at twist $t=(p+q-4)$.  
This means we project onto the $Y$ dependence $Y_{12}^P$ with $P=2$. 
The only surviving diagram is the first one on the second line of (\ref{ME4ptdiags}) 
and we obtain the following equation,
\be
\tilde{\alpha}_1 = \sum_{\underline{t},\underline{t}' \, \vdash t} 
		\langle \mathcal{O}_p \mathcal{O}_q \mathcal{O}_{\underline{t}}\rangle (g^{-1})_{\underline{t} \underline{t}'} \langle \mathcal{O}_{\underline{t}'} \mathcal{O}_r \mathcal{O}_s \rangle 
\ee
If $t>0$, the three-point functions $\langle  \mathcal{O}_{\underline{t}'} \mathcal{O}_r \mathcal{O}_s \rangle$ all vanish. 
This is because $\mathcal{O}_s$ is a SPO of charge $s=p+q+r-4=t+r$ and the three-point function is therefore extremal. 
We conclude $\tilde{\alpha_1}= 0$ and a similar analysis of the $(\mathcal{O}_p \times \mathcal{O}_r)$ 
and $(\mathcal{O}_q \times \mathcal{O}_r)$ OPEs reveals that also $\tilde{\alpha}_2 = \tilde{\alpha}_3=0$. 
This same conclusion was already reached in \cite{Aprile:2018efk} where we argued that free theory 
propagator topologies in all SPO four-point functions are absent when one of the operators is only 
connected to one other. The assumption $t>0$ excludes the identity operator, and implies that the 
diagrams we considered are connected. 

Now let us perform the half-BPS OPE $(\mathcal{O}_p \times \mathcal{O}_q)$ at twist $t=(p+q-2)$. 
We have to project the $Y$ dependence onto $Y_{12}^P$ with $P=1$, giving us
\be
\alpha_1+\alpha_2 = 
\langle \mathcal{O}_p \mathcal{O}_q \mathcal{O}_{t}\rangle \langle \mathcal{O}_t \mathcal{O}_t \rangle^{-1} \langle \mathcal{O}_{t} \mathcal{O}_r \mathcal{O}_s \rangle\,.
\ee
In the second equality we have used the fact that the three-point functions 
$\langle  \mathcal{K}_{\underline{t}} \mathcal{O}_r \mathcal{O}_s \rangle$ 
can be splitted at least on four-point auxiliary points with $k=s-r-(p+q+2)=1$, 
thus are near-extremal and vanish according to the discussion in section \ref{more_cases}. 
Using our previous computation for $\langle \mathcal{O}_{t} \mathcal{O}_r \mathcal{O}_s \rangle$ (see  \eqref{n})
we arrive at
\begin{align}
	\alpha_1+\alpha_2 = p q r (p+q-2) \langle \cO_s \cO_s \rangle\ .
\end{align}
Repeating the analysis above in the other two OPE channels we obtain two more 
similar equations with the unique solution
\begin{align}
	\alpha_1 = pqr(p-1)\langle \cO_s \cO_s \rangle, \quad \alpha_2 = pqr(q-1)\langle \cO_s \cO_s \rangle,\quad \alpha_3 = pqr(r-1)\langle \cO_s \cO_s \rangle\ ,
\label{N2E4pt}	
\end{align}
This is exactly the same solution we derived in equation~\eqref{4-point} with different means. 

As we anticipated, 
performing the half-BPS OPE at twist $t=p+q$ does not yield further constraints on the coefficients $\alpha_i$. 
Instead we obtain the relation
\begin{align}
\alpha_3 + \langle \mathcal{O}_p \mathcal{O}_r \rangle \langle \mathcal{O}_q \mathcal{O}_s \rangle +  \langle \mathcal{O}_p \mathcal{O}_s \rangle \langle \mathcal{O}_q \mathcal{O}_r \rangle &
= \sum_{\underline{t},\underline{t}' \, \vdash t} \langle \mathcal{O}_p \mathcal{O}_q \mathcal{K}_{\underline{t}}\rangle (\tilde{g}^{-1})_{\underline{t} \underline{t}'} \langle \mathcal{K}_{\underline{t}'} \mathcal{O}_r \mathcal{O}_s \rangle\,.
\end{align}
On the l.h.s.~we have been careful to include the disconnected contributions that  can be present. 
The contribution $\langle \mathcal{O}_p \mathcal{O}_q \mathcal{O}_{t}\rangle$  is extremal and vanishing,  
thus only multi-particles exchanged appear in the second equality above. Moreover, from the fact that
there are no bridges between $\mathcal{O}_p$ and $\mathcal{O}_q$ due to extremality,  we find that the three point function
$\langle \mathcal{O}_p \mathcal{O}_q \mathcal{K}_{\underline{t}}\rangle$  is equal to the two-point 
function $\langle [\mathcal{O}_p \mathcal{O}_q] \mathcal{K}_{\underline{t}}\rangle = \tilde{g}_{(p,q), \underline{t}}$. 
The r.h.s.~above thus simplifies and we obtain
\be
\alpha_3 + \langle \mathcal{O}_p \mathcal{O}_r \rangle \langle \mathcal{O}_q \mathcal{O}_s \rangle +  \langle \mathcal{O}_p \mathcal{O}_s \rangle \langle \mathcal{O}_q \mathcal{O}_r \rangle = \langle [\mathcal{O}_p \mathcal{O}_q] \mathcal{O}_r \mathcal{O}_s \rangle.
\ee
This is equivalent to taking the coincidence limit $x_1 \rightarrow x_2$ on the original 
four-point function $\langle \mathcal{O}_p \mathcal{O}_q \mathcal{O}_r \mathcal{O}_s \rangle$ 
which yields immediately
\be
\langle [\mathcal{O}_p \mathcal{O}_q] \mathcal{O}_r \mathcal{O}_s \rangle = pqr(r-1)\langle \mathcal{O}_s \mathcal{O}_s \rangle + \langle \mathcal{O}_p \mathcal{O}_r \rangle \langle \mathcal{O}_q \mathcal{O}_s \rangle +  \langle \mathcal{O}_p \mathcal{O}_s \rangle \langle \mathcal{O}_q \mathcal{O}_r \rangle\,.
\label{N2Ecolim}
\ee
where $s=p+q+r-4$ as before and we have used the derived result (\ref{N2E4pt}) for $\alpha_3$. 
The other coincidence limit $x_3 \rightarrow x_4$ gives us
\be
\langle \mathcal{O}_p \mathcal{O}_q [\mathcal{O}_r \mathcal{O}_s]\rangle = \delta_{2r} 2pq \langle \mathcal{O}_s \mathcal{O}_s \rangle + \langle \mathcal{O}_p \mathcal{O}_r \rangle \langle \mathcal{O}_q \mathcal{O}_s \rangle + \langle \mathcal{O}_p \mathcal{O}_s \rangle \langle \mathcal{O}_q \mathcal{O}_r \rangle\,.
\ee
The connected part only contributes for $r=2$. The double coincidence limit 
$(x_1\rightarrow x_2, x_3 \rightarrow x_4)$ then gives two-point functions of product operators. 
\be
\langle [\mathcal{O}_p \mathcal{O}_q] [\mathcal{O}_r \mathcal{O}_s] \rangle = \delta_{2r} 2 pq \langle \mathcal{O}_s \mathcal{O}_s \rangle + \langle \mathcal{O}_p \mathcal{O}_r \rangle \langle \mathcal{O}_q \mathcal{O}_s \rangle +  \langle \mathcal{O}_p \mathcal{O}_s \rangle \langle \mathcal{O}_q \mathcal{O}_r \rangle\,.
\ee

\subsubsection*{The case of $\langle \mathcal{O}_2 \mathcal{O}_q \mathcal{O}_r \mathcal{O}_s\rangle$}
At four points, whenever one of the charges takes (the smallest possible) value $p=2$, 
the correlator contributes to a single $su(4)$ representation in each OPE channel. 
Such four-point functions are also called next-to-next-to extremal, and in principle 
have six propagator structures. Here we focus on the three topologies shown below,\\[.2cm]
\begin{equation}
\label{nn4pt}
\braket{\mathcal{O}_2 \mathcal{O}_q \mathcal{O}_r \mathcal{O}_s}_c=
\alpha_1\,\,
\vcenter{\hbox{\begin{tikzpicture}[scale=1.4]
		\draw[very thick] (1,1) -- (0,1);
		\draw[very thick] (1,0) -- (0,1);
		\draw[very thick] (1,0) -- (1,1);
		\draw [very thin,gray] (0,0) -- (0,1) ;
		\draw [very thin,gray] (0,0) -- (1,1) ;
		\end{tikzpicture}
}}+
\alpha_2\,\,
\vcenter{\hbox{\begin{tikzpicture}[scale=1.4]
		\draw[very thick] (1,1) -- (0,1);
		\draw[very thick] (1,0) -- (0,1);
		\draw[very thick] (1,0) -- (1,1);
		\draw [very thin,gray] (0,0) -- (0,1) ;
		\draw [very thin,gray] (0,0) -- (1,0) ;
		\end{tikzpicture}
}}+
\alpha_3\,\,
\vcenter{\hbox{\begin{tikzpicture}[scale=1.4]
		\draw[very thick] (1,1) -- (0,1);
		\draw[very thick] (1,0) -- (0,1);
		\draw[very thick] (1,0) -- (1,1);
		\draw [very thin,gray] (0,0) -- (1,1) ;
		\draw [very thin,gray] (0,0) -- (1,0) ;
		\end{tikzpicture}
}}\,.
\end{equation}\\[.2cm]
Diagrams where one of the operators is connected to only one are omitted, since 
the color factor vanish by exactly the same arguments as above. We also omit 
cases in which diagrams are disconnected. 

Performing the half BPS OPE $(\mathcal{O}_2 \times \mathcal{O}_q)$ at twist $q$ gives
\begin{align}
\alpha_1 +\alpha_2 = \frac{\langle \cO_2 \cO_q \cO_{q} \rangle \langle \cO_{q} \cO_r \cO_{s} \rangle}{\langle \cO_{q} \cO_{q} \rangle}  = 2 q \langle \mathcal{O}_q \mathcal{O}_r \mathcal{O}_s \rangle\,,
\end{align}
where we simplified the result using our previous results for maximally extremal three-point functions.
Again we obtain two similar equations from the other crossing channels and finally we arrive at
\begin{align}
\alpha_1 &= (q+r-s) \langle \mathcal{O}_q \mathcal{O}_r \mathcal{O}_s \rangle\,,\, \notag \\
\alpha_2 &= (q+s-r) \langle \mathcal{O}_q\mathcal{O}_r \mathcal{O}_s \rangle\,,\, \notag \\
\alpha_3 &= (r+s-q) \langle \mathcal{O}_q \mathcal{O}_r \mathcal{O}_s \rangle\,.
\end{align}
Note that setting $s=q+r-2$ above we find agreement with (\ref{N2E4pt}) in the case $p=2$. 
The coincidence limit $x_1 \rightarrow x_2$ then gives three-point functions involving products of single-particle operators,
\be
\langle [\mathcal{O}_2 \mathcal{O}_q] \mathcal{O}_r \mathcal{O}_s \rangle  = (r+s-q) \langle \mathcal{O}_q \mathcal{O}_r \mathcal{O}_s \rangle + \langle \mathcal{O}_2 \mathcal{O}_r \rangle \langle \mathcal{O}_q \mathcal{O}_s \rangle +  \langle \mathcal{O}_2 \mathcal{O}_s \rangle \langle \mathcal{O}_q \mathcal{O}_r \rangle \,.
\label{O2dbltrace3pt}
\ee

\subsection{N${}^3$E correlators}
\label{NME4pt}
Let us now consider $N^3$-extremal 4-point functions where $s=p+q+r-6$. 
We focus on the seven connected diagrams (out of ten) depicted below,
\begin{align}
\braket{\mathcal{O}_p \mathcal{O}_q \mathcal{O}_r \mathcal{O}_s}_c  =\quad
&\alpha_1\,\,
\vcenter{\hbox{\begin{tikzpicture}[scale=1.4]
\draw[very thick] (1,0) -- (0,0);
\draw [very thick](1,0) -- (0,1);
\draw [very thick](1,0) -- (1,1);
\draw [bend left=20][very thin,gray] (0,0) -- (0,1) ;
\draw [bend right=20][very thin,gray] (0,0) to (0,1);
\draw [very thin,gray] (0,0) -- (1,1) ;
\end{tikzpicture}
}}
+
\alpha_2\,\,
\vcenter{\hbox{\begin{tikzpicture}[scale=1.4]
\draw [very thick](1,0) -- (0,0);
\draw [very thick](1,0) -- (0,1);
\draw [very thick](1,0) -- (1,1);
\draw [bend left=20][very thin,gray] (0,0) -- (0,1) ;
\draw [bend right=20][very thin,gray] (0,0) to (0,1);
\draw [very thin,gray] (0,1) -- (1,1) ;
\end{tikzpicture}
}}
+
\alpha_3\,\,
\vcenter{\hbox{\begin{tikzpicture}[scale=1.4]
\draw [very thick](1,0) -- (0,0);
\draw [very thick](1,0) -- (0,1);
\draw [very thick](1,0) -- (1,1);
\draw [bend right=10][very thin,gray] (0,0) to (1,1) ;
\draw [bend left=10][very thin,gray] (0,0) to (1,1) ;
\draw [very thin,gray] (0,0) to (0,1) ;
\end{tikzpicture}}}
+\alpha_4\,\,
\vcenter{\hbox{\begin{tikzpicture}[scale=1.4]
\draw [very thick](1,0) -- (0,0);
\draw [very thick](1,0) -- (0,1);
\draw [very thick](1,0) -- (1,1);
\draw [bend right=10][very thin,gray] (0,0) to (1,1) ;
\draw [bend left=10][very thin,gray] (0,0) to (1,1) ;
\draw [very thin,gray] (1,1) to (0,1) ;
\end{tikzpicture}
}}
\notag \\[.2cm]
 +
&\alpha_5\,\,
\vcenter{\hbox{\begin{tikzpicture}[scale=1.4]
\draw [very thick](1,0) -- (0,0);
\draw [very thick](1,0) -- (0,1);
\draw [very thick](1,0) -- (1,1);
\draw [bend left=10][very thin,gray] (1,1) to (0,1) ;
\draw [bend right=10][very thin,gray] (1,1) to (0,1);
\draw [very thin,gray] (0,0) -- (0,1) ;
\end{tikzpicture}
}}
+\alpha_6\,\,
\vcenter{\hbox{\begin{tikzpicture}[scale=1.4]
\draw [very thick](1,0) -- (0,0);
\draw [very thick](1,0) -- (0,1);
\draw [very thick](1,0) -- (1,1);
\draw [bend left=10][very thin,gray] (1,1) to (0,1) ;
\draw [bend right=10][very thin,gray] (1,1) to (0,1);
\draw [very thin,gray] (0,0) -- (1,1) ;
\end{tikzpicture}
}}
+
		\alpha_7
\vcenter{\hbox{
		\begin{tikzpicture}[scale=1.4]
\draw [very thick](1,0) -- (0,0);
\draw [very thick](1,0) -- (0,1);
\draw [very thick](1,0) -- (1,1);
\draw [very thin,gray] (0,0) -- (1,1) ;
\draw [very thin,gray] (1,1) -- (0,1) ;
\draw [very thin,gray] (0,0) -- (0,1) ;
\end{tikzpicture}
}}\,.
\end{align}
Again, we are omitting three diagrams where one operator is connected to only one other, which would vanish. 
Recall that the operator $\mathcal{O}_s$ sits at the right corner on the bottom of each square.

Performing the half-BPS OPE $(\mathcal{O}_p \times \mathcal{O}_q)$ at twist $t=p+q-4$, 
i.e.~by projection onto $Y_{12}^P$ with $P=2$, yields
\begin{align}
\alpha_1+\alpha_2&= \frac{\braket{\cO_p \cO_q \cO_{p+q-4}}\braket{\cO_{p+q-4} \cO_r \cO_{s}}}{\braket{\cO_{p+q-4}\cO_{p+q-4}}} =(p+q-4)r \frac{\braket{\cO_p \cO_q \cO_{p+q-4}}}{\braket{\cO_{p+q-4}\cO_{p+q-4}}} \braket{ \cO_s\cO_s}\,.
\label{twistp+q-4}
\end{align} 
The multi-particle term vanishes by extremality and the contribution 
$\langle \cO_{p+q-4} \cO_r \cO_{s} \rangle$ is 
ME, thus it simplifies to give the second equality above.
We obtain two similar equations from the other OPEs $(\mathcal{O}_p \times \mathcal{O}_r)$ 
and $(\mathcal{O}_q \times \mathcal{O}_r)$. These only involve the pairs $\alpha_{3}+\alpha_4$ 
and $\alpha_{5}+\alpha_6$. The color factor $\alpha_7$ does not enter any of these constraints

From the OPE $(\mathcal{O}_p \times \mathcal{O}_q)$ at twist $t=p+q-2$ we obtain
\begin{align}
\alpha_3+\alpha_5+\alpha_7&=   pq\braket{\cO_{p+q-2} \cO_r \cO_{s}}+ 
		\sum_{\underline{t},\underline{t}' \, \vdash t} \braket{\cO_p \cO_q \mathcal{K}_{\underline{t}}} (\tilde{g}^{-1})_{\underline{t} \underline{t}'} \braket{\mathcal{K}_{\underline{t}'} \cO_r \cO_{s}}\,.
\label{twistp+q-2}
\end{align}
In general both single-particle and multi-particle terms contribute. 
The multi-particle contribution have more structure than previous cases, but
let us note that the three-point function $\langle \mathcal{K}_{\underline{t}'} \cO_r \cO_{s} \rangle$ 
is only non-zero if $\mathcal{K}_{\underline{t}'} = [\mathcal{O}_{t_1^\prime} \mathcal{O}_{t-t_1^\prime}]$, 
namely a product of two single-particle operators. A contribution with more than two single-particle operators 
can be understood as a near-extremal five-point function and vanishes according to the discussion in section \ref{more_cases}. Then, we 
are in a favourable position since the three point function $\langle  [\mathcal{O}_{t_1^\prime} 
\mathcal{O}_{t-t_1^\prime}] \cO_r \cO_{s} \rangle$ can be computed as 
the coincidence limit of the N${}^2$E four-point functions described in (\ref{N2Ecolim}). 
In particular, 
\be
\langle \mathcal{K}_{\underline{t}'} \cO_r \cO_{s} \rangle = 
		\langle [\mathcal{O}_{t_1^\prime} \mathcal{O}_{t-t_1^\prime}] \cO_r \cO_{s} \rangle = t_1^\prime(t-t_1^\prime)r (r-1)\langle \mathcal{O}_s \mathcal{O}_s \rangle\,.
\ee
We obtain two similar equations from the other OPEs $(\mathcal{O}_p \times \mathcal{O}_r)$ 
and $(\mathcal{O}_q \times \mathcal{O}_r)$. These involve $\alpha_2+ \alpha_4$ with $\alpha_7$, and,  
$\alpha_1+\alpha_6$ with $\alpha_7$. 

In total we have six equations with particular structure. If we consider the last three equations 
that we derived, and we subtract the first three equation, we obtain a single equation for $3\alpha_7$. 
Thus we determine $\alpha_7$ completely,
\begin{align}
\alpha_7 = \frac{1}{3}\biggl(& 
pq\braket{\cO_{p+q-2} \cO_r \cO_{s}} + \sum_{\underline{t},\underline{t}' \, \vdash t} 
		\braket{\cO_p \cO_q \mathcal{K}_{\underline{t}}} (\tilde{g}^{-1})_{\underline{t} \underline{t}'} \braket{\mathcal{K}_{\underline{t}'} \cO_r \cO_{s}} \notag  \\
&
- (p+q-4)r \frac{\braket{\cO_p \cO_q \cO_{p+q-4}}}{\braket{\cO_{p+q-4}\cO_{p+q-4}}} \braket{ \cO_s\cO_s}\biggr) + \text{ cyc}(p,q,r) \,.
\end{align}
The remaining equations determine five of the remaining coefficients in terms of 
one remaining, $\alpha_1$ say. In specific cases additional conditions can come from 
crossing symmetry constraints.  In Appendix \ref{HBPSOPE-app} we discuss various 
examples with different degree of crossing symmetry. 

In all examples discussed in Appendix \ref{HBPSOPE-app}, the coefficients 
$\alpha_1,\ldots,\alpha_6$ are consistent with
the following solution
\begin{align}
&\alpha_1=F(p,q,r) \braket{\cO_s \cO_s} \,, \quad \alpha_2= F(q,p,r) \braket{\cO_s \cO_s } \,, \quad \alpha_3= F(p,r,q) \braket{\cO_s \cO_s}\, , \notag \\
&\alpha_4=F(r,p,q) \braket{\cO_s \cO_s} \,, \quad \alpha_5=F(q,r,p) \braket{\cO_s \cO_s} \,, \quad \alpha_6= F(r,q,p) \braket{\cO_s \cO_s} \,,
\label{solu_michele1}
\end{align}
where
\begin{equation}
\label{solu_michele2}
F(p,q,r)= (p-2)r \frac{\braket{\cO_p \cO_q \cO_{p+q-4}}}{\braket{ \cO_{p+q-4} \cO_{p+q-4}}} 
\end{equation}
The bootstrap problem allows the deformation $F(p,q,r)\rightarrow F(p,q,r)+\tilde{F}(p,q,r)$,
 where $\tilde{F}(p,q,r)$ is totally antisymmetric and unconstrained.
In all cases we studied we have found that the deformation $\tilde{F}=0$. \footnote{If we consider the 
correlators $\langle T_p T_q T_r \mathcal{O}_s \rangle$  instead of 
$\braket{\mathcal{O}_p \mathcal{O}_q \mathcal{O}_r \mathcal{O}_s}$ we find that 
all coefficients are a simple Laurent polynomial in $N$ multiplied by 
$\langle \mathcal{O}_s \mathcal{O}_s \rangle$. For example, from the results 
presented in Appendix \ref{HBPSOPE-app} with $p=3 \leq q \leq r$ we can see that we have
$
\alpha_7 = 6qr (N - 3(q-1)(r-1)/N).
$} 

\section{Conclusions}

We have considered a new basis of half-BPS operators in $\cal N$=4 SYM, namely the single particle operators (and products of these). 
These are the natural duals to single particle supergravity states on $AdS_5\times S^5$. In particular we have seen that they naturally 
interpolate between point-like gravitons and giant gravitons, as they should. We have also considered the free theory, all $N$,  correlators 
of these operators. Interestingly, all near extremal correlators of SPOs vanish, and this is presumably tied to the conjecture in~\cite{DHoker:2000xhf} 
that the corresponding supergravity couplings vanish. The near extremal correlators are $n$-point correlators with extremality degree strictly less than $n-2$.  
Going away from this case, the maximally extrema correlators then have a very simple form, similarly for the next-to-maximally extremal correlators. The complexity then increases by lowering the extremality, but even in such a case we have found additional simplicity, compared to the single-trace correlators.

It is interesting to revisit  past discussions involving half-BPS operators, especially concerning large $N$ limits and the relation to string theory computations via AdS/CFT, in the light of this basis. As we have seen in section~\ref{largeN} the SPOs correctly interpolate between single trace operators and the operators conjectured to be dual to $S^5$ giant graviton operators. 
In~\cite{Bissi:2011dc} a comparison of half-BPS three-point functions of two giant graviton operators and one point-like graviton was performed and compared with the analogous computation in gauge theory. The gauge theory was computed using two large Schur polynomial operators and one single trace operator. The results were found to not quite agree and it was conjectured this was to do with the inability of the Schur polynomials to correctly interpolate between giant and point-like gravitons. The SPOs on the other hand do precisely interpolate between the two as show in section~\ref{largeN}.  On the other hand,the extremal correlators of SPOs  simply  vanish! 
In~\cite{Caputa:2012yj} this issue was revisited and it was argued that indeed there were subtleties in the extremal case which are not present in the N-extremal case. The NE with two giant gravitons and one point-like graviton were computed in gauge theory (using Schur polynomials for the giant gravitons and single trace operators for the pointlike  operator) as well as in string theory and this time found agreement. Since we have explicit formulae for the NE 3-point functions we can check this agrees here also. 
 
We start with the next-to-extremal three-point function of unit normalised single particle operators. From~\eqref{n} we have 
\begin{align}
\label{eq:pqrFirst}
\frac{\langle \cO_p \cO_q \cO_r \rangle}{\sqrt{\langle \cO_p\cO_p \rangle \langle \cO_q\cO_q \rangle \langle \cO_r\cO_r \rangle}} = pq \sqrt{\frac{\langle \cO_r\cO_r \rangle}{\langle \cO_p\cO_p \rangle \langle \cO_q\cO_q \rangle}},, \ \ \ p+q=r+2.
\end{align} 
Now consider the limit $N\rightarrow \infty$ with $p$ staying finite, but $q,r \rightarrow \infty$ such that $q'=q/N,r'=r/N$ are fixed.  
Taking the appropriate limits of the two point functions~\eqref{OpOpexp} we find
\begin{align}
\label{NELimit}
\frac{\langle \cO_p\cO_q\cO_r \rangle}{\sqrt{\langle \cO_p\cO_p \rangle \langle \cO_q\cO_q \rangle \langle \cO_r\cO_r \rangle}} \rightarrow
			 \sqrt{p}\frac{r}{N} \left( 1- \frac{r}{N} \right)^{\frac{p-2}{2} } , \ \ \ p+q=r+2
\end{align}
which is in precise agreement with~\cite{Caputa:2012yj}.

We can also compute the normalised next-to-next-to-extremal three-point function  given by \eqref{conclusion1}-\eqref{conclusion2} 
in the same limit, $N\rightarrow \infty$ with $p,q'=q/N,r'=r/N$ fixed 
\begin{align}
\label{NNELimit}
\frac{\langle \cO_p\cO_q\cO_r \rangle}{\sqrt{\langle \cO_p\cO_p \rangle \langle \cO_q\cO_q \rangle \langle \cO_r\cO_r \rangle}} 
\rightarrow \sqrt{p}\frac{r}{N} \left( 1- \frac{r}{N} \right)^{\!\!\frac{p-4}{2}} \left(1-\frac{(p-1)r}{2N}\right) , \ \ \ p+q=r+4.
\end{align}
It would be interesting to compare with the corresponding string theory computation.

%

There are a number of further avenues one could go down from here. Firstly one could try to generalise 
our story beyond the half-BPS sector. 
Bases for more general operators in $\cN=4$ SYM have been given, 
for example~\cite{Brown:2007xh,Bhattacharyya:2008rb,Brown:2008ij,deMelloKoch:2007rqf,
deMelloKoch:2007nbd,Bekker:2007ea,Lewis-Brown:2020nmg} and it would be interesting to look again at these from the perspective of 
single-particle operators. 

The same definition of SPOs presumably holds also for orthogonal and symplectic gauge groups in 
$\cN=4$ SYM which can be obtained via a $\mathbb{Z}_2$ orientifold projection of the standard 
AdS${}_5\times S^5$ setup~\cite{Witten:1998xy}. Half-BPS operators  in these theories  have been 
studied in~\cite{Aharony:2002nd,Caputa:2013hr,Caputa:2013vla,Lewis-Brown:2018dje} and it would 
be interesting to consider single-particle operators in that context.
It would also be interesting to study-single particle states for other AdS$\times S$ backgrounds, 
for example in AdS${}_3$, as it was pointed out in \cite{Taylor:2007hs}, ABJM, 
and for the mysterious six-dimensional (2,0) theory on AdS${}_7\times S^4$.

Finally, it would be very interesting  to consider aspects of the dynamics of the single-particle 
operators that have not been explored yet, and go beyond the computation of the one-loop amplitudes 
in \cite{Aprile:2019rep}, along the lines suggested in \cite{ioepedro}.
For instance, the trace basis is widely used in the context of integrability 
and in turn integrability based techniques allow the computation of exact correlators. 
The simplest four-point correlator one could study, i.e.~the octagon configuration of 
\cite{Coronado:2018ypq},\footnote{See \cite{vasco} for a five-point analog, called the decagon.} 
was firstly obtained by using hexagonalization from weak coupling, and later re-derived 
in other beautiful ways \cite{Coronado:2018cxj,Belitsky:2019fan,Belitsky:2020qrm,Belitsky:2020qir}.
From the OPE point of view at weak coupling, many properties of the correlators are due to 
single-trace stringy states acquiring an anomalous dimension with universal features. 
This mechanism is quite democratic and perhaps the distinction between single-traces and 
single-particle external states does not matter at weak coupling. On the other hand, the situation in the supergravity 
regime is quite different, and the half-BPS single-particle operators are properly the dual of the KK modes, 
beyond the planar limit. It would be very interesting to understand how integrability based techniques 
\cite{Bargheer:2019kxb,Bargheer:2019exp} modify or adapt when correlators of single-particle operators are considered.

\section*{Acknowledgements}
FA would like to thank Pedro Vieira, Frank Coronado, and Till Bargheer for stimulating discussions. 
FA is partially supported by the ERC-STG grant 637844- HBQFTNCER.  FS is supported  by  the Italian  Ministry  of 
Research  under  grant  PRIN  20172LNEEZ  and by the INFN  under GRANT73/CALAT.
JD, HP were supported in part by the ERC Consolidator grant 648630 IQFT. MS is supported
by a Mayflower studentship from the University of Southampton.
AS is supported by an STFC studentship and PH acknowledges support from STFC grant 
ST/P000371/1 and the  European Union's Horizon 2020 research and innovation programme 
under the Marie Sk\l{}odowska-Curie grant agreement No. 764850 ``SAGEX''.

\appendix

\section{Trace Sector Formulae}\label{trace_sector_app}

In section \ref{sec_gen_formulas} and \ref{trace_sec} we obtained the result
\begin{align}
	\cO_p& = \sum_{\{q_1..q_m\}\vdash p}  C_{q_1,..,q_m}T_{q_1,..q_m} \\
	C_{q_1,..q_m} &= 
	\frac{|[\sigma_{q_1..q_m}]|}{(p-1)!} \sum_{s \in \mathcal{P}(\{q_1,..,q_m\})}\frac{{(-1)^{|s|+1}}{(N+1-p)_{p-\Sigma(s)}(N+p- \Sigma(s))_{\Sigma(s)}}}{{(N)_{p}}-{(N+1-p)_{p}}}\label{coeff_here}
	\end{align}
which is explicit in $p$ and $q_1\ldots q_m$, and depends on group theory 
data which we explained in section \ref{trace_sec}.

The value of $m$ distinguishes the splitting of $\cO_p$ in number of traces and below 
we give explicit examples for the double trace sector $m=2$, and the triple trace sector $m=3$. 

\subsection{Double Trace Sector}

Consider the partition $q_1 + q_2 = p$.  The powerset in the sum is
\begin{equation}
\mathcal{P}\left( \{q_1q_2\} \right) = \{\{ \}, \{q_1 \}, \{q_2 \}, \{q_1,q_2 \} \}
\end{equation}
and the corresponding values of $\Sigma$ are
\begin{equation}
\Sigma(\{ \}) = 0\; ,\; \Sigma(\{q_1 \}) = q_1 \;,\; \Sigma(\{q_2 \}) = q_2 \; , \; \Sigma(\{q_1,q_2 \}) = q_1+q_2 = p.
\end{equation}
Furthermore the size of the conjugacy class is $|[q_1q_2]|=p!/(q_1q_2)$ as long as $q_1 \neq q_2$. Otherwise $q_1=q_2=p/2$ and $|[q_1q_2]|=p!/(2q_1q_2)=(p-1)!/(2p)$. 
With these informations, the coefficient of $T_{q_1} T_{q_2}$ in $\cO_p$,  from~\eqref{coeff_here},  is

\begin{align}
\begin{split}
\label{eq:DoubleTraceCoeffs}
C_{q_1q_2} =\left\{  \begin{array}{ll}
\displaystyle
\frac{p}{q_1q_2} \times  \frac{
-(N{-}p{+}1)_{p} - (N)_{p} + (N{-}p{+}1)_{q_2} (N{+}q_2)_{p-q_2} + (N{-}p{+}1)_{q_1} (N{+}q_1)_{p-q_1}   }{ {(N)_{p}}-{(N-p+1)_{p}} } \\[.5cm]
\displaystyle
\ \ \ \, \frac{2}{p} \times \frac{-(N{-}p{+}1)_{p} - (N)_{p} + 2(N{-}p{+}1)_{p/2} (N{+}p/2)_{p/2}  }{{(N)_{p}}-{(N-p+1)_{p}}} 
\end{array}
\right.
\end{split}
\end{align}
As we pointed out, the above formula holds for the coefficients 
of the double trace contributions to the single particle operator of any weight.

\subsection{Triple Trace Sector}

Consider the partition $q_1 + q_2 +q_3 = p$. By making explicit~\eqref{coeff_here} we find,
\begin{align}
C_{q_1q_2q_3}=& \frac{p}{q_1q_2q_3}\frac{ -(N-p+1)_p +(N)_p }{ {(N)_{p}}-{(N-p+1)_{p}} }+\\
& \frac{p}{q_1q_2q_3}\frac{ - \sum_{i=1}^3 (N-p+1)_{q_i}(N+q_i)_{p-q_i}  + \sum_{i=1}^3 (N-p+1)_{p-q_i} (N+p-q_i)_{q_i} }{ {(N)_{p}}-{(N-p+1)_{p}} } \notag
\end{align}
The other two possible cases, in which $q_{i}=q_{j}$ and $q_1=q_2=q_3=p/3$, 
only differ compared to the result above by the the size of the conjugacy class. 
In the first case we have to further divide by $2$, and in the second case by $6$.
This formula thus cover all possible triple trace contributions to single particle operators of any weight. 

For any value of $m$, i.e.~trace sector, our function $C_{\underline{q}}$ can be made very explicit, as in the examples discussed above.

%
%
%
%

\section{Wick Contractions}\label{Wick_sec}

\def\x{\mathrm{\phi}}
\def\pphi{ {\pmb \phi} }

Given the set of all admissible propagator structures for a correlator, say 
\beq
{\tt PropStruct}=\Big\{ \mathcal{P}_{1},\ldots \Big\}
\eeq
we will now determine the associated Wick contractions. 

Recall that an elementary fields of $\mathcal{N}=4$  transforms under 
$SO(6)$ of R-symmetry, i.e.\! $\phi^I$, but itself is an $N\times N$ matrix 
in the adjoint of the gauge group. For the rest of this section we will then 
assume the replacement
\begin{align}
\big( \phi^I(X)\big)_{\alpha\beta} \qquad \rightarrow \qquad \sum_{i} \phi^I_i (X) \, \mathcal{T}_{\alpha\beta}^i
\end{align}
where $\mathcal{T}_{\alpha\beta}^i$ are a basis of generators for the 
representation under the gauge group. This replacement on the trace basis becomes, 
\begin{align}\label{gauge_field_rep}
T_{p}(X) &\quad \rightarrow \quad Y_{I_1} \ldots Y_{I_p} \sum_{i_1,\ldots i_p } \phi^I_{i_1} \ldots \phi^{I_p}_{i_p} (X) \, 
						\times\, \mathcal{T}_{\alpha_1\alpha_2}^{i_1} \ldots \mathcal{T}_{\alpha_{2p-1} \alpha_1 }^{i_p}  \\  
T_{\{p_1,p_2\}}(X) &\quad \rightarrow \quad Y_{I_1} \ldots Y_{I_p} \sum_{i_1,\ldots i_p } \phi^I_{i_1} \ldots \phi^{I_p}_{i_p} (X) \, \times
\mathcal{T}_{\alpha_1\alpha_2}^{i_1} \ldots  \mathcal{T}_{\alpha_{2p_1-1}\alpha_1}^{i_{p_1}} \times \mathcal{T}^{i_{p_1+1}}_{\beta_1\beta_2} \ldots \mathcal{T}_{\beta_{2p_2-1}\beta_1}^{i_p}  \notag\\
\vdots &  \notag
\end{align}
and so on so forth, i.e.\! if the operator is multi-trace, the trace is splitted accordingly.  

For what concerns finding the Wick contractions of an assigned propagator structure, 
the trace structure of the operators can be ignored. (Even though sometimes we can 
use the cyclic symmetry of the traces.) If we draw a blob to represent the operator, 
the only relevant information at this stage will be the number of $SO(6)$ indexes, 
i.e.\! legs attached to the blob. Single- or multi-trace, we will draw the same blob. 
For each Wick contraction, or bridge between legs (belonging to different blobs), 
say $\alpha_1$ and $\beta_2$, we insert a $\delta_{\alpha_1 \beta_2}$ which links the generators. 
For $SU(N)$ the sum over generators is then replaced by
\begin{align}\label{identity}
\sum_{i=1}^{N-1} 
\mathcal{T}^i_{\alpha_1\alpha_2} \mathcal{T}^i_{\alpha_3\alpha_4} =
\delta_{\alpha_1\alpha_4}\delta_{\alpha_2\alpha_3} -\frac{1}{N} \delta_{\alpha_1\alpha_2}\delta_{\alpha_3\alpha_4}
\end{align}
The color factor depends crucially on the trace structure of the operators and the 
type of propagator, whether $SU(N)$ or $U(N)$.

To enumerate the Wick contractions we need two standard combinatorial objects :

{\bf 1.} Define $\mathscr{C}(k\subseteq n)$ as the \emph{combinations} of $k$ integers 
in the set $\{1,\dots, n\}$. These are ordered sets. For example, if $n=3$ and $k=2$ the 
combinations are $12,13,23$. The number of the combinations is the binomial coefficient
\beq\label{formula_combinations}
|\mathscr{C}(k\subseteq n)|= \left(\!\begin{array}{c} n \\ k\end{array}\!\right)= \frac{n!}{(n-k)!k!}
\eeq

{\bf 2.} Define $\mathscr{D}(k\subseteq n)$ as the \emph{dispositions} of $k$ integers 
in the set $\{1,\ldots, n\}$. These are not ordered sets. For example, if $n=3$ and $k=2$, 
the dispositions are $12, 13, 23, 21, 31, 32$. The number of dispositions is indeed the 
number of $k$-combinations acted with $k$-permutations, 
\beq\label{formula_dispositions}
|\mathscr{D}(k\subseteq n)|= \left(\!\begin{array}{c} n \\ k\end{array}\!\right) k! = \frac{n!}{(n-k)!}
\eeq

To begin with, consider a simple example, the two-point function $\langle T_3, T_3\rangle$. 
There is obviously a single propagator structure, with three brigdes, $b_{(12)}=3$, 
Starting with $\phi^{I_1}\phi^{I_2}\phi^{I_3}(X_1) \phi^{I_4}\phi^{I_5}\phi^{I_6}(X_2)$, 
there are six Wick contractions, and we can rearrange them as follows, 
\bea
\underbracket{\ \phi^{I_1}(X_1)\phi^{I_4}(X_2)}\quad  \underbracket{\ \phi^{I_2}(X_1) \phi^{I_5}(X_2)}\quad \underbracket{\ \phi^{I_3}(X_1) \phi^{I_6}(X_2)}\notag\\
\vdots
\eea
These six Wick contractions are indexed by $\mathscr{C}(3\subseteq 3)$ (legs-out), 
which has dimension 1, paired with $\mathscr{D}(3\subseteq 3)$ (legs-in), which has 
dimension 6. In other words,
\beq
\begin{array}{l} 1^{\phantom{'}}2^{\phantom{'}}3^{\phantom{'}} \\[-.1cm] 1'2'3'\end{array}\quad
\begin{array}{l} 1^{\phantom{'}}2^{\phantom{'}}3^{\phantom{'}}\\[-.1cm]  1'3'2'\end{array}\quad
\begin{array}{l} 1^{\phantom{'}}2^{\phantom{'}}3^{\phantom{'}}\\[-.1cm]  2'1'3'\end{array}\quad
\begin{array}{l} 1^{\phantom{'}}2^{\phantom{'}}3^{\phantom{'}}\\[-.1cm]  2'3'1'\end{array}\quad
\begin{array}{l} 1^{\phantom{'}}2^{\phantom{'}}3^{\phantom{'}}\\[-.1cm]  3'1'2'\end{array}\quad
\begin{array}{l} 1^{\phantom{'}}2^{\phantom{'}}3^{\phantom{'}}\\[-.1cm]  3'2'1'\end{array}
\eeq
where the $'$ alphabet is $\{1'\rightarrow 4, \ 2'\rightarrow 5, \ 3'\rightarrow 6\}$.

Generating Wick contractions for a propagator structure $\mathcal{P}=\{ b_{(12)},\ldots b_{(ij)},\ldots \}$ 
is slightly more involved but the logic is similar to the example above. 
First organise  $\mathcal{P}=\{ b_{(12)},\ldots b_{(ij)},\ldots \}$ as a matrix of the following form
\beq
\left[
\begin{array}{cccccccccccc} 
               			 p_1&    b_{(12)}     & \dots & {\color{red} b_{(1i)}	}	     & &    \ldots             &     \ldots                & \ldots              &  {\color{red} b_{(1j)}  }  &\ldots &&\ldots \\
			                &         \ddots   & & \\
                                 	        &  		      & & {\color{red} p_i }   &&     \ldots             &     \ldots                 &   \ldots           &{\color{red} b_{(ij)} }  & \ldots  &   &\vdots \\[.2cm]
                                          & &                     &                             & &  { p_{i+1}}         &    b_{(i+1,i+2)}       &                       &   \vdots        &  & & \vdots \\
 				        & &                  & &                             &  &  { \ddots}          &                              &                       &                        & &    \vdots \\
				        & &                  &  &                            &  &  		            &         { p_{j-1}}       & { b_{(j-1,j)}     }  &                        & & \vdots \\
                                          & &                 & &&                           &  &                         &  {\color{red} p_j }   & b_{(jj+1)}                                 & &   \vdots \\ 
                                          & &    & &             &                 &  &                                   &                              & { {} p_{j+1} }             &   &   \vdots \\
                                          & &    & &              &                &  &                                   &                               &    &    { {} \ddots  }         & \vdots \\
                                           & &   & &              &                &  &                                   &                               &  & &                                          { {} p_{n}} \\
\end{array}
\right]
\eeq 
 Conservation of charge at the blob $p_i$ means  
 \begin{align}
 p_i= \sum_{k=1}^{i-1} b_{(ki)} + \sum_{k=i+1}^{n} b_{(ik)}
 \end{align}
pictorially the r.h.s is the sum over all the numbers on the hook having $p_i$ at the corner. 
 
 We will start enumerating the Wick contractions going along the rows of the matrix, 
 starting from the $b_{(12)}$.  Upon visiting a $b_{(ij)}$, we define the \emph{updated} 
 values of $p_i$ and $p_j$ given by
\bea
p_{[ij]}^{}&\equiv& 
 - \underbracket{ \ \ \sum_{k=1}^{i-1} b_{(ki)} }_{  \substack{ {\rm \#\ on\ the\ rows\ above \ } (i,i) } } \ +\ p_i\ \
 - \underbracket{\ \ \sum_{k=i+1}^{j-1} b_{(ik)} }_{\substack{ {\rm \#\ on\ the\ cln\ before\ }(i,j)  } }\\
p_{[ji]}^{}&\equiv&  - \underbracket{ \ \ \sum_{k=1}^{i-1} b_{(kj)} }_{  \substack{ {\rm \#\ on\ the\ rows\ above \ } (i,i)} }\  +\  p_j
\eea
The idea is that when we visit an entry $(ij)$ we assign $b_{ij}$ of the free legs of $p_{[ij]}$. These are legs out, 
and we link them with free legs of $p_{[ji]}$. The latter are legs in. 
Thus changing row index we consider legs in, and changing column index we consider legs out.  

%
Define now the Wick contractions $\mathcal{W}[b_{(ij)}]$ for the bridges going from blob 
$i$ with $p^{}_{[ij]}$ legs-out, reaching blob $j$ with $p^{}_{[ji]}$ legs-in. These are enumerated by
\beq
\mathcal{W}[ b_{(ij)} ] =  \mathscr{C}( b_{(ij)}\subseteq p^{}_{[ij]})  \otimes \mathscr{D}( b_{(ij)}\subseteq p^{}_{[ji]} )
\eeq
The dimension of  $\mathcal{W}[ b_{(ij)} ]$  is simply
\beq
| \mathcal{W}[ b_{(ij)} ] | = |\mathscr{C}( b_{(ij)}\subseteq p^{}_{[ij]}) | \times |\mathscr{D}( b_{(ij)}\subseteq p^{}_{[ji]} )|
\eeq
%
The set of all possible Wick contractions for $\mathcal{P}=\{ b_{(12)},b_{(13)},\ldots \}$ is given by 
\beq
\mathcal{W}[\mathcal{P}]=\otimes_{1\leq i<j\leq n} \mathcal{W}[b_{(ij)}]
\eeq
with dimension $|\mathcal{W}[\mathcal{P}]| =\prod_{i<j} |\mathcal{W}[b_{(ij)}]|$. 
By using the explicit formulas,  
\begin{align}
|\mathscr{C}(k\subseteq n)|= \frac{n!}{(n-k)!k!}\qquad;\qquad
|\mathscr{D}(k\subseteq n)|= \frac{n!}{(n-k)!}
\end{align}
we find
\begin{align}\label{Wick_contate}
|\mathcal{W}[\mathcal{P}]|=
\prod_{1 \leq i<j\leq n} \ \frac{1}{b_{ij}!} \left( -\sum_{k=1}^{i-1} b_{ki}+p_i - \sum_{k=i+1}^j b_{ik}+1\right)_{\!\! b_{ij} }\!\!\!\times 
\left( p_j-\sum_{k=1}^{i} b_{kj}+1\right)_{  \!\! b_{ij} }
\end{align}
where $(a)_b$ is the Pochhammer symbol.




Consider some explicit examples to fix ideas. 

At three points we find that
\begin{align}
|\mathcal{W}[\langle \cO_p\cO_q\cO_r\rangle]|= \frac{p!q!r!}{ (\frac{q+r-p}{2})! (\frac{p+r-q}{2})!  (\frac{p+q-r}{2})!  }
\end{align}
which is indeed fully symmetric. 

At four points we find
\beq
\begin{array}{rl}
{\tt Legs\_out}=&\overbrace{ \rule{0pt}{.3cm} c_1\ldots c_{d_{(12)}} }^{\substack{ \mathscr{C}(b_{(12)}\subseteq p_{[12]}) \\ \cup} } 
			     \overbrace{ \rule{0pt}{.3cm} c_1\ldots c_{d_{(13)}} }^{\substack{ \mathscr{C}(b_{(13)}\subseteq p_{[13]}) \\ \cup} }  
			     \overbrace{  \rule{0pt}{.3cm} c_1 \ldots c_{d_{(14)}} }^{\substack{ \mathscr{C}(b_{(14)}\subseteq p_{[14]}) \\ \cup}  } 
			     \overbrace{\color{red} \rule{0pt}{.3cm} c_{1} \ldots c_{d_{(23)}}}^{\substack{ \mathscr{C}(b_{(23)}\subseteq p_{[23]}) \\ \cup} }
			     \overbrace{ \color{red} \rule{0pt}{.3cm} c_{1} \ldots c_{d_{(24)}}}^{\substack{ \mathscr{C}(b_{(24)}\subseteq p_{[24]}) \\ \cup} }
			     \overbrace{ \color{blue}  \rule{0pt}{.3cm} c_{1} \ldots c_{d_{(34)}}}^{\substack{ \mathscr{C}(b_{(34)}\subseteq p_{[34]}) \\ \cup} }  \\[.2cm]
{\tt Legs\_in}\,=& \underbrace{ \color{red} d_1\ldots d_{d_{(12)}} }_{\substack{ \cap\\ \mathscr{D}( b_{(12)}\subseteq p_{[21]}) } }
			   \underbrace{ \color{blue}  d_1\ldots d_{d_{(13)}} }_{\substack{ \cap\\ \mathscr{D}( b_{(13)}\subseteq p_{[31]}) } } 
			   \underbrace{\color{olive} d_1\ldots d_{d_{(14)}} }_{\substack{ \cap\\ \mathscr{D}( b_{(14)}\subseteq p_{[41]}) } } 
			   \underbrace{ \color{blue}  d_{1} \ldots d_{d_{(23)}}}_{\substack{ \cap\\ \mathscr{D}(b_{(23)}\subseteq p_{[32]})} }
			   \underbrace{ \color{olive} d_1 \ldots d_{d_{(24)}} }_{\substack{ \cap\\ \mathscr{D}( b_{(24)}\subseteq p_{[42]}) } }
			   \underbrace{ \color{olive} d_1 \ldots d_{d_{(34)}} }_{\substack{ \cap\\ \mathscr{D}( b_{(34)}\subseteq p_{[43]}) } }
\\
\end{array}
\quad\nonumber
\eeq
%
%

Combinations $c_i$ and dispositions $d_i$ generated so far, do not refer to the 
specific numbering of the legs of the correlator. What we have done so far has been to enumerate Wick contractions. 
The final task is to assign them to the indexes from $1$ to  $\sum_{i=1}^n p_i$.
In the example above the color-code denotes which subset of numbers belong to the same blob.

\subsection{Digits}

Let us introduce the characteristic vector of a combination in $\mathscr{C}(k\subseteq n)$. 
This is a digit of $n$ bits, i.e.\! $0$ or $1$, with $k$ values of $1$ distributed in the digit. 
The positions of the $1$ give the combination, for example
\beq
\{ 1,3,5\}\in \mathscr{C}(3\subseteq 5)\qquad \longleftrightarrow \qquad [10101]
\eeq

It is simple to think of the combinations in this way, since the only operation one has to 
do is to sequenciate the $1$s to the right or the $0$s to the left, starting from the 
configuration $[111100\ldots]$ with $k$ values of $1$s at the beginning. 
The integer corresponding to a given combination is computed by the Gosper's 
hack in {\tt C++}.\footnote{See for example \url{https://en.wikipedia.org/wiki/Combinatorial_number_system }}
This is the fastest way to do it, and it works well for low charges (since there is an 
obvious bound provided by the number of bits of the machine.) 

The idea of sequenciating is useful to generate dispositions as well. 
Recall that $\mathscr{D}(k\subseteq n)$ is obtained by acting with the 
symmetry group $S_k$ on each combination in $\mathscr{C}(k\subseteq n)$.
So we pick a $k-$combination $C=\{c_1,\ldots c_k\}$. (It is ordered, but 
otherwise it is not necessary). Starting form the subset of length one 
${C}_1=\{c_1\}$ we generate the set
\begin{align}
{C}_2=\{ \{c_1,0\},\{0,c_1\}\}/.\{ 0 \rightarrow c_2 \}
\end{align} 
Then for each element in ${ C}_2$ we generate all sets obtained by sequenciating an extra $0$. For example, 
\begin{align}
{C}_{3,1}=\{ \{c_1,c_2,0\},   \{c_1,0,c_2\},  \{0,c_1,c_2\}\}/.\{0 \rightarrow c_3\} \\
{C}_{3,2}=\{ \{c_2,c_1,0\},   \{c_2,0,c_1\},  \{0,c_2,c_1\}\}/.\{0 \rightarrow c_3\}
\end{align}
and so on. This recursive procedure on the length of the subsets in $C$ gives all the dispositions of $C$. 

For the Wick contractions, there are no repetitions in the $\{c_i\}$,  
since they come from the combinations. 

Instead, for generating the sequences $|d_1,\ldots d_{n-1}\rangle$, which 
we encountered in section \ref{NEnpoint}, it is useful to consider cases with 
repetitions. Indeed, the possible values of $|d_1,\ldots d_{n-1}\rangle$ 
can be obtained by taking all partitions of $2(n-2)=\sum_{i=1}^{n-1} d_i$, 
because of the constraint on the tree, and then by considering all dispositions 
of these partitions. The parts can have repetitions. 

The minor modification of the algorithm above is the following. The idea is to collect the 
repetitions in the original sequence as vectors $\vec{c}_i$ with $\vec{c}_i\neq \vec{c}_j$ if $i\neq j$. 
The recursive procedure will now work upon replacing the $0$ with these vectors 
at each step. For example, start from $C=\{ c_1 c_1,c_2c_2,c_3\ldots \}$.  Then we generate
\begin{align}
{C}_2=\{ \{c_1,c_1,0\},\{c_1,0,c_1\}, \{0,c_1,c_1\}\}/.\{ 0 \rightarrow \vec{c}_2 \}
\end{align} 
We go on with
\begin{align}
&
{C}_{3,1}=\{ 
\{c_1,c_1,c_2, c_2,0\},   \{c_1,c_1,c_2, 0,c_2\},  \notag \\[.2cm]
&
\rule{1.5cm}{0pt} \{c_1,c_1,0,c_2 ,c_2\} ,  \{c_1,0,c_1,c_2 ,c_2\}, \{0,c_1,c_1,c_2, c_2\} \}/.\{0 \rightarrow c_3\}  \notag\\
&
\ \vdots 
\end{align}

\subsection{Pr\"{u}fer sequences and Trees}\label{app_Prufer}

A unique way to label a tree is to use a Pr\"{u}fer sequence $s=(s_1\ldots )$, 
constructed as follows \cite{Prufer}. Consider the tree made of points at positions  
${1,\ldots n-1}$, each point with $d_i$ legs attached, as described above. For the example 
in \eqref{example_tree1}, this is
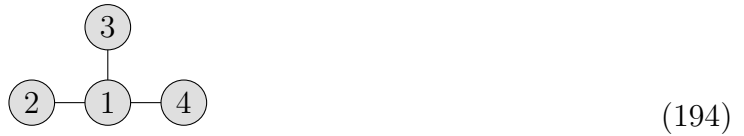
\be
\begin{tikzpicture}
		\def\lato 	{1}
		
		\def\xuno	{-2}
		\def\yuno	{0}
		
		\def\xdue {\xuno+\lato}
		\def\ydue {\yuno}

		\def\xtre {\xdue}
		\def\ytre {\yuno+\lato}
		
		\def\xqua {\xdue+\lato}
		\def\yqua {\yuno}	
		
		\def\rad {.3cm}
		\def\color{gray!25}
		
		\draw[thin]  (\xuno,\yuno)-- (\xdue,\ydue);
		\draw[thin]  (\xdue,\ydue)--(\xtre,\ytre);
		\draw[thin]  (\xdue,\ydue)--(\xqua,\yqua);		
		
		
		\draw[fill=\color,draw=black] (\xuno,\yuno) circle (\rad);
		\draw[fill=\color,draw=black] (\xdue,\ydue) circle (\rad);
		\draw[fill=\color,draw=black] (\xtre,\ytre) circle (\rad);
		\draw[fill=\color,draw=black] (\xqua,\yqua) circle (\rad);	
		
		
		\draw (\xuno,\yuno) 				 node[] 							{$2$};
		\draw (\xdue,\ydue) 				 node[] 							{$1$};
		\draw (\xtre,\ytre) 				 node[] 							{$3$};
		\draw (\xqua,\yqua) 				 node[] 							{$4$};

		\draw (\xqua+.4*\lato,\yqua)		 node[] 							{$$};
\end{tikzpicture}		
\ee
We define a~\emph{ leaf }in the tree as a pair of positions $\ell_1-\ell_2$ in 
which $\ell_1$ is connected to $\ell_2$ with one and only one bridge. At step $k$ of 
the Pr\"{u}fer algorithm, remove the leaf $\ell_1-\ell_2$ with the \emph{smallest} labelled 
position ${\ell_1}$ and assign $s_k=\ell_2$ to the sequence. 
Then, write 
\begin{align}
|\mathcal{W}[\ell_1-\ell_2]|=(q_{\ell_1}-d_{\ell_1}+1)\times (q_{\ell_2}-\#(\ell_2))
\end{align} 
to count the Wick contractions corresponding to that leaf, where $\#(\ell_2)$ 
is the number of times $\ell_2$ appeared in the sequence 
previously (step $<k$). We stop when there is only one leaf left. 
%
Here below we give another example, (rich enough to illustrate the various steps)

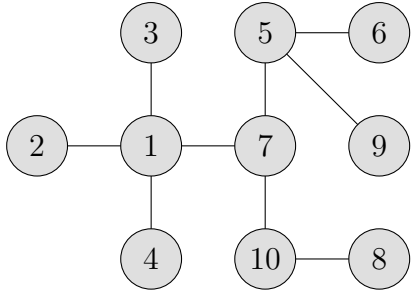
\be
\begin{tikzpicture}
		\def\lato 	{1.5}
		
		\def\xuno	{-2}
		\def\yuno	{0}
		
		\def\xdue {\xuno+\lato}
		\def\ydue {\yuno}

		\def\xtre {\xdue}
		\def\ytre {\yuno+\lato}
		
		\def\xqua {\xdue}
		\def\yqua {\yuno-\lato}	
		
		\def\xcin {\xdue+\lato}
		\def\ycin {\yuno}

		\def\xsei {\xcin}
		\def\ysei{\yuno+\lato}	
		
		\def\xset {\xsei+\lato}
		\def\yset {\yuno+\lato}			
		
		\def\xott {\xsei+\lato}
		\def\yott {\ycin}
		
		\def\xnov {\xcin}
		\def\ynov {\ycin-\lato}
		
		\def\xdie {\xsei+\lato}
		\def\ydie {\ycin-\lato}
		
		\def\rad {.4cm}
		\def\color{gray!25}

		\draw[thin]  (\xuno,\yuno)-- (\xdue,\ydue);
		\draw[thin]  (\xdue,\ydue)--(\xtre,\ytre);
		\draw[thin]  (\xdue,\ydue)--(\xqua,\yqua);
		\draw[thin]  (\xdue,\ydue)--(\xcin,\ycin);
		\draw[thin]  (\xcin,\ycin)--(\xsei,\ysei);
		\draw[thin]  (\xsei,\ysei)--(\xset,\yset);
		\draw[thin] (\xsei,\ysei)-- (\xott,\yott);
		\draw[thin] (\xcin,\ycin)-- (\xnov,\ynov);
		\draw[thin] (\xnov,\ynov)-- (\xdie,\ydie);


		\draw[fill=\color,draw=black] (\xuno,\yuno) circle (\rad);
		\draw[fill=\color,draw=black] (\xdue,\ydue) circle (\rad);
		\draw[fill=\color,draw=black] (\xtre,\ytre) circle (\rad);
		\draw[fill=\color,draw=black] (\xqua,\yqua) circle (\rad);
		\draw[fill=\color,draw=black] (\xcin,\ycin) circle (\rad);
		\draw[fill=\color,draw=black] (\xsei,\ysei) circle (\rad);
		\draw[fill=\color,draw=black] (\xset,\yset) circle (\rad);	
		\draw[fill=\color,draw=black] (\xott,\yott) circle (\rad);	
		\draw[fill=\color,draw=black] (\xnov,\ynov) circle (\rad);
		\draw[fill=\color,draw=black] (\xdie,\ydie) circle (\rad);

		\draw (\xuno,\yuno) 				 node[] 							{$2$};
		\draw (\xdue,\ydue) 				 node[] 							{$1$};
		\draw (\xtre,\ytre) 				 node[] 							{$3$};
		\draw (\xqua,\yqua) 				 node[] 							{$4$};
		\draw (\xcin,\ycin) 				 node[] 							{$7$};
		\draw (\xsei,\ysei) 				 node[] 							{$5$};
		\draw (\xset,\yset) 				 node[] 							{$6$};		
		\draw (\xott,\yott) 				 node[] 							{$9$};		
		\draw (\xnov,\ynov) 				 node[] 							{$10$};	
		\draw (\xdie,\ydie) 				 node[] 							{$8$};

		\draw[]	(\xdie+4.5*\lato,\ytre-\lato)  node[left,font=\footnotesize] {$
									\begin{array}{ll} 
									s_1=1 & (q_1) (q_2)\\[.1cm]
									s_2=1 & (q_3) (q_1-1)\\[.1cm]
									s_3=1 & (q_4)(q_1-2)\\[.1cm]
									s_4=7 & (q_1-3)(q_7)\\[.1cm]
									s_5=5 & (q_6)(q_5)\\[.1cm]
									s_5=10 & (q_8)(q_{10})\\[.1cm]
									s_6=5 & (q_9)(q_5-1)\\[.1cm]
									s_6=7 & (q_5-2)(q_7-1)\\[.1cm]
									s_7=10 & (q_7-2)(q_{10}-1)
									\end{array}$};
		
\end{tikzpicture}
\ee

The nice feature of the Pr\"{u}fer algorithm is that it reduces the pairwise computation 
of Wick contractions, which follows from our results in \eqref{Wick_contate},
\begin{align}
|\mathcal{W}[\mathcal{T}]|
&=\prod_{\substack{ i<j \\ b_{ij}=1}} \ \left( q_i-d_i+ \sum_{k=j}^{n-1} b_{ik}\right)\times 
\left( q_j-\sum_{k=1}^{i-1} b_{kj}\right)
\label{Wick_tree1}
\end{align}
(where we used conservation of charge on the tree)
%
to a sequence. In particular, it gives a 
more efficient ordering of the $b_{ij}$ to count  $|\mathcal{W}[\mathcal{T}]|$.
For instance, the pairs $(ij)=\{ (56), (57), (59)\}$ in the example above would count differently,
 even though the total result does not change.
Thus we can use the Pr\"{u}fer algorithm to rearrange the counting.

In the Pr\"{u}fer algorithm it is clear that 
the first time two operators $i$ and $j$ appear in the sequence, 
they count with $q_i q_j$, because necessarily $d_i=1$ or $d_j=1$. The second time one 
of this operators appears again, it counts with $q_i-1$ or $q_j-1$, and so on. The total number 
of Wick contractions is then,
\begin{align}
|\mathcal{W}[\mathcal{T}]|=\prod_{i=1}^{n-1}  q_i (q_i-1) \dots (q_i-d_i+1)
\end{align}
which is the result we quoted in \eqref{tree_wick_counting}.

\section{Low charge examples for multipoint orthogonality}\label{app_determinant_multipoint}

In this section we reconsider the idea of phrasing the multi-point orthogonality theorem as an identity of the form $det=0$, 
and we discuss some examples to have an idea about how the combinatorics works. 
Let us recall our setup of section \ref{section_3}. We want to study, 
\begin{align}\label{corr_F}
\mathcal{F}_{p|\underline{q_1}\ldots \underline{q_{n-1}} }(x,x_1\ldots x_{n-1})=
						\langle \cO_{p}(x) T_{\underline{q_1}}(x_1) \dots T_{\underline{q_{n-1}}}(x_{n-1})  \rangle 
\end{align}
where $\underline{q_i}$ can be a partition of $q_i$ and $n\ge3$. 
We can use the arrangement of $\cO_{p}(x)$ in \eqref{det_rep} as a determinant, 
to rewrite our correlator as another determinant 
\be\label{det_correlator_F}
\mathcal{F}_{p|\underline{q_1},\ldots ,\underline{q_{n-1}} }=  \frac{1}{\mathcal{N}_p} \det\left( \begin{array}{llcl}  
\mathcal{C}_{\lambda_1| {\lambda_1} }			& 	\mathcal{C}_{\lambda_2| {\lambda_1} }				& \ldots 		\ \	&\mathcal{C}_{ {p}|\lambda_1 }  \\ 
		\vdots  							& 	\vdots										& \ldots 		\ \	& \vdots \\  
\mathcal{C}_{{\lambda_1}|\lambda_{P-1}  }		&  	\mathcal{C}_{\lambda_{2}| {\lambda_{P-1}} }			& \ldots 		\ \	&   \mathcal{C}_{p|\lambda_{P-1} } \\[.2cm]
\hline
\rule{0pt}{.5cm}
\mathcal{C}_{\lambda_1|	\underline{q_1}\ldots \underline{q_{n-1}}}		
										& 	\mathcal{C}_{\lambda_2|	\underline{q_1}\ldots \underline{q_{n-1}}}		
																						& \ldots	\ \ & \mathcal{C}_{p|	\underline{q_1}\ldots \underline{q_{n-1}}}	\end{array}
\right)   
\ee
Notice that differently from $\mathcal{F}_{p|\underline{q_1}\ldots \underline{q_{n-1}} }$, 
the (color factor of the) correlator $\mathcal{C}_{\lambda_i |	\underline{q_1}\ldots \underline{q_{n-1}}}$ 
now involves only traces, 
\begin{align}
\mathcal{C}_{\lambda_i |	\underline{q_1},\ldots ,\underline{q_{n-1}}}
=\langle T_{\lambda_i}(x) T_{\underline{q_1}}(x_1) \dots T_{\underline{q_{n-1}}}(x_{n-1})  \rangle 
\end{align}


As we noticed in the main text of section \ref{section_3}, a feature of writing the correlator 
as a determinant is that if a propagator structure is  such that the rows of the matrix 
are not independent, the determinant will vanish. The two-point functions from which 
we started our story are an obvious example, but also a trivial multipoint generalisation of that.  
Take the list of $\underline{q_1},\ldots ,\underline{q_{n-1}}$ 
to coincide with a partition of $p$, say $\lambda_I$. Then,  passing to the double line notation,  
it is clear that $\mathcal{C}_{\lambda_j |	\underline{q_1},\ldots ,\underline{q_{n-1}}}$ 
has the same color factor of the corresponding two-point function regardless of the number of points.

More generally, the property we want to assign to the propagator structure in such 
a way that the determinant vanishes is that
\begin{align}\label{cond_det0}
\mathcal{C}_{\lambda_j |	\underline{q_1}, \ldots, \underline{q_{n-1}}} = \sum_{I=1,\ldots P-1} K_I\, \mathcal{C}_{\lambda_j |	\lambda_I }\qquad j=1,2\ldots  P
\end{align}
for some $K_I$. Namely, the vector made by  
$\mathcal{C}_{\lambda_j |	\underline{q_1}, \ldots , \underline{q_{n-1}}}$, 
is a linear combination of the rows in \eqref{det_correlator_F}.
This translates into a requirement about the topology of the diagram, 
since \emph{all partitions $\lambda_{I=1,\ldots P-1}$} involved in  \eqref{cond_det0} 
have at least length two, i.e.\! $\lambda_P=\{p\}$ is excluded, 
and there cannot be self-contractions within $T_{\lambda_I}$ in the two point 
functions $ \mathcal{C}_{\lambda_j |	\lambda_I }$.  We conclude that a necessary 
condition for $\mathcal{F}_{p|\underline{q_1},\ldots ,\underline{q_{n-1}} }$ to vanish is that
the diagram disconnects as soon as we remove $\cO_p$.

The reasoning above leaves a free parameter. In fact, the assignment 
$\underline{q_1},\ldots ,\underline{q_{n-1}}$ can be such that 
\begin{align}
	  \tfrac12 \left(-p + \sum_{i=1}^{n-1} q_i\right)=k\ge 0
\end{align}
and yet the diagram disconnects as soon as we remove $\cO_p$. The value 
of $k$ measures the excess of $\sum q_i$ to be a partition of $p$. This is possible 
precisely because  differently from a two-point function, in a multipoint function 
there can be $k\ge 0$ Wick contractions  distributed among the 
$T_{\underline{q_1}}(x_1) \dots T_{\underline{q_{n-1}}}(x_{n-1})$, 
such that when $\cO_p$ is removed, the diagram still disconnects.

\subsubsection*{Example (I)}

Consider the $SU(N)$ single-particle operator $\cO_4$, which is the SPO with 
the simplest admixture,  $T_{\{22\}}$ and $T_4$, and take the four-point function 
$\langle \cO_4(x) T_2(x_1) T_2(x_2) T_2(x_3)\rangle$. 
This is an example with $k=1$, and there is only one possible propagator structure, 
        \be
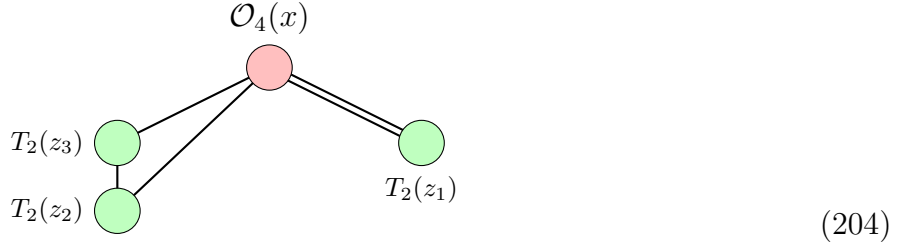
             \begin{tikzpicture}
		\def\lato 	{2}
		
		\def\xuno	{-.5}
		\def\yuno	{2}
		
		\def\xdue {\xuno+\lato}
		\def\ydue {1}
		
		\def\xtre {\xuno-\lato}
		\def\ytre {1}

		\def\xquat{\xuno-\lato}
		\def\yquat {0.1}


		\draw[thick] (\xuno,\yuno+.05)  --  (\xdue,\ydue+.05);
		\draw[thick] (\xuno,\yuno-.05)  --  (\xdue,\ydue-.05);
		\draw[thick] (\xuno,\yuno)  --  (\xtre,\ytre);
		\draw[thick] (\xuno,\yuno)  --  (\xquat,\yquat);
		\draw[thick] (\xtre,\ytre)  --  (\xquat,\yquat);


		\draw[fill=red!25,draw=black] (\xuno,\yuno) circle (.3cm);
		\draw[ fill=green!25,draw=black] (\xdue,\ydue) circle (.3cm);
		\draw[ fill=green!25,draw=black] (\xtre,\ytre) circle (.3cm);
		\draw[ fill=green!25,draw=black] (\xquat,\yquat) circle (.3cm);

		\draw (\xuno,\yuno+.3) 				 node[above] 							{$ \mathcal{O}_{4}(x)$};
		\draw (\xdue,\ydue-.3) 			 	node[below,font=\footnotesize] 				{$T_2(z_1)$};
		\draw (\xquat-.3,\yquat) 				 node[left,font=\footnotesize] 				{$T_2(z_2)$};
		\draw (\xtre-.3,\ytre) 				 	node[left,font=\footnotesize]				{$T_2(z_3)$};

		\draw[]	(\xdue+1.7,\ydue)  node[left] {$\qquad$};

\end{tikzpicture}		
\ee
namely, 
$\mathcal{F}_{4|2,2,2}\simeq g_{x1}^2 g_{x2} g_{x3} g_{23}$. 
In formulas we will now find that $\mathcal{F}_{4\mid 2,2,2}$ vanishes,
\begin{align}
\mathcal{F}_{4|222}\simeq \frac{1}{\mathcal{N}_4} \det\left( \begin{array}{ll} \mathcal{C}_{\{22\}|\{22\}} & \mathcal{C}_{4|\{22\} } \\    
																		\mathcal{C}_{\{22\}|2,2,2} &  \mathcal{C}_{4|2,2,2 } \end{array} \!\!\right)=0
\end{align}
where
\begin{align}	\label{example_4222}																
\begin{array}{l} 		 \mathcal{C}_{\{22\}|\{22\}} = 	8(N^4-1) \\ [.2cm]
					\mathcal{C}_{\{22\}|2,2,2} = 	32(N^4-1) \end{array}\qquad;\qquad
\begin{array}{l}					
					\mathcal{C}_{4|\{22\} } = 8  (N^2-1)(2N^2-3)/N 	\\[.2cm]
					\mathcal{C}_{4|2,2,2 } = 32  (N^2-1)(2N^2-3) /N						
\end{array}						
\end{align}

We see here that  $\mathcal{F}_{4|2,2,2}=0$ is a consequence 
of the four-point functions in the second line of \eqref{example_4222}  being 
proportional by a factor of four to  the two-point functions in the first line.
The figure below provides some more intuition. We draw $\mathcal{F}_{4|2,2,2}$, and 
by looking at a single contribution to the color factor of $\mathcal{C}_{4|2,2,2 }$, 
we illustrate that its double line notation is the same as that of the two-point function 
$\mathcal{C}_{4|\{22\} }$, i.e.\! in double line notation  $T_2(x_1) T_2(x_2) T_2(x_3)$  
``behaves'' like the operator $T_{22}$,

        \be
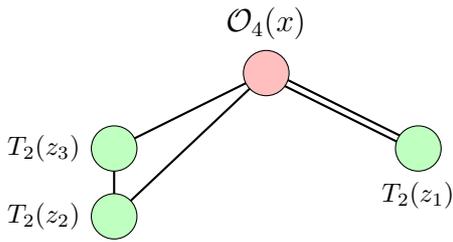
\label{figura_doubleline1}
             \begin{tikzpicture}
		\def\lato 	{2}
		
		\def\xuno	{-.5}
		\def\yuno	{2}
		
		\def\xdue {\xuno+\lato}
		\def\ydue {1}
		
		\def\xtre {\xuno-\lato}
		\def\ytre {1}

		\def\xquat{\xuno-\lato}
		\def\yquat {0.1}


		\draw[thick] (\xuno,\yuno+.05)  --  (\xdue,\ydue+.05);
		\draw[thick] (\xuno,\yuno-.05)  --  (\xdue,\ydue-.05);
		\draw[thick] (\xuno,\yuno)  --  (\xtre,\ytre);
		\draw[thick] (\xuno,\yuno)  --  (\xquat,\yquat);
		\draw[thick] (\xtre,\ytre)  --  (\xquat,\yquat);


		\draw[fill=red!25,draw=black] (\xuno,\yuno) circle (.3cm);
		\draw[ fill=green!25,draw=black] (\xdue,\ydue) circle (.3cm);
		\draw[ fill=green!25,draw=black] (\xtre,\ytre) circle (.3cm);
		\draw[ fill=green!25,draw=black] (\xquat,\yquat) circle (.3cm);

		\draw (\xuno,\yuno+.3) 				 node[above] 							{$ \mathcal{O}_{4}(x)$};
		\draw (\xdue,\ydue-.3) 			 	node[below,font=\footnotesize] 				{$T_2(z_1)$};
		\draw (\xquat-.3,\yquat) 				 node[left,font=\footnotesize] 				{$T_2(z_2)$};
		\draw (\xtre-.3,\ytre) 				 	node[left,font=\footnotesize]				{$T_2(z_3)$};

		
		\def\step 	{0}
		
		\def\xunoA	{5}
		\def\yunoA	{2}
	
		\def\xunoB	{5}
		\def\yunoB	{0}

		\def\xunoC	{6.3}
		\def\yunoC	{0}
		
		\def\xunoD	{8}
		\def\yunoD	{0}

		\def\color{gray!60}
		\def\rad{.05cm}
		\def\spesso{thin}

	
\shadedraw [shading=axis,top color=red!25, draw=white] (\xunoA+2,\yunoA+.25) ellipse (2.5cm and .8cm)	;
\shadedraw [shading=axis, top color=green!25,draw=white ] (\xunoB+.5,\yunoB) circle  (.6cm)	;
\shadedraw [shading=axis, top color=green!25,draw=white ] (\xunoC+.5,\yunoB) circle  (.6cm)	;
\shadedraw [shading=axis, top color=green!25,draw=white ] (\xunoD+.5,\yunoB) circle  (.6cm)	;

				
		\draw[\spesso] (\xunoA,\yunoA) -- (\xunoB,\yunoB);
		\draw[\spesso] (\xunoA+4,\yunoA) -- (\xunoD+1,\yunoD);
		
		\draw[\spesso] (\xunoA,\yunoA) -- (\xunoB,\yunoB);
		\draw[\spesso] (\xunoA+4,\yunoA) -- (\xunoD+1,\yunoD);
		
         	\draw[\spesso] (\xunoB+.7-.2*2,\yunoB+.1) -- (\xunoA+1.1-.2*2,\yunoA);
	     	\draw[\spesso] (\xunoA+1.1,\yunoA) -- (\xunoC+.7,\yunoC+.1);
		
		\draw[\spesso] (\xunoB+1,\yunoB)-- (\xunoC,\yunoC);
		\draw[\spesso] (\xunoB+.7,\yunoB+.125)  --  (\xunoC+.7-.2*2,\yunoC+.125);
		
		\draw[\spesso] (\xunoC+1,\yunoC)--  (\xunoA+2.1-.2*2,\yunoA) ;

		\draw[\spesso] (\xunoA+3.1,\yunoA)  -- 		(\xunoD+.7,\yunoD+.1);
		\draw[\spesso] (\xunoA+3.1-.2*2,\yunoA)  -- 		(\xunoD+.7-.2*2,\yunoD+.1);
		
		\draw[\spesso] (\xunoD,\yunoD)  --  (\xunoA+2.1,\yunoA);

		\draw[\spesso] (\xunoA,\yunoA)  .. controls (\xunoA+.2,\yunoA+1) and (\xunoA+4-.2,\yunoA+1) ..  (\xunoA+4,\yunoA);	
		
		\draw[fill=\color] (\xunoA,\yunoA)  circle (\rad);
		\draw[fill=\color] (\xunoA+4,\yunoA)  circle (\rad);
		
		\draw[\spesso] 	(\xunoA+1.1,\yunoA) arc (0:180: .2);
		\draw[fill=\color] (\xunoA+1.1,\yunoA)  circle (\rad);
		\draw[fill=\color] (\xunoA+1.1-.2*2,\yunoA)  circle (\rad);

		\draw[\spesso]  	 (\xunoA+2.1,\yunoA) arc (0:180: .2);				
		\draw[fill=\color] (\xunoA+2.1,\yunoA)  circle (\rad);
		\draw[fill=\color] (\xunoA+2.1-.2*2,\yunoA)  circle (\rad);
		
		\draw[\spesso]  	(\xunoA+3.1,\yunoA) arc (0:180: .2);
		\draw[fill=\color] (\xunoA+3.1,\yunoA)  circle (\rad);
		\draw[fill=\color] (\xunoA+3.1-.2*2,\yunoA)  circle (\rad);


		\draw[\spesso] (\xunoB,\yunoB)  .. controls (\xunoB+.1,\yunoB-.5) and (\xunoB+.9,\yunoB-.5) ..  (\xunoB+1,\yunoB);	
		\draw[fill=\color]  (\xunoB,\yunoB)  circle (\rad);
		\draw[fill=\color] (\xunoB+1,\yunoB)  circle (\rad);

		\draw[\spesso] 	(\xunoB+.7,\yunoB+.1)  arc (0:-180: .2);
		\draw[fill=\color] (\xunoB+.7,\yunoB+.1)  circle (\rad);
		\draw[fill=\color]	(\xunoB+.7-.2*2,\yunoB+.1)  circle (\rad);

		\draw[\spesso] (\xunoC,\yunoC)  .. controls (\xunoC+.1,\yunoC-.5) and (\xunoC+.9,\yunoC-.5) ..  (\xunoC+1,\yunoC);	
		\draw[fill=\color]  (\xunoC,\yunoC)  circle (\rad);
		\draw[fill=\color] (\xunoC+1,\yunoC)  circle (\rad);
		
		\draw[\spesso] 	(\xunoC+.7,\yunoC+.1)  arc (0:-180: .2);
		\draw[fill=\color] (\xunoC+.7,\yunoC+.1)  circle (.05cm);
		\draw[fill=\color]	(\xunoC+.7-.2*2,\yunoC+.1)  circle (.05cm);

		\draw[\spesso] (\xunoD,\yunoD)  .. controls (\xunoD+.1,\yunoD-.5) and (\xunoD+.9,\yunoD-.5) ..  (\xunoD+1,\yunoD);	
		\draw[fill=\color]  (\xunoD,\yunoD)  circle (\rad);
		\draw[fill=\color] (\xunoD+1,\yunoD)  circle (\rad);
		
		\draw[\spesso] 	(\xunoD+.7,\yunoD+.1)  arc (0:-180: .2);
		\draw[fill=\color] (\xunoD+.7,\yunoD+.1)  circle (.05cm);
		\draw[fill=\color]	(\xunoD+.7-.2*2,\yunoD+.1)  circle (.05cm);

		\draw[fill=\color]	(\xunoD+2.5,\yunoD)  node[left] {$\qquad$};
		\draw[fill=\color]	(\xunoC+.8,\yunoD-1)  node[below, font=\footnotesize] {A contribution to the color factor};
		\draw[fill=\color]	(\xuno,\yunoD-1)  node[below, font=\footnotesize] { $\mathcal{F}_{4\mid 222}\simeq g_{x1}^2 g_{x2} g_{x3} g_{23}$ };

	   \end{tikzpicture}
	\ee

Notice also that the factor of four in \eqref{example_4222} is the number of ways we 
can contract $T_2(x_2)T_2(x_3)$ with $k=1$ Wick contractions. Precisely because 
of this Wick contraction, we can see  in \eqref{figura_doubleline1} that the loop 
counting is the same for both the four- and the two- point functions.

\subsubsection*{Example (II)}

In the next example we consider the four-point function $\langle \cO_4(x) T_2(z_1) T_2(z_2) T_4(z_3)\rangle$, 
which corresponds to a case with $k=2$. There are six propagator structures, but we are 
interested in the one drawn below, since it will have a vanishing color factor,

\be\label{figura_doubleline2}
             \begin{tikzpicture}
		\def\lato 	{2}
		
		\def\xuno	{-0.5}
		\def\yuno	{1.5}
		
		\def\xdue {\xuno+\lato}
		\def\ydue {1.5}
		
		\def\xtre {\xuno-\lato}
		\def\ytre {0.5}

		\def\xquat{\xuno-0.3*\lato}
		\def\yquat {0.5}

		\draw[thick] (\xuno,\yuno+.05)  --  (\xdue,\ydue+.05);
		\draw[thick] (\xuno,\yuno-.05)  --  (\xdue,\ydue-.05);

		\draw[thick] (\xuno,\yuno+.05)  --  (\xtre,\ytre+.05);
		\draw[thick] (\xuno,\yuno-.05)  --  (\xtre,\ytre-.05);
		\draw[thick] (\xtre,\ytre+.05)  --  (\xquat+.05,\yquat+.05);
		\draw[thick] (\xtre,\ytre-.05)  --  (\xquat+.05,\yquat-.05);


		\draw[fill=red!25,draw=black] (\xuno,\yuno) circle (.3cm);
		\draw[ fill=green!25,draw=black] (\xdue,\ydue) circle (.3cm);
		\draw[ fill=green!25,draw=black] (\xtre,\ytre) circle (.3cm);
		\draw[ fill=green!25,draw=black] (\xquat,\yquat) circle (.3cm);

		\draw (\xuno,\yuno+.3) 				 node[above] 							{$ \mathcal{O}_{4}(x)$};
		\draw (\xdue,\ydue-.3) 			 	node[below,font=\footnotesize] 				{$T_2(z_1)$};
		\draw (\xquat-.3,\yquat-.25) 				 node[below,font=\footnotesize] 				{$T_2(z_2)$};
		\draw (\xtre-.3,\ytre) 				 	node[left,font=\footnotesize]				{$T_4(z_3)$};

		\draw (\xuno,\yuno+.3) 			 node[above] {$\mathcal{O}_{4}(x)$};

		\def\step 	{0}
		
		\def\xunoA	{5}
		\def\yunoA	{2}

		\def\xunoC	{6.3}
		\def\yunoC	{0}
			
		\def\xunoBprime{\xunoC-2.25}
		
		\def\xunoB	{5.15}
		\def\yunoB	{0}
		
		\def\xunoD	{7.75}
		\def\yunoD	{0}
		
		\def\xunoE 	{9.5}
		\def\yunoE	{0}

		\def\color{gray!60}
		\def\rad{.05cm}
		\def\spesso{thin}

	
\shadedraw [shading=axis,top color=red!25, draw=white] (\xunoA+2,\yunoA+.25) ellipse (2.5cm and .8cm)	;
\shadedraw [shading=axis, top color=green!25,draw=white ] (\xunoC-.8,\yunoC) ellipse (2cm and .8cm)	;
\shadedraw [shading=axis, top color=green!25,draw=white ] (\xunoE+.5,\yunoE) circle  (.6cm)	;
\shadedraw [shading=axis, top color=green!25,draw=white ] (\xunoD+.5,\yunoB) circle  (.6cm)	;

%
		\draw[\spesso] (\xunoA,\yunoA) 			-- (\xunoC-2.5,\yunoC);
		\draw[\spesso] (\xunoA+1.1-.2*2,\yunoA) 		-- (\xunoBprime+.7-.2*2,\yunoB+.1);
		\draw[\spesso] (\xunoA+1.1,\yunoA) 			-- (\xunoBprime+.7,\yunoB+.1);
		\draw[\spesso] (\xunoA+2.1-.2*2,\yunoA) 		-- (\xunoB+.7-.2*2,\yunoB+.1);
		
		\draw[\spesso] (\xunoC+1,\yunoC)  			-- (\xunoD,\yunoD) ;
		
		\draw[\spesso]  (\xunoB+.7,\yunoB+.1) 		 .. controls (\xunoD-.5,\yunoB+1) and  (\xunoD+.8,\yunoB+.7) .. (\xunoD+1,\yunoD);
		\draw[\spesso]  	(\xunoC+.6-.2*2,\yunoC+.1) 	.. controls (\xunoD-.7,\yunoB+.6) and  (\xunoD+.6,\yunoE+.6) .. (\xunoD+.7,\yunoE);
		\draw[\spesso]  	(\xunoC+.6,\yunoC+.1) 		.. controls (\xunoD-.2*2-.5,\yunoB+.3) and  (\xunoD-.2*2+.6,\yunoE+.4) .. (\xunoD+.7-.2*2,\yunoE);

		\draw[\spesso] (\xunoA+3.1,\yunoA)  -- 		(\xunoE+.7,\yunoE+.1);
		\draw[\spesso] (\xunoA+3.1-.2*2,\yunoA)  -- 		(\xunoE+.7-.2*2,\yunoE+.1);
		\draw[\spesso] (\xunoE,\yunoE)  --  (\xunoA+2.1,\yunoA);
		\draw[\spesso] (\xunoA+4,\yunoA) -- (\xunoE+1,\yunoE);

		\draw[\spesso] (\xunoA,\yunoA)  .. controls (\xunoA+.2,\yunoA+1) and (\xunoA+4-.2,\yunoA+1) ..  (\xunoA+4,\yunoA);	
		
		\draw[fill=\color] (\xunoA,\yunoA)  circle (\rad);
		\draw[fill=\color] (\xunoA+4,\yunoA)  circle (\rad);
		
		\draw[\spesso] 	(\xunoA+1.1,\yunoA) arc (0:180: .2);
		\draw[fill=\color] (\xunoA+1.1,\yunoA)  circle (\rad);
		\draw[fill=\color] (\xunoA+1.1-.2*2,\yunoA)  circle (\rad);

		\draw[\spesso]  	 (\xunoA+2.1,\yunoA) arc (0:180: .2);				
		\draw[fill=\color] (\xunoA+2.1,\yunoA)  circle (\rad);
		\draw[fill=\color] (\xunoA+2.1-.2*2,\yunoA)  circle (\rad);
		
		\draw[\spesso]  	(\xunoA+3.1,\yunoA) arc (0:180: .2);
		\draw[fill=\color] (\xunoA+3.1,\yunoA)  circle (\rad);
		\draw[fill=\color] (\xunoA+3.1-.2*2,\yunoA)  circle (\rad);


		\draw[\spesso] 	(\xunoC-2.5,\yunoC)  .. controls (\xunoC-2.5+.2,\yunoC-1) and (\xunoC+1-.2,\yunoB-1) ..  (\xunoC+1,\yunoC);	
		\draw[fill=\color]   (\xunoC-2.5,\yunoC)  circle (\rad);
		\draw[fill=\color]		(\xunoC+1,\yunoC)  circle (\rad);

		\draw[\spesso] 	(\xunoBprime+.7,\yunoB+.1)  arc (0:-180: .2);
		\draw[fill=\color] (\xunoBprime+.7,\yunoB+.1)  circle (\rad);
		\draw[fill=\color]	(\xunoBprime+.7-.2*2,\yunoB+.1)  circle (\rad);

		\draw[\spesso] 	(\xunoB+.7,\yunoB+.1)  arc (0:-180: .2);
		\draw[fill=\color] (\xunoB+.7,\yunoB+.1)  circle (\rad);
		\draw[fill=\color]	(\xunoB+.7-.2*2,\yunoB+.1)  circle (\rad);

		
		\draw[\spesso] 	(\xunoC+.6,\yunoC+.1)  arc (0:-180: .2);
		\draw[fill=\color] (\xunoC+.6,\yunoC+.1)  circle (.05cm);
		\draw[fill=\color]	(\xunoC+.6-.2*2,\yunoC+.1)  circle (.05cm);

		\draw[\spesso] (\xunoD,\yunoD)  .. controls (\xunoD+.1,\yunoD-.5) and (\xunoD+.9,\yunoD-.5) ..  (\xunoD+1,\yunoD);	
		\draw[fill=\color]  (\xunoD,\yunoD)  circle (\rad);
		\draw[fill=\color] (\xunoD+1,\yunoD)  circle (\rad);
		
		\draw[\spesso] 	(\xunoD+.7,\yunoD)  arc (0:-180: .2);
		\draw[fill=\color] (\xunoD+.7,\yunoD)  circle (.05cm);
		\draw[fill=\color]	(\xunoD+.7-.2*2,\yunoD)  circle (.05cm);

		\draw[\spesso] (\xunoE,\yunoE)  .. controls (\xunoE+.1,\yunoE-.5) and (\xunoE+.9,\yunoE-.5) ..  (\xunoE+1,\yunoE);	
		\draw[fill=\color]  (\xunoE,\yunoE)  circle (\rad);
		\draw[fill=\color] (\xunoE+1,\yunoE)  circle (\rad);
		
		\draw[\spesso] 	(\xunoE+.7,\yunoE+.1)  arc (0:-180: .2);
		\draw[fill=\color] (\xunoE+.7,\yunoE+.1)  circle (.05cm);
		\draw[fill=\color]	(\xunoE+.7-.2*2,\yunoE+.1)  circle (.05cm);

		\draw[fill=\color]	(\xunoD+2.5,\yunoD)  node[left] {$\qquad$};
		\draw[fill=\color]	(\xunoC+.8,\yunoD-1)  node[below, font=\footnotesize] {A contribution to the color factor};
		\draw[fill=\color]	(\xuno,\yunoD-1)  node[below, font=\footnotesize] { $\mathcal{F}_{4\mid422}\simeq g_{x1}^2g_{x3}^2 g^2_{23}$ };

	   \end{tikzpicture}
\ee

By a mechanism similar to the previous example, we will now see that $\mathcal{F}_{4\mid4,2,2}=0$. 
However, this new example shows that the combinatorics in general is subtle/complicated. 

In formulas we find that $\mathcal{F}_{4\mid4,2,2}$ gives
\begin{align}
\!\!\!
\mathcal{F}_{4|4,2,2}\simeq \frac{1}{\mathcal{N}_4} \det\left( \begin{array}{ll} \mathcal{C}_{\{22\}|\{22\}} & \mathcal{C}_{4|\{22\} } \\    
																		\mathcal{C}_{\{22\}|4,2,2} &  \mathcal{C}_{4|4,2,2 } \end{array} \!\!\right)=0
\end{align}
where
\begin{align}	\label{example_2_combinat}																	
\begin{array}{l} 		 \mathcal{C}_{\{22\}|\{22\}} = 	8(N^4-1) \\ [.2cm]
					\mathcal{C}_{\{22\}|4,2,2} = 	32(N^4-1)(2N^2-3)/N \end{array}\quad;\quad
\begin{array}{l}		 
					\mathcal{C}_{4|\{22\} } = 8  (N^2-1)(2N^2-3)/N 	\\[.2cm]
					\mathcal{C}_{4|4,2,2 } = 32  (N^2-1)(2N^2-3)^2 /N^2						
\end{array}						
\end{align}

This time the four-point functions in the second line of \eqref{example_2_combinat} are 
proportional to the two-point functions on the first line by an $N$-dependent factor. 
This feature can be seen in the double line diagram \eqref{figura_doubleline2}, in which 
we can count four loops instead of three, and three was the counting of the previous 
example \eqref{figura_doubleline1}. From the combinatorial point of view is then non 
trivial that $\mathcal{F}_{4|4,2,2}$ vanishes. 

In general, the determinant argument we started with, only leads to a relation between 
rows which eventually can get complicated. It would be very interesting to have a more 
direct combinatorial argument, alternative to the proof we gave in section \ref{proof}.

\section{More N$^3$E correlators at 4-pt}\label{HBPSOPE-app}


Here we give a few more details on the N$^{3}$E four-point functions 
analysed with the half-BPS OPE as in section \ref{NME4pt}. Thus focussing on the following diagrams, 
\begin{align}
\braket{\mathcal{O}_p \mathcal{O}_q \mathcal{O}_r \mathcal{O}_s}_c  =\quad
&\alpha_1\,\,
\vcenter{\hbox{\begin{tikzpicture}[scale=1.4]
\draw[very thick] (1,0) -- (0,0);
\draw [very thick](1,0) -- (0,1);
\draw [very thick](1,0) -- (1,1);
\draw [bend left=20][very thin,gray] (0,0) -- (0,1) ;
\draw [bend right=20][very thin,gray] (0,0) to (0,1);
\draw [very thin,gray] (0,0) -- (1,1) ;
\end{tikzpicture}
}}
+
\alpha_2\,\,
\vcenter{\hbox{\begin{tikzpicture}[scale=1.4]
\draw [very thick](1,0) -- (0,0);
\draw [very thick](1,0) -- (0,1);
\draw [very thick](1,0) -- (1,1);
\draw [bend left=20][very thin,gray] (0,0) -- (0,1) ;
\draw [bend right=20][very thin,gray] (0,0) to (0,1);
\draw [very thin,gray] (0,1) -- (1,1) ;
\end{tikzpicture}
}}
+
\alpha_3\,\,
\vcenter{\hbox{\begin{tikzpicture}[scale=1.4]
\draw [very thick](1,0) -- (0,0);
\draw [very thick](1,0) -- (0,1);
\draw [very thick](1,0) -- (1,1);
\draw [bend right=10][very thin,gray] (0,0) to (1,1) ;
\draw [bend left=10][very thin,gray] (0,0) to (1,1) ;
\draw [very thin,gray] (0,0) to (0,1) ;
\end{tikzpicture}}}
+\alpha_4\,\,
\vcenter{\hbox{\begin{tikzpicture}[scale=1.4]
\draw [very thick](1,0) -- (0,0);
\draw [very thick](1,0) -- (0,1);
\draw [very thick](1,0) -- (1,1);
\draw [bend right=10][very thin,gray] (0,0) to (1,1) ;
\draw [bend left=10][very thin,gray] (0,0) to (1,1) ;
\draw [very thin,gray] (1,1) to (0,1) ;
\end{tikzpicture}
}}
\notag \\[.2cm]
 +
&\alpha_5\,\,
\vcenter{\hbox{\begin{tikzpicture}[scale=1.4]
\draw [very thick](1,0) -- (0,0);
\draw [very thick](1,0) -- (0,1);
\draw [very thick](1,0) -- (1,1);
\draw [bend left=10][very thin,gray] (1,1) to (0,1) ;
\draw [bend right=10][very thin,gray] (1,1) to (0,1);
\draw [very thin,gray] (0,0) -- (0,1) ;
\end{tikzpicture}
}}
+\alpha_6\,\,
\vcenter{\hbox{\begin{tikzpicture}[scale=1.4]
\draw [very thick](1,0) -- (0,0);
\draw [very thick](1,0) -- (0,1);
\draw [very thick](1,0) -- (1,1);
\draw [bend left=10][very thin,gray] (1,1) to (0,1) ;
\draw [bend right=10][very thin,gray] (1,1) to (0,1);
\draw [very thin,gray] (0,0) -- (1,1) ;
\end{tikzpicture}
}}
+
		\alpha_7
\vcenter{\hbox{
		\begin{tikzpicture}[scale=1.4]
\draw [very thick](1,0) -- (0,0);
\draw [very thick](1,0) -- (0,1);
\draw [very thick](1,0) -- (1,1);
\draw [very thin,gray] (0,0) -- (1,1) ;
\draw [very thin,gray] (1,1) -- (0,1) ;
\draw [very thin,gray] (0,0) -- (0,1) ;
\end{tikzpicture}
}}\,.
\end{align}
Recall that the operator $\mathcal{O}_s$ sits at the right corner on the bottom of each square.

\subsubsection*{Example: $\langle \mathcal{O}_3 \mathcal{O}_3 \mathcal{O}_3 \mathcal{O}_3 \rangle$}

In this case crossing symmetry implies 
\be
\alpha_1 = \alpha_2 = \alpha_3 = \alpha_4 = \alpha_5 = \alpha_6
\label{crossing3333}
\ee
The relevant equations, (\ref{twistp+q-4}) and (\ref{twistp+q-2}), respectively, simplify to become
\begin{align}
\alpha_1 &= 18 \frac{\langle \mathcal{O}_3 \mathcal{O}_3 \rangle^2}{\langle \mathcal{O}_2 \mathcal{O}_2 \rangle}\, = 81 \frac{(N^2-1)(N^2-4)^2}{N^2}.\\
2\alpha_1 + \alpha_7 &= 9 \langle \mathcal{O}_3 \mathcal{O}_3 \mathcal{O}_4 \rangle + \frac{\langle \mathcal{O}_3 \mathcal{O}_3 [\mathcal{O}_2 \mathcal{O}_2] \rangle^2}{\langle [\mathcal{O}_2 \mathcal{O}_2] [\mathcal{O}_2 \mathcal{O}_2] \rangle} = 324\frac{(N^2-1)(N^2-4)(N^2-8)}{N^2}\,.
\end{align}
In the latter, which comes from (\ref{twistp+q-2}), we find that only  
$\mathcal{K}=[\mathcal{O}_2 \mathcal{O}_2]$ is relevant, and also that the corresponding
three-point function is simply obtained from the results in (\ref{O2dbltrace3pt}).
Combining the above results we find the remaining coefficient,
\be
\alpha_ 7 = 162 \frac{(N^2-1)(N^2-4)(N^2-12)}{N^2} = 54 \frac{N^2-12}{N}  \langle \mathcal{O}_3 \mathcal{O}_3 \rangle\,.
\ee
Taking coincidence limits then gives
\be
\langle [\mathcal{O}_3 \mathcal{O}_3] \mathcal{O}_3 \mathcal{O}_3 \rangle = \langle [\mathcal{O}_3 \mathcal{O}_3] [\mathcal{O}_3 \mathcal{O}_3] \rangle = 2 \alpha_1 + 2 \langle \mathcal{O}_3 \mathcal{O}_3 \rangle^2 = 18 \frac{(N^2-1)(N^2-4)^2(N^2+8)}{N^2}\,,
\ee
where the $\langle \mathcal{O}_3 \mathcal{O}_3 \rangle^2$ term comes from the disconnected contributions to the four-point function.

\subsubsection*{Example: $\langle \mathcal{O}_4 \mathcal{O}_4 \mathcal{O}_4 \mathcal{O}_6 \rangle$}

This example obeys the maximal crossing symmetry conditions (\ref{crossing3333}). We have
\be
\alpha_1 = 8 \frac{\langle \mathcal{O}_4 \mathcal{O}_4 \mathcal{O}_4 \rangle \langle \mathcal{O}_6 \mathcal{O}_6 \rangle}{\langle \mathcal{O}_4 \mathcal{O}_4 \rangle}\,
\ee
and
\be
2 \alpha_1 + \alpha_7 = 16 \langle \mathcal{O}_4 \mathcal{O}_6 \mathcal{O}_6 \rangle + \sum_{\underline{t},\underline{t}' \, \vdash 6} \langle \mathcal{O}_4 \mathcal{O}_4 \mathcal{K}_{\underline{t}} \rangle (\tilde{g}^{-1})_{\underline{t} \underline{t}'} \langle \mathcal{K}_{\underline{t}'} \mathcal{O}_4 \mathcal{O}_6 \rangle\,.
\ee
The solution for $\alpha_7$ reads
\be
\alpha_7 = 64 \frac{(2N^4-187N^2+81)}{N(N^2+1)} \langle \mathcal{O}_6 \mathcal{O}_6 \rangle\,.
\ee

For $\langle T_4 T_4 T_4 \mathcal{O}_6 \rangle$ we have
\be
\alpha_7\rightarrow 64 \frac{2N^2-81}{N} \langle \mathcal{O}_6 \mathcal{O}_6 \rangle\,.
\ee

\subsubsection*{Example: $\langle \mathcal{O}_3 \mathcal{O}_3 \mathcal{O}_4 \mathcal{O}_4 \rangle$}

In this case crossing symmetry requires
\be
\alpha_1 = \alpha_2\,, \quad \alpha_3 = \alpha_5\, \quad \alpha_4 = \alpha_6\,,
\label{crossing33rr}
\ee
So we only have four independent coefficients.
From equation (\ref{twistp+q-4}) and its crossing we find
\begin{align}
\alpha_1 &= 24 \frac{\langle \mathcal{O}_3 \mathcal{O}_3 \rangle \langle \mathcal{O}_4 \mathcal{O}_4 \rangle}{\langle \mathcal{O}_2 \mathcal{O}_2 \rangle} \\
\alpha_3 + \alpha_4 &= 9 \frac{\langle \mathcal{O}_3 \mathcal{O}_4 \mathcal{O}_3 \rangle \langle \mathcal{O}_4 \mathcal{O}_4 \rangle}{\langle \mathcal{O}_3 \mathcal{O}_3 \rangle} = 81\frac{\langle \mathcal{O}_4 \mathcal{O}_4 \rangle^2}{\langle \mathcal{O}_3 \mathcal{O}_3 \rangle} 
\end{align}
From equation (\ref{twistp+q-2}) and its crossing we obtain
\begin{align}
2\alpha_3 + \alpha_7 &= 9 \langle \mathcal{O}_4 \mathcal{O}_4 \mathcal{O}_4 \rangle + \frac{\langle \mathcal{O}_3 \mathcal{O}_3 [\mathcal{O}_2 \mathcal{O}_2]\rangle \langle [\mathcal{O}_2 \mathcal{O}_2]\mathcal{O}_4 \mathcal{O}_4 \rangle}{\langle [\mathcal{O}_2 \mathcal{O}_2][\mathcal{O}_2 \mathcal{O}_2]\rangle}\\
\alpha_1 + \alpha_4 +\alpha_7 & = 34 \langle  \mathcal{O}_3 \mathcal{O}_4 \mathcal{O}_5 \rangle + \frac{\langle [\mathcal{O}_2 \mathcal{O}_3] \mathcal{O}_3 \mathcal{O}_4  \rangle^2}{\langle [\mathcal{O}_2 \mathcal{O}_3] [\mathcal{O}_2 \mathcal{O}_3]\rangle}\,.
\end{align}
The two-particle operators entering the above equations are $\mathcal{K} = [\mathcal{O}_2 \mathcal{O}_2]$ and $\mathcal{K} = [\mathcal{O}_2 \mathcal{O}_3]$.
Notice also the appereance of $\langle \mathcal{O}_4 \mathcal{O}_4 \mathcal{O}_4 \rangle$ which is NME three point function computed in section \ref{NME_trepoint}. 
In sum we have four equations and thus we can determine the independent coefficients,
\begin{align}
\alpha_1&= 
36\frac{(N^2-4)}{N}\langle \mathcal{O}_4 \mathcal{O}_4 \rangle = 24\frac{\langle \mathcal{O}_3 \mathcal{O}_3 \rangle\langle \mathcal{O}_4 \mathcal{O}_4 \rangle}{\langle \mathcal{O}_2 \mathcal{O}_2 \rangle}\,, \notag \\
2\alpha_3=\alpha_4 &= 
72 \frac{N(N^2-9)}{N^2+1} \langle \mathcal{O}_4 \mathcal{O}_4 \rangle  = 54 \frac{ \langle \mathcal{O}_4 \mathcal{O}_4 \rangle^2}{ \langle \mathcal{O}_3 \mathcal{O}_3 \rangle}\,, \notag \\
\alpha_7 &= 
72 \frac{N^4-25N^2-6}{N(N^2+1)} \langle \mathcal{O}_4 \mathcal{O}_4 \rangle\,. \notag
\end{align}
Having obtained the four-point correlator we then find the coincidence limits,
\begin{align}
\langle [\mathcal{O}_3 \mathcal{O}_3] \mathcal{O}_4 \mathcal{O}_4\rangle &= 144 \frac{N(N^2-9)}{N^2+1} \langle \mathcal{O}_4 \mathcal{O}_4 \rangle\,. \notag \\
\langle [\mathcal{O}_3 \mathcal{O}_4] \mathcal{O}_3 \mathcal{O}_4\rangle = \langle [\mathcal{O}_3 \mathcal{O}_4] [\mathcal{O}_3 \mathcal{O}_4] \rangle &= 72 \frac{N^4-6N^2-2}{N(N^2+1)} \langle \mathcal{O}_4 \mathcal{O}_4 \rangle + \langle \mathcal{O}_3 \mathcal{O}_3 \rangle \langle \mathcal{O}_4 \mathcal{O}_4 \rangle \,.
\end{align}

If we consider instead $\langle T_3 T_3 T_4 \mathcal{O}_4 \rangle$ we find $\alpha_1$ is unchanged while
\be
2\alpha_3=\alpha_4 \rightarrow 72 \frac{N^2-6}{N} \langle \mathcal{O}_4 \mathcal{O}_4 \rangle\,, \quad  \alpha_7 \rightarrow 72 \frac{N^2-18}{N} \langle \mathcal{O}_4 \mathcal{O}_4 \rangle \,.
\ee

\subsubsection*{Example: $\langle \mathcal{O}_3 \mathcal{O}_3 \mathcal{O}_5 \mathcal{O}_5 \rangle$}

This is the first case where we need to consider more than one possible exchanged multi-particle 
operator in an OPE channel. The crossing symmetry conditions (\ref{crossing33rr}) 
still apply but now from equation (\ref{twistp+q-4}) and crossing we have
\begin{align}
\alpha_1 &= 30 \frac{\langle \mathcal{O}_3 \mathcal{O}_3 \rangle \langle \mathcal{O}_5 \mathcal{O}_5 \rangle}{\langle \mathcal{O}_2 \mathcal{O}_2 \rangle}\,, \notag \\
\alpha_3+\alpha_4 &= \frac{\langle \mathcal{O}_3 \mathcal{O}_4 \mathcal{O}_5 \rangle^2}{\langle \mathcal{O}_4 \mathcal{O}_4 \rangle} = 144 \frac{\langle \mathcal{O}_5 \mathcal{O}_5\rangle^2}{\langle \mathcal{O}_4 \mathcal{O}_4 \rangle} \,.
\end{align}
From equation (\ref{twistp+q-2}) we find,
\be
2\alpha_3 + \alpha_7 = 9\langle \mathcal{O}_4 \mathcal{O}_5 \mathcal{O}_5 \rangle + \frac{\langle \mathcal{O}_3 \mathcal{O}_3 [\mathcal{O}_2 \mathcal{O}_2] \rangle \langle [\mathcal{O}_2 \mathcal{O}_2] \mathcal{O}_5 \mathcal{O}_5 \rangle}{\langle [\mathcal{O}_2 \mathcal{O}_2] [\mathcal{O}_2 \mathcal{O}_2] \rangle }\,,
\ee
Its crossing transformation has more structure:
\be
\alpha_1 + \alpha_4 +\alpha_7= 15 \langle \mathcal{O}_3 \mathcal{O}_5 \mathcal{O}_6 \rangle + \sum_{\underline{t},\underline{t}' \, \vdash 6} \langle \mathcal{O}_3 \mathcal{O}_5 \mathcal{K}_{\underline{t}} \rangle (\tilde{g}^{-1})_{\underline{t} \underline{t}'} \langle \mathcal{K}_{\underline{t}'} \mathcal{O}_3 \mathcal{O}_5 \rangle\,.
\ee
Here $\mathcal{K}$ runs over the set of twist six multiparticle operators 
$\{[\mathcal{O}_2 \mathcal{O}_4],[\mathcal{O}_3 \mathcal{O}_3], [\mathcal{O}_2 \mathcal{O}_2 \mathcal{O}_2]\}$. 
The matrix of their two-point functions and hence its inverse can be determined from results on coincidence limits presented earlier. We find
\be
\langle \mathcal{K}_{\underline{t}} \mathcal{K}_{\underline{t}'} \rangle = 
\begin{pmatrix}
2(N^2+7)\langle \mathcal{O}_4 \mathcal{O}_4 \rangle & 18 \langle \mathcal{O}_4 \mathcal{O}_4 \rangle & 0 \\
18 \langle \mathcal{O}_4 \mathcal{O}_4 \rangle  & 4(N^2+8)\frac{\langle \mathcal{O}_3 \mathcal{O}_3 \rangle^2}{\langle \mathcal{O}_2 \mathcal{O}_2 \rangle} & 48 \langle \mathcal{O}_3 \mathcal{O}_3 \rangle \\
0 & 48 \langle \mathcal{O}_3 \mathcal{O}_3 \rangle & 24(N^2+1)(N^2+3) \langle \mathcal{O}_2 \mathcal{O}_2 \rangle
\end{pmatrix}\,. 
\label{multimat}
\ee
In sum, the above equations determine all the coefficients,
\begin{align}
\alpha_1 & = 30 \frac{\langle \mathcal{O}_3 \mathcal{O}_3 \rangle \langle \mathcal{O}_5 \mathcal{O}_5 \rangle}{\langle \mathcal{O}_2 \mathcal{O}_2 \rangle}\,, \notag \\
3\alpha_3=\alpha_4 &= 108 \frac{\langle \mathcal{O}_5 \mathcal{O}_5 \rangle^2}{\langle \mathcal{O}_4 \mathcal{O}_4 \rangle}\,, \notag \\
\alpha_7 &= 90 \frac{(N^4 - 43 N^2 -72)}{N(N^2+5)} \langle \mathcal{O}_5 \mathcal{O}_5 \rangle\,.
\end{align}

If we consider instead $\langle T_3 T_3 T_5 \mathcal{O}_5 \rangle$ we again find $\alpha_1$ is unchanged while
\be
3\alpha_3=\alpha_4 \rightarrow 135\frac{N^2-8}{N} \langle \mathcal{O}_5 \mathcal{O}_5 \rangle\,, \quad  \alpha_7 \rightarrow 90 \frac{N^2-24}{N} \langle \mathcal{O}_5 \mathcal{O}_5 \rangle \,.
\ee

\subsubsection*{Example: $\langle \mathcal{O}_3 \mathcal{O}_3 \mathcal{O}_6 \mathcal{O}_6 \rangle$}

This example is very similar to the previous one with the crossing conditions (\ref{crossing33rr}) still valid 
and three twist seven multiparticle operators participating in the crossing transformation of equation (\ref{twistp+q-2}). We simply quote the results here,
\begin{align}
\alpha_1 &= 36 \frac{\langle \mathcal{O}_3 \mathcal{O}_3 \rangle \langle \mathcal{O}_6 \mathcal{O}_6 \rangle}{\langle \mathcal{O}_2 \mathcal{O}_2 \rangle}\,, \notag \\
4 \alpha_3 = \alpha_4 &= 180 \frac{\langle \mathcal{O}_6 \mathcal{O}_6 \rangle^2}{\langle \mathcal{O}_5 \mathcal{O}_5 \rangle}\,, \notag \\
\alpha_7 &= 108\frac{(N^6-65N^4-408N^2-80)}{N(N^4+15N^2+8)} \langle \mathcal{O}_6 \mathcal{O}_6 \rangle\,.
\end{align}

If we consider instead $\langle T_3 T_3 T_6 \mathcal{O}_6 \rangle$ we find
\be
4 \alpha_3 = \alpha_4 \rightarrow 216 \frac{N^2-10}{N} \langle \mathcal{O}_6 \mathcal{O}_6 \rangle\,, \quad \alpha_7 \rightarrow 108 \frac{N^2-30}{N} \langle \mathcal{O}_6 \mathcal{O}_6 \rangle\,.
\ee

\subsubsection*{Example: $\langle \mathcal{O}_3 \mathcal{O}_4 \mathcal{O}_4 \mathcal{O}_5 \rangle$}

In this case crossing symmetry implies a different set of conditions,
\be
\alpha_1=\alpha_3\,, \quad \alpha_2=\alpha_4\,, \quad \alpha_5=\alpha_6\,.
\ee
equation (\ref{twistp+q-4}) nd its crossing transformation become
\begin{align}
\alpha_1+\alpha_2 & = 108 \frac{\langle \mathcal{O}_4 \mathcal{O}_4 \rangle \langle \mathcal{O}_5 \mathcal{O}_5 \rangle}{\langle \mathcal{O}_3 \mathcal{O}_3 \rangle}\,
2\alpha_5 &= 12 \frac{\langle \mathcal{O}_4 \mathcal{O}_4 \mathcal{O}_4 \rangle \langle \mathcal{O}_5 \mathcal{O}_5 \rangle}{\langle \mathcal{O}_4 \mathcal{O}_4 \rangle}\,.
\end{align}
Equation (\ref{twistp+q-2}) becomes
\be
\alpha_1+\alpha_5+\alpha_7 = 12 \langle \mathcal{O}_4 \mathcal{O}_5 \mathcal{O}_5 \rangle + \frac{\langle \mathcal{O}_3 \mathcal{O}_4 [\mathcal{O}_2 \mathcal{O}_3] \rangle \langle [\mathcal{O}_2 \mathcal{O}_3] \mathcal{O}_4 \mathcal{O}_5 \rangle}{\langle [\mathcal{O}_2 \mathcal{O}_3] [\mathcal{O}_2 \mathcal{O}_3] \rangle}\,,
\ee
Its crossing involves the two-particle operators
$\{[\mathcal{O}_2 \mathcal{O}_4],[\mathcal{O}_3 \mathcal{O}_3], [\mathcal{O}_2 \mathcal{O}_2 \mathcal{O}_2]\}$. 
These are precisely the same operators of the previous example. Thus
\be
2\alpha_2 + \alpha_7 =15 \langle \mathcal{O}_4 \mathcal{O}_4 \mathcal{O}_6 \rangle + \sum_{\underline{t},\underline{t}' \, \vdash 6} \langle \mathcal{O}_4 \mathcal{O}_4 \mathcal{K}_{\underline{t}} \rangle (\tilde{g}^{-1})_{\underline{t} \underline{t}'} \langle \mathcal{K}_{\underline{t}'} \mathcal{O}_3 \mathcal{O}_5 \rangle\,.
\ee
The solution obtained is
\begin{align}
\alpha_1&=36\frac{\langle \mathcal{O}_4 \mathcal{O}_4 \rangle \langle \mathcal{O}_5 \mathcal{O}_5 \rangle}{\langle \mathcal{O}_3 \mathcal{O}_3 \rangle}\,, \quad \alpha_2 = 72 \frac{\langle \mathcal{O}_4 \mathcal{O}_4 \rangle \langle \mathcal{O}_5 \mathcal{O}_5 \rangle}{\langle \mathcal{O}_3 \mathcal{O}_3 \rangle}\,, \notag \\
\alpha_5&= 6 \frac{\langle \mathcal{O}_4 \mathcal{O}_4 \mathcal{O}_4 \rangle \langle \mathcal{O}_5 \mathcal{O}_5 \rangle}{\langle \mathcal{O}_4 \mathcal{O}_4 \rangle}\,, \quad \alpha_7 = 96 \frac{(N^4-50N^2+9)}{N(N^2+1)} \langle \mathcal{O}_5 \mathcal{O}_5 \rangle\,.
\end{align}

If we consider instead $\langle T_3 T_4 T_4 \mathcal{O}_5 \rangle$ we find
\be
2\alpha_1=\alpha_2\rightarrow 96\frac{N^2-6}{N}\langle \mathcal{O}_5 \mathcal{O}_5 \rangle\,, \quad \alpha_5 \rightarrow 96\frac{N^2-9}{N}\langle \mathcal{O}_5 \mathcal{O}_5 \rangle\,, \quad  \alpha_7 \rightarrow 96\frac{N^2-27}{N} \langle \mathcal{O}_5 \mathcal{O}_5 \rangle\,.
\ee

\subsubsection*{Example: $\langle \mathcal{O}_3 \mathcal{O}_4 \mathcal{O}_5 \mathcal{O}_6 \rangle$}

This is the first example with no crossing symmetry conditions. We have confirmed with explicit Wick contraction computation
that our solution in \eqref{solu_michele1}-\eqref{solu_michele2} is reproduced. 
We quote the coefficient $\alpha_7$
\be
\alpha_7 = 120 \frac{(N^6-82N^4-231N^2+12)}{N(N^2+1)(N^2+5)}\langle \mathcal{O}_6\mathcal{O}_6\rangle
\ee

For $\langle T_3 T_4 T_5 \mathcal{O}_6 \rangle$ we have
\be
\alpha_7 \rightarrow 120 \frac{N^2-36}{N} \langle \mathcal{O}_6 \mathcal{O}_6 \rangle\,.
\ee

	\end{document}